\documentclass[preprint,floatfix] {revtex4} 
\newcommand{\rvec}{\mathrm {\mathbf {r}}} 
 
\usepackage{graphicx}
\usepackage{tikz}
\usepackage{subfigure}
\usepackage{xcolor}
\usepackage{amsmath}
\usepackage{enumerate}
\usepackage{tabularx}
\usepackage{nicefrac}

\usepackage{color, soul}
\definecolor{darkblue}{rgb}{0,0,0.5}
\setulcolor{darkblue}

\begin{document}

\title{Hydrogen-like ions in plasma environment}

\author{Neetik Mukherjee$^{a}$}
\email{pchem.neetik@gmail.com}
%%%\altaffiliation{Email: pchem.neetik@gmail.com}

\author{Chandra N.~Patra$^{b}$}
\email{patra@barc.gov.in}

\author{Amlan K.~Roy$^{a}$}
\altaffiliation{Corresponding author. Email: akroy@iiserkol.ac.in, akroy6k@gmail.com.}

\affiliation{$^{a}$Department of Chemical Sciences, 
IISER Kolkata, Mohanpur-741246, Nadia, WB, India\\
$^{b}$Theoretical Chemistry Section, Chemistry Group, Bhabha Atomic Research Centre, Mumbai-400085, India}

\begin{abstract}
%%1234567890 %%1234567890 %%1234567890 %%1234567890 %%1234567890 %%1234567890 %%1234567890 %%1234567890 %%1234567890 %%1234567890
The behavior of H-like ions embedded in astrophysical plasmas in the form of \emph{dense, strongly and weakly coupled} 
plasmas are investigated. In these, the increase and decrease in temperature is impacted with a change in confinement radius $(r_{c})$. 
Two independent and generalized scaling ideas have been applied to modulate the effect of plasma screening constant ($\lambda$) and charge of ion ($Z$) 
on such systems. Several new relations are derived to interconnect the original Hamiltonian and two scaled Hamiltonians.
In exponential cosine screened Coulomb potential (ECSCP) (dense) and weakly coupled plasma (WCP) these scaling relations have provided 
a linear equation connecting the critical screening constant $(\lambda^{(c)})$ and $Z$. Their ratio offers a state-dependent constant, beyond which, 
a particular state vanishes. Shannon entropy has been employed to understand the plasma effect on the ion. With increase in $\lambda$, the 
accumulation of opposite charge surrounding the ion increases leading to a reduction in number of bound states. However, with rise in ionic charge $Z$, 
this effect can be delayed. The competing effect of plasma charge density ($n_e$) and temperature in WCP and ECSCP is investigated. A recently proposed
simple virial-like theorem has been established for these systems. Multipole ($k=1-4$) oscillator strength (OS) and polarizabilities for these 
are studied considering $1s, 2s$ states. As a bonus, analytical closed-form expressions are derived for $f^{(k)}$ and $\alpha^{(k)} (k=1-4)$ 
involving $1s$ and $2s$ state, for \emph{free H-like ion}. 


\vspace{2mm}
\end{abstract}
\maketitle
\section{Introduction}
The discovery and development of quantum confinement \cite{sen14} has triggered the study of influence of environment on quantum 
systems. In confined condition, rearrangement of orbitals may occur in atoms/molecules, leading to some fascinating changes in physical, chemical properties. 
Especially this leads to an increase in coordination number of atoms \cite{grochala07}, enhanced reactivity of atoms/molecules, room temperature superconductivity
\cite{snider20}, etc. The environment driven confinement has profound application in 
condensed matter, semiconductor physics, astrophysics, nanotechnology etc. In this context, the influence of plasma environment 
\cite{ichimaru82,Weishett89,murillo98} in astrophysical systems is a subject of topical interest. Particularly, the impact of charge cloud and temperature on 
bound quantum states can be determined by investigating atoms and ions trapped inside various plasma environments 
\cite{zhu20,bhattacharyya15,jiao21}.  

In such conditions, the competing effect of plasma free electron density ($n_{e}$) and temperature (T) play a pivotal role in 
stabilizing the bound states of a given system. The plasma coupling parameter $(\Gamma)$ is expressed as \cite{das14};
\begin{equation}\label{eq:1}
\begin{aligned}
\Gamma =\frac{E_{\mathrm{coulomb}}}{E_{\mathrm{thermal}}}=\frac{Q^{2}}{4\pi\epsilon_{0}ak_{b}T}.
\end{aligned}
\end{equation}              
Here, $Q$ denotes the charge on the particle, inner particle separation is given by $a=\left(\frac{3}{4\pi n_{e}}\right)^{\frac{1}{3}}$, 
$k_{b}$ signifies Boltzmann constant, and $n_{e}$ refers to plasma electron density. Depending on the value of $\Gamma$, following 
two situations may be envisaged.  
\begin{enumerate}
\item 
$\Gamma < 1$: This arises for low dense and high temperature or weakly coupled plasma (WCP). The thermal energy is higher 
than coulomb energy in this case.  
\item
$\Gamma > 1$: This occurs for strongly coupled plasma (SCP). They have high density and low temperature. The thermal energy 
is now lower than coulomb energy. This type of plasma has been produced experimentally. 
\end{enumerate}
In hot WCP, the collective screening effect of plasma on the electron-charged particle interaction is assumed 
to behave as Debye-H\"uckel potential, expressed in the form, $V_{1}(r)=-\frac{Z}{r}e^{-\lambda_{1} r}$. Here, 
$\lambda_{1}=\sqrt{\frac{4\pi e^2 n_{e}}{k_{b}T}}$ corresponds to the inverse of Debye radius ($D$). The screening parameter arises 
due to surrounding plasma cloud. In last two decades, this system has been studied vigorously with immense interest. The impact of plasma screening effect 
on energy spectrum \cite{solyu12,paul09a,bahar14,bahar16}, inelastic electron-ion scattering \cite{gutierrez94, yoon96}, two proton transitions \cite{paul08,paul09} and
transition probabilities involving electron-impact excitation \cite{jung95,jung96,song03}, etc., have been investigated. The dynamic plasma screening effect was 
considered in \cite{jung97,qi09b,liu08,liu08a}. The relativistic correction on plasma screening effect was also explored \cite{poszwa15}.  
Various spectroscopic properties including multipole oscillator strength (OS) and static multipole polarizabilities were calculated for 
H-like atoms embedded in WCP \cite{saha02,qi08,qi09,qi09a,bassi12} using several numerical methods. A time-dependent variation perturbation 
method was employed to calculate transition probabilities, OS, static dipole polarizabilities for ground state at different $\lambda_{1}$ 
values \cite{saha02a}. Numerical symplectic integration method \cite{qi08,qi09,qi09a}, mean excitation energy based approximation formula 
\cite{bassi12}, integration based shooting technique \cite{das12}, 
linear variation method \cite{kang13}, etc., were also employed to extract these spectroscopic properties. The hyperpolarizability of H atom under 
spherically confined Debye plasma was reported in \cite{saha11}. Closed form expression for critical screening constant in ground state of WCP was 
proposed in \cite{montgomery18}. Numerical values for ground and low-lying excited states were reported in \cite{stubbins93}. Recently, 
generalized pseudo spectral (GPS) method was used in computing OS and polarizabilities in ground and excited states ($\ell=0$) 
\cite{zhu20}. In all these cases, calculations were mostly concentrated in estimating the dipole OS and polarizabilities considering $1s$ as the 
initial state. However, WCP in a \emph{confined} condition with varying $\lambda_{1}$ has not yet been well explored. This remains one of the 
primary objectives of this communication.                 

The composite screening and wake effect around a slow moving test charge in low density quantum plasma is mimicked by using an exponential cosine 
screened Coulomb potential (ECSCP), having the form, $V_{2}(r)=-\frac{Z}{r}e^{-\lambda_{2} r} \cos{\lambda_{2}r}$. Here, 
$\lambda_{2}=\frac{k_{q}}{\sqrt{2}}=\sqrt{\frac{n_{e}\omega_{pe}}{\hbar}}$ signifies the screening parameter, whereas $k_{q}$ is the electron 
plasma wave number connected to electron plasma frequency and number density. The cosine term in this model is introduced under 
the assumption that the quantum force acting on plasma electrons predominates over statistical pressure of plasmas \cite{jiao21}. A 
variety of theoretical methods like perturbation and variation method \cite{lam72}, Pad\'e scheme \cite{lai82}, shooting method \cite{singh83}, 
SUSY perturbation method \cite{dutt86}, asymptotic iteration \cite{bayrak07}, variation using hydrogenic wave functions \cite{paul11}, 
J-matrix \cite{nasser11}, symplectic integration \cite{qi16}, GPS \cite{roy13}, basis expansion method with  
Slater-type orbitals \cite{lai13}, Laguerre polynomials \cite{lin10}, etc, were employed to extract the eigenvalue and eigenfunctions of this 
system. Similarly, the influence of $\lambda_{2}$ on energy spectrum \cite{bahar14,bahar16}, electron-impact excitation \cite{song03}, 
photoionization cross-section \cite{lin10,lin11}, etc., were discussed in appreciable detail. Relativistic correction to the screening effect was also 
explored. Further, the laser-induced excitation on confined H atom (CHA) in ECSCP was pursued using Bernstein-polynomial method \cite{lumb14}. In this 
context, the impact of shape of laser pulse, $r_{c}, \lambda_{2}$, as well as various laser parameters on the dynamics of the system has been
examined and analyzed. Several attempts were made to estimate the characteristic value of $\lambda_{2}$ at which a bound state designated by  
quantum numbers $n, \ell$ disappears \cite{montgomery18}. The critical screening parameters for $n \le 6$ and $0 \le \ell \le n-\ell$
was accurately estimated in \cite{diaz91}. The dipole OS and polarizabilities at various $\lambda_{2}$ values were reported before 
in \cite{singh83,dutt86,qi16,lai13,roy16}. Recently, the utility of GPS method in ECSCP \cite{jiao21} has been examined by evaluating OS and 
polarizabilities. But here again, barring a few exceptions, majority calculations have focused in ground 
state. Moreover, to the best of our knowledge, ECSCP under a \emph{confined} environment has not been probed so far in a sufficiently thorough manner. 

In SCP \cite{ichimaru82}, an ion experiences plasma effect within the ion sphere radius (R). Thus, no electron 
current moves through the boundary surface. It is generally described by a potential of the form \cite{das12}, 
\begin{equation}
\begin{aligned}    
V_{3}(r)=\begin{cases}
=-\frac{Z}{r}+\left(\frac{Z-N_{e}}{2R}\right)\left[3-\left(\frac{r}{R}\right)^{2}\right] \\
=0 \ \ \ r > r_{c}=R,  
\end{cases}
\end{aligned}
\end{equation}
where $R=\left[\frac{3(Z-N_{e})}{4\pi n_{e}}\right]^{\frac{1}{3}}$. The free electrons in an ion sphere distribute uniformly. This model is 
extensively used and expected to be valid in the limit of low temperature and high density. Several theoretical methods have been employed to 
understand the effect of SCP on energy levels and wave functions of H-like atoms \cite{murillo03,belkhiri15, chen18}. Moreover, atomic 
transition probabilities \cite{salzmann87}, transition energies and polarizabilities \cite{das12}, photoionization and photoionization cross 
section \cite{qi17,das14}, OS and static polarizabilities \cite{sen13}, etc., in this case, were studied previously. However, akin to the 
earlier two cases (WCP and ECSCP), most of the works have been restricted to ground state only.

We have a number of objectives in this article. At first, a detailed investigation is made on the three plasma conditions, \emph{viz.}, WCP, 
ECSCP and SCP, with special emphasis on their confinement situation and excited states, where, literature results are quite scarce. It may be noted 
that, the influence of a physical situation governed by a potential of the form, $V=\infty$, at $r > r_{c}$, in the context of plasma has not been 
considered before. Besides, its significance and relation to the plasma environment is also not very clear. Here the confined condition is mapped with 
\emph{plasma temperature}. It may be noted that, the multipole OS and polarizabilities of H atom in various plasmas have been reported in a number of 
publications. However,
such works in the confined scenario, as implied above, have not been considered before. Thus a secondary objective is to examine the \emph{effect of 
confinement} on multipole OS and polarizabilities for WCP, ECSCP and SCP. Two different scaling ideas connecting $\lambda$ 
and $Z$ are formulated. The relation between these two individual concepts are derived and explained. Additionally, Shannon entropy ($S$) has been 
invoked to determine the critical screening constant in \emph{free} WCP and ECSCP. This has been attempted for the first time and our results 
show this can be an interesting and novel route. Beyond this critical parameter (the binding energy of a given 
state disappears), no bound states could be found. After some debate, It is now a well accepted fact that, the 
standard form of virial theorem (VT) does not ordinarily obey in enclosed conditions. An appropriate modified form is invoked in 
\cite{mukherjee19}, which holds good in both free and confined conditions. The utility and efficiency of this newly derived relation is examined 
in the context of plasma environment. 

Thus we have performed detailed calculations of multipole OS ($k=1-4$) and polarizabilities in $1s,~2s$ states of WCP, ECSCP and SCP employing 
the accurate GPS wave functions. Here $k=1-4$ represent dipole, quadrupole, octupole and hexadecapole transitions respectively. In WCP and ECSCP, 
we have demonstrated the spectroscopic properties in two different ways. At first, these are calculated by varying $\lambda$, keeping $r_{c}$ fixed. 
Secondly, the impact of variation of $r_{c}$ on these properties at fixed $\lambda$ is also verified. Analogous calculations are done in SCP, with 
change in $r_{c}$. As a bonus, some analytical closed-form expressions of multipole OS (up to hexadecapole) and polarizabilities (up to 
hexadecapole) are derived for $1s,~2s$ states of \emph{free H atom} (FHA). In literature, these forms are available in \emph{dipole case only}. The 
article is organized in following parts: Sec.~II presents a brief description about the formalism employed in the current work. In Section III, 
the connection between plasma temperature and quantum confinement is proposed and explained. Section 
IV provides a detailed discussion of the results for WCP, ECSCP and SCP. Finally, we conclude with a few remarks and future prospects, in Sec.~V.
 
\section{Theoretical Formalism}
The time-independent radial Schr\"odinger equation (SE) for the spherically confined plasma system is expressed as (in a.u.), 
\begin{equation}\label{eq:2}
	\left[-\frac{1}{2} \ \frac{d^2}{dr^2} + \frac{\ell (\ell+1)} {2r^2} + V_{c}(r) + V_{0} \ \theta(r-r_{c}) \right] \psi_{n,\ell}(r)=
	\mathcal{E}_{n,\ell}\ \psi_{n,\ell}(r).
\end{equation}
Here $V_{0}$ is a positive number with numerical value approaching $\infty$ and $\theta(r-r_{c})$ is a Heaviside function that 
reaches 1 at $r=r_{c}$, while \emph{zero} otherwise, whereas $V_c(r)$ represents the various plasma potential discussed later in this section.  
To calculate energy and spectroscopic properties, the GPS method has been exploited. Over the time, its accuracy and efficiency in calculating 
various bound-state properties in several central potentials in both free and confined condition have been verified and established 
(see \cite{roy2014,roy2014a,roy2014b,mukherjee19,mukherjee20} and references therein).    

%%%start from here
\subsection{Virial-like Theorem}
Recently a virial-like relation has been proposed for free and confined quantum systems, by invoking the time-independent 
non-relativistic SE and Hypervirial theorem \cite{mukherjee19}. The generalized form of this equation is expressed as,
\begin{equation}\label{eq:3}
\langle \hat{T}^{2} \rangle_{n}-\langle \hat{T} \rangle^{2}_{n} = \langle \hat{V}^{2} \rangle_{n}-\langle \hat{V} \rangle^{2}_{n} 
\end{equation}
It can be used as a necessary condition for an exact quantum system to obey. Further, it has been proved that, the following equation, 
\begin{equation}\label{eq:4}
\begin{aligned}
(\Delta \hat{T}_{n})^{2} & = & \langle \hat{V} \rangle_{n} \langle \hat{T} \rangle_{n}
-\langle \hat{T}\hat{V} \rangle_{n} = (\Delta \hat{V}_{n})^{2} = \langle \hat{T} \rangle_{n} \langle \hat{V} \rangle_{n}
-\langle \hat{V}\hat{T} \rangle_{n},
\end{aligned}
\end{equation}
can act as a sufficient condition for a bound, stationary state \cite{mukherjee19}. Moreover, an alteration in boundary condition does not 
influence the general form. These are applicable in all coordinate systems, such as ellipsoidal, parabolic, cylindrical, spheroidal, etc. 
This also holds good in unconfined and confined systems (including angular confinement). In the present endeavor, this has been extended to 
the plasma environment.  

\subsection{Multipole polarizabilities}
The static multipole polarizabilities can be expressed in following form,
\begin{equation}\label{eq:5}
\alpha^{(k)}_{i}=\alpha^{(k)}_{i}(\mathrm{bound})+\alpha^{k}_{i}(\mathrm{continuum}).
\end{equation} 
It is customary to write $\alpha^{(k)}_{i}$ in terms of compact sum-over states form \cite{das12}. However it can also be directly 
computed by adopting the standard perturbation theory framework \cite{dalgarno62}. In the former procedure, Eq.~(\ref{eq:4}) modifies to,
\begin{equation}\label{eq:6}
\begin{aligned}
\alpha^{(k)}_{i} & = \sum_{n}\frac{f^{(k)}_{ni}}{(\mathcal{E}_{n}-\mathcal{E}_{i})^{2}}
-c\int\frac{|\langle R_{i}|r^{k}Y_{kq}(\rvec)|R_{\epsilon n}\rangle|^{2}}{(\mathcal{E}_{\epsilon n}-\mathcal{E}_{i})} \ \mathrm{d}\epsilon, \\
\alpha^{(k)}_{i}(\mathrm{bound}) & = \sum_{n}\frac{f^{(k)}_{ni}}{(\Delta \mathcal{E}_{ni}}, \ \ \ \ \ 
\alpha^{k}_{i}(\mathrm{continuum}) = c\int\frac{|\langle R_{i}|r^{k}Y_{kq}(\rvec)|R_{\epsilon n}\rangle|^{2}}{(\mathcal{E}_{\epsilon n}-\mathcal{E}_{i})} \  \mathrm{d}\epsilon.
\end{aligned}
\end{equation}  
In Eq.~(\ref{eq:5}), the summation and integral terms represent the bound and continuum contributions respectively, $f^{(k)}_{ni}$ signifies 
the multipole OS ($k$ is a positive integer), $c$ is a constant which depends on $\ell$ quantum number. $f^{(k)}_{ni}$
measures the mean probability of transition between an initial ($i$) to final ($n$) state, which is normally expressed as, 
\begin{equation}\label{eq:7}
f^{(k)}_{ni}=\frac{8\pi}{(2k+1)}\Delta\mathcal{E}_{ni}|\langle r^{k} Y_{kq}(\rvec)\rangle|^{2}. 
\end{equation}    
Designating the initial and final states as $|n \ell m\rangle$ and $|n^{\prime}\ell^{\prime}m^{\prime}\rangle$, one can easily derive,  
\begin{equation}\label{eq:8}
f^{(k)}_{ni}=\frac{8\pi}{(2k+1)} \ \Delta\mathcal{E}_{ni} \ \frac{1}{2\ell+1}\sum_{m}\sum_{m^{\prime}} |\langle n^{\prime}\ell^{\prime}m^{\prime}|r^{k}Y_{kq}(\rvec)|n \ell m \rangle|^{2}.
\end{equation}
The application of Wigner-Eckart theorem and sum rule for \emph{3j} symbol further leads to,  
\begin{equation}\label{eq:9}
f^{(k)}_{ni}=2 \ \frac{(2\ell^{\prime}+1)}{(2k+1)} \ \Delta\mathcal{E}_{ni} \ |\langle r^{k}\rangle^{n^{\prime}\ell^{\prime}}_{n \ell}|^{2} \
\left\{\begin{array}{c} \ell^{\prime} \ \ k \ \ \ell\\ 0 \ \ 0 \ \ 0 \end{array}\right\}^{2}. 
\end{equation}
The transition matrix element is expresses by the radial integral,
\begin{equation}\label{eq:10}
\langle r^{k} \rangle = \int_{0}^{\infty} R_{n^{\prime} \ell^{\prime}}(r) r^{k} R_{n \ell} (r) r^{2} \mathrm{d}r.
\end{equation}
Thus it is clear that $f^{(k)}_{ni}$ depends on $n, \ell$ quantum numbers, while being independent of magnetic quantum number $m$.
In this article, we aim to compute multipole ($k=1-4$) polarizabilities and OS for $1s, 2s$ states. The corresponding selection rule for  
dipole OS ($k=1$) for these two states are ($i=1$ or 2),
\begin{equation}\label{eq:11}
\begin{aligned}
f^{(1)}_{np-is} & = 2 \ \Delta \mathcal{E}_{np-is} |\langle r \rangle^{np}_{is}|^{2} \ \left\{\begin{array}{c} 1 \ \ 1 \ \ 0\\ 0 \ \ 0 \ \ 0 \end{array}\right\}^{2} & = \frac{2}{3} \ 
\Delta \mathcal{E}_{np-is} \  |\langle r \rangle^{np}_{is}|^{2}.
\end{aligned} 
\end{equation}
The quadrupole OS ($k=2$) can be written as below, 
\begin{equation}\label{eq:12}
\begin{aligned}
f^{(2)}_{nd-is} & = 2 \ \Delta \mathcal{E}_{nd-is} |\langle r^{2} \rangle^{nd}_{is}|^{2} \left\{\begin{array}{c} 2 \ \ 1 \ \ 0\\ 0 \ \ 0 \ \ 0 \end{array}\right\}^{2} & = \frac{2}{5} \ 
\Delta \mathcal{E}_{nd-is} \  |\langle r^{2} \rangle^{nd}_{is}|^{2}.
\end{aligned} 
\end{equation}
Similarly, for the octupole OS ($k=3$), the expression becomes, 
\begin{equation}\label{eq:13}
\begin{aligned}
f^{(3)}_{nf-is} & = 2 \ \Delta \mathcal{E}_{nf-is} |\langle r^{3} \rangle^{nf}_{is}|^{2} \left\{\begin{array}{c} 3 \ \ 1 \ \ 0\\ 0 \ \ 0 \ \ 0 \end{array}\right\}^{2} & = \frac{2}{7} \ 
\Delta \mathcal{E}_{nf-is} \  |\langle r^{3} \rangle^{nf}_{is}|^{2}.
\end{aligned} 
\end{equation}
And for the hexadecapole OS ($k=4$), one gets, 
\begin{equation}\label{eq:14}
\begin{aligned}
f^{(4)}_{ng-is} & = 2 \ \mathcal{E}_{ng-is} |\langle r^{4} \rangle^{ng}_{is}|^{2} \left\{\begin{array}{c} 4 \ \ 1 \ \ 0\\ 0 \ \ 0 \ \ 0 \end{array}\right\}^{2} & = \frac{2}{9} \ 
\Delta \mathcal{E}_{ng-is} \  |\langle r^{4} \rangle^{ng}_{is}|^{2}.  \\
\end{aligned} 
\end{equation}
The analytical closed from expressions for multipole oscillator strength ($k=1-4$) for all possible transitions, and polarizabilities in FHA are 
collected in Appendix~A. It is important to mention that, there exists a multipole OS sum rule as follows, 
\begin{equation}\label{eq:a}
S^{(k)}=\sum_{m}f^{(k)}=k\langle \psi_{i}|r^{(2k-2)}|\psi_{i}\rangle, 
\end{equation}
where the summation includes all the bound states. 

\subsection{Plasma Characteristics}
Plasma is a statistical system of mobile charged particles, which interact with each other through electromagnetic forces. Here,  
the coupling occurs between quantum states and plasma density. Now we briefly discuss the characteristics of various H-atom plasmas.
\begin{figure}                         %%%Fig. 1, CHA
\begin{minipage}[c]{0.48\textwidth}\centering
\includegraphics[scale=0.75]{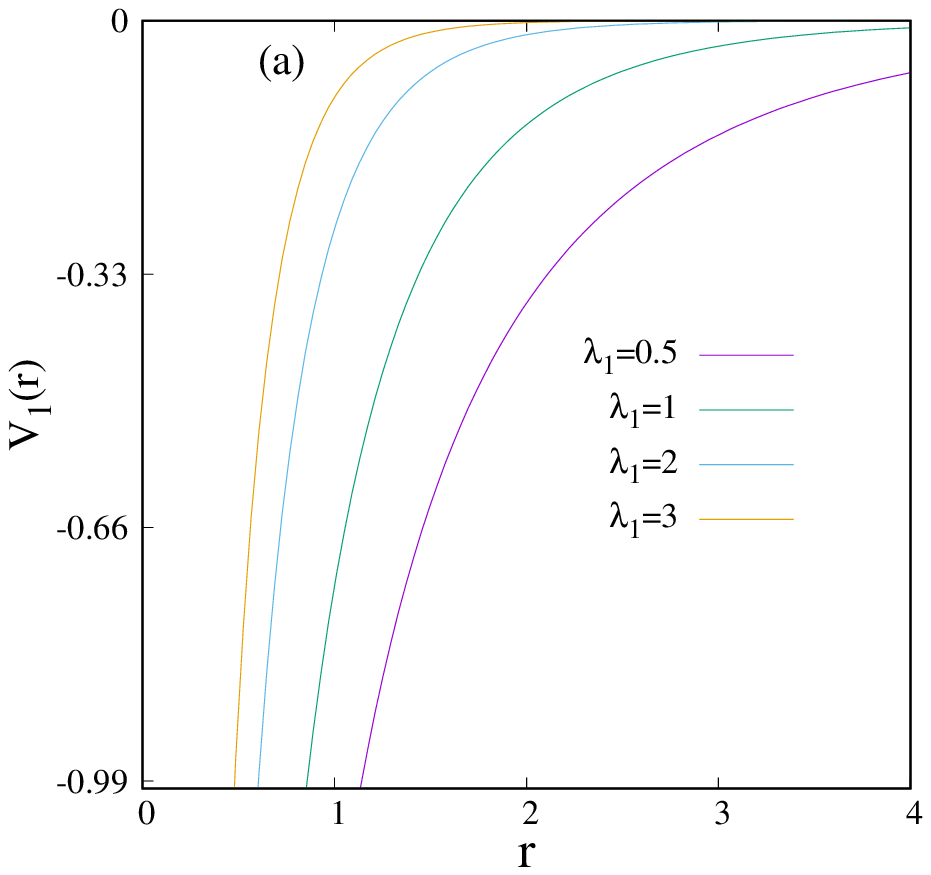}
\end{minipage}%
\vspace{1mm}
\begin{minipage}[c]{0.48\textwidth}\centering
\includegraphics[scale=0.75]{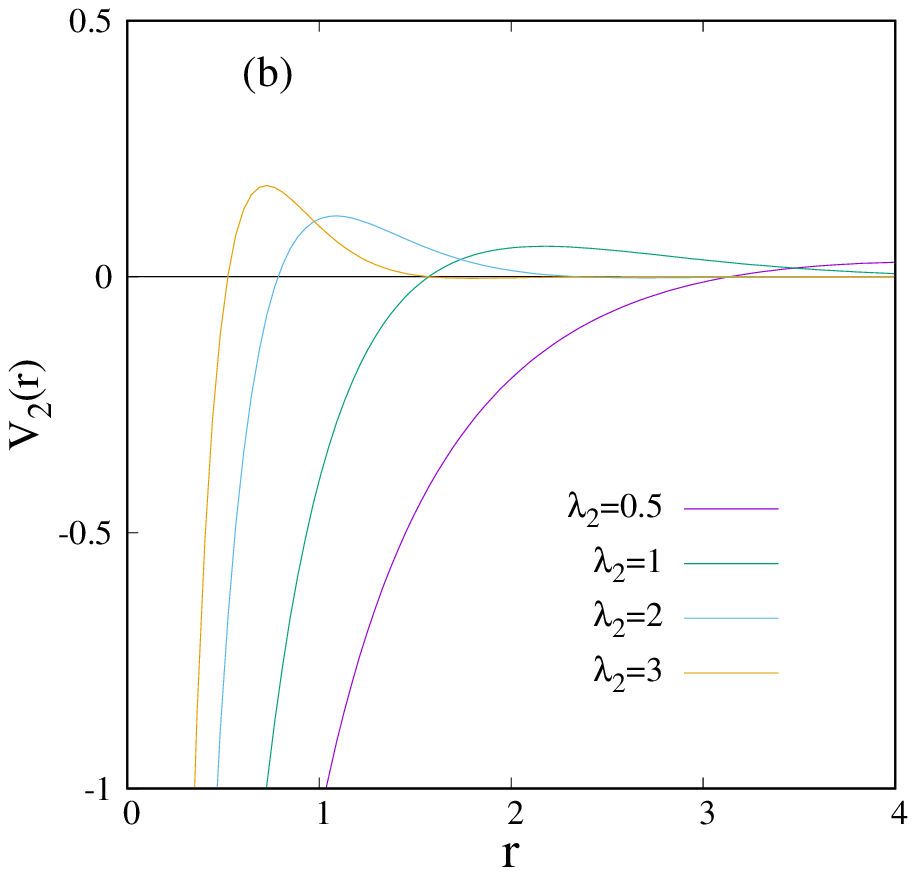}
\end{minipage}%
\caption{Plot of (a) $V_{1}(r)$, Eq.~\ref{eq:15} (b) $V_{2}(r)$, Eq.~\ref{eq:16}, at selected $\lambda$ values, namely, $0.5,~1,~2,~3$ 
keeping $Z=2$. For details, see text.}
\end{figure} 

In a hot plasma, the collective plasma screening effect on H atom is normally mapped by using Debye-H\"uckel potential of the form \cite{ichimaru82},     
\begin{equation}\label{eq:15}
\begin{aligned}    
V_{1}(r)=\begin{cases}
=-\frac{Z}{r}e^{-\lambda_{1} r}, \ \ \ \ r \le r_{c} \\
=0, \ \ \ \ \ \  r > r_{c}.  
\end{cases}
\end{aligned}
\end{equation}
In this form of potential, the probability of finding plasma particles inside the Debye sphere is negligible. In addition to screening 
effect, here it is assumed that the charge cloud is confined in spherical enclosure. This situation provides an alternate boundary 
condition for such systems. However, at $r_{c} \rightarrow \infty$ this restriction vanishes. The Debye radius ($D=\frac{1}{\lambda_{1}}$) 
plays an important role in WCP. For example (i) at a fixed $n_{e}$, $D \propto \sqrt{T}$ and (ii) at a certain $T$, 
$D \propto \frac{1}{\sqrt{n_{e}}}$. Most importantly, at a constant $D$, $n_{e} \propto T$. It means that, to keep $\lambda_{1}$ or $D$ fixed, 
with rise in $T$, $n_{e}$ increases. Further, with increase in $n_{e}$, the plasma tail effect declines. Conversely, with rise in $T$, it 
enhances. But, here incorporation of radial confinement is indirectly controls the tail effect. When $r_{c}$ is large, then $T$ 
predominates over $n_{e}$. On the other side, at low $r_{c}$ region, the effect of $n_{e}$ prevails. Therefore, in this work, we have 
probed WCP in two different motives: (i) firstly with the variation of $r_{c}$ at a fixed $\lambda_{1}$ and (ii) secondly, the effect of 
$\lambda_{1}$ at a certain $r_{c}$. Figure~1(a) portrays that, an enhancement in $\lambda_{1}$ leads to a growth in plasma electron 
density surrounding the positive ion.            
  
With increase in plasma density, the multi-particle cooperative interaction enhances. Thus, $D$ becomes comparable to de Broglie wave 
length, and hence quantum effect appears \cite{ghosal10}. In this context, Debye-H\"uckel model becomes inappropriate to explain the 
plasma properties. In ECSCP, $\lambda_{2}$ is connected to plasma frequency as $\lambda_{2} \propto \sqrt{\omega_{pe}}$. 
It has the form,  
\begin{equation}\label{eq:16}
\begin{aligned}    
V_{2}(r)=\begin{cases}
=-\frac{Z}{r}e^{-\lambda_{2} r}\ \cos(\lambda_{2}r), \ \ \ \ r \le r_{c} \\
=0, \ \ \ \ \ \ r > r_{c},  
\end{cases}
\end{aligned}
\end{equation}
Due to the incorporation of cosine term, ECSCP exhibits stronger screening effect compared to WCP. There occurs a maximum at 
$r_{\mathrm{max}}=\frac{\pi}{2\lambda_{2}}$. The temperature connection to $\lambda_{2}$ is not known. However, like WCP, here also 
$r_{c}$ plays same role: with progress in $r_{c}$, temperature effect enhances. Figure~1(b) imprints that, with rise in $\lambda_{2}$
the position of maximum gets left shifted and hence, plasma density advances. Like WCP, here too the effects of both $\lambda_{2}$ and 
$r_{c}$ are explored. At $\lambda =0$, both WCP and ECSCP modify to FHA-like systems.                

In case of SCP, the ion experiences a spherically symmetric environment within a radius $R$, commonly known as Wigner-Seitz radius. Beyond $R$, 
the effect of the potential vanishes. Hence the potential is
expressed as,    
\begin{equation}\label{eq:17}
\begin{aligned}    
V_{3}(r)=\begin{cases}
=-\frac{Z}{r}+\frac{Z-N_{e}}{2R}\left[3-\left(\frac{r}{R}\right)^{2}\right], \\
=0, \ \ \ \ \ \ r > r_{c}=R.
\end{cases}
\end{aligned}
\end{equation}
With decrease in $R$, $n_{e}$ increases and \emph{vice versa}. $T$ does not appear directly in this case. However, it is implicit that the 
change in $R$ exerts the effect of $T$. At $r_{c} \rightarrow \infty$,
Eq.~(\ref{eq:17}) reduces to FHA. It is necessary to mention that, in SCP $Z \ge 2$ condition needs to be obeyed.   

\subsection{Scaling transformation}
In case of plasma potentials, scaling concept has been implied previously in \cite{zhu20,das12,qi17,montgomery18}. This work employs two independent 
scaling ideas and attempts to derive a single equation connecting the original and scaled Hamiltonians. Thus, starting from an arbitrary set of 
$Z$ and $\beta$, one can easily estimate a given desired property for a series of $Z$ and $\beta$, connected by the scaling relation. To proceed further, 
one can write Eq.~(\ref{eq:2}) as follows, 
\begin{equation}\label{eq:18}
\begin{aligned}
-\frac{\hbar^{2}}{2m} \ \frac{d^2}{dr^2} \ \psi_{n,\ell}(r) + V_{c}(Z;\beta;r) \ \psi_{n,\ell}(r) +V_{0}\theta(r-r_{c}) \ \psi_{n,\ell}(r) = 
\mathcal{E}_{n,\ell} \ \psi_{n,\ell}(r), \\
\theta(r-r_{c})=0,  \ \ \ \mathrm{at} \ \ \ r \leq r_{c},  \ \ \ \ \ \ \theta(r-r_{c}) =1, \ \ \ \mathrm{at} \ \ \ r > r_{c}. 
\end{aligned}  
\end{equation}
Here $V_{c}(Z,\beta,r)$ is the potential that describes a H atom under the influence of plasma environment, $\theta(r-r_{c})$ is 
Heaviside theta function and $V_{0}$ is taken to be an infinitely large positive constant. The use of atomic unit, $\hbar=m=1$, transforms 
Eq.~(\ref{eq:18}) as below,  
\begin{equation}\label{eq:19}
-\frac{1}{2} \ \frac{d^2}{dr^2} \ \psi_{n,\ell}(r) + V_{c}(Z;\beta;r) \ \psi_{n,\ell}(r) +V_{0}\theta(r-r_{c}) \  \psi_{n,\ell}(r) = 
\mathcal{E}_{n,\ell} \ \psi_{n,\ell}(r)
\end{equation}
For H-isoelectronic series, it is interesting to probe the impact of $Z$ as well as $\beta$, on the properties of a given system. Now 
analytical relations among $\langle T^{n} \rangle, \langle V^{n} \rangle, \langle TV \rangle$, $f^{(k)}_{ni}, \alpha^{(k)}_{ni}$ 
with $Z$ and $\beta$ will be established, by employing two independent, parallel scaling transformations. 
\begin{enumerate}[(i)]
\item
In the first case, we apply a transformation $(r=Z r_{1})$. The Hamiltonian can then be modified in the following form,
\begin{equation}\label{eq:20}
H(Z;\beta;r_{c};r) \rightarrow H\left(1;\frac{\beta}{Z};Zr_{c};r_{1}\right).
\end{equation}
Thus, the $Z$-containing part of the potential becomes independent of it.  

This substitution transforms the Hamiltonian in Eq.~(\ref{eq:18}) in following form, 
\begin{equation}\label{eq:22}
\begin{aligned}
-\frac{1}{2} \nabla^{2}_{1} \psi_{n,\ell}(r_{1}) + V_{c}\left(1,\frac{\beta}{Z},r_{1}\right) \psi_{n,\ell}(r_{1}) + Z^{2} V_{0} \theta\left(r_{1}
-Z r_{c} \right) \psi_{n,\ell}(r_{1}) \\
= Z^{2} \ \mathcal{E}_{n,\ell} \ \psi_{n,\ell}(r_{1}).
\end{aligned}
\end{equation}
The eigenfunctions, eigenvalues of initial and modified Hamiltonians are connected as,
\begin{equation}\label{eq:23}
\begin{aligned}
\mathcal{E}_{n,\ell}\left[1;Z;\beta;r_{c}\right] & =Z^{2} \mathcal{E}_{n,\ell}\left[1;1;\frac{\beta}{Z};Zr_{c}\right], \\
\psi_{n,\ell}(1;Z;\beta;r_{c};r) & =\frac{1}{Z^\frac{3}{2}}\psi_{n,\ell}\left(1;1;\frac{\beta}{Z};Zr_{c};r_{1}\right).
\end{aligned}  
\end{equation} 
Then $\langle T^{n} \rangle, \langle V^{n} \rangle, \langle TV \rangle$ and $Z$ are found to be related as,
\begin{equation}\label{eq:24}
\begin{aligned}
\langle V^{n} \rangle\left[1;Z;\beta;r_{c} \right] & = Z^{2n} \  \langle V^{n} \rangle\left[1;1;\frac{\beta}{Z}; Zr_{c}\right],  \  
\langle T^{n} \rangle\left[1;Z;\beta; r_{c} \right]  = Z^{2n} \  \langle T^{n} \rangle\left[1;1;\frac{\beta}{Z}; Zr_{c}\right] \\
\langle TV \rangle\left[1;Z;\beta; r_{c} \right] & = Z^{4} \  \langle TV \rangle\left[1;1;\frac{\beta}{Z}; Zr_{c}\right], \ 
\langle VT \rangle\left[1;Z;\beta; r_{c} \right]  = Z^{4} \  \langle VT \rangle\left[1;1;\frac{\beta}{Z}; Zr_{c}\right] \\
\end{aligned}
\end{equation}
The multipole OS now takes the form,
\begin{equation}\label{eq:25}
f^{(k)}_{ni}\left[1;Z;\beta;r_{c}\right] =  \frac{f^{(k)}_{ni}\left[1;1;\frac{\beta}{Z};Zr_{c}\right]}{Z^{2(k-1)}}.
\end{equation}
This equation suggests that, dipole ($k=1$) OS is independent of this scaling transformation. However, quadrupole ($k=2$), octupole ($k=3$) and 
hexadecapole ($k=4$) OSs depend on $Z$. Now, some simple mathematical manipulation provides the modified expression 
of $\alpha^{(k)}_{i}(\mathrm{bound})$ as follows,
\begin{equation}\label{eq:26}
\alpha^{(k)}_{i}(\mathrm{bound})\left[1;Z;\beta;r_{c}\right] =  \frac{\alpha^{(k)}_{i}(\mathrm{bound})\left[1;1;\frac{\beta}{Z};Zr_{c}\right]}{Z^{2k+2}}.
\end{equation} 

\item
Another transformation $(r=\frac{r_{2}}{\beta})$, can be applied to alter the same Hamiltonian as,
\begin{equation}\label{eq:21}
H(Z;\beta;r_{c};r) \rightarrow H\left(\frac{Z}{\beta};1;\beta r_{c};r_{2}\right).
\end{equation}
Now the potential is mapped such that, the $\beta$-containing part becomes free of it. 

The substitution of $r=\frac{r_{2}}{\beta}$ transforms the Hamiltonian in Eq.~(\ref{eq:18}) in the form, 
\begin{equation}\label{eq:27}
\begin{aligned}
-\frac{1}{2} \nabla^{2} \psi_{n,\ell}(r_{2}) + V_{c}\left(\frac{Z}{\beta};1;r_{2}\right) \psi_{n,\ell}(r_{2}) + \frac{1}{\beta^{2}} V_{0} \theta\left(r_{2}
-\beta r_{c}\right) \psi_{n,\ell}(r_{2}) \\
= \left(\frac{\mathcal{E}_{n,\ell}}{\beta^{2}}\right) \psi_{n,\ell}(r_{2}).
\end{aligned}
\end{equation}
The eigenfunctions, eigenvalues of initial and modified Hamiltonians are related as,
\begin{equation}\label{eq:28}
\begin{aligned}
\mathcal{E}_{n,\ell}\left[1;Z;\beta;r_{c}\right] & = \beta^{2} \ \ \mathcal{E}_{n,\ell}\left[1;\frac{Z}{\beta};1;\beta r_{c}\right], \\
\psi_{n,\ell}(1;Z;\beta;r_{c};r) & = \beta^{\frac{3}{2}} \ \psi_{n,\ell}\left(1;\frac{Z}{\beta};1;\beta r_{c};r_{2}\right).
\end{aligned}  
\end{equation} 
Then $\langle T^{n} \rangle, \langle V^{n} \rangle, \langle TV \rangle$ and $\beta$ are connected as,
\begin{equation}\label{eq:29}
\begin{aligned}
\langle V^{n} \rangle\left[1;Z;\beta; r_{c} \right]  = \beta^{2n} \ \ \langle V^{n} \rangle\left[1;\frac{Z}{\beta};1;\beta r_{c}\right], \ \ \ \ 
\langle T^{n} \rangle\left[1;Z;\beta; r_{c} \right]  = \beta^{2n} \ \ \langle T^{2} \rangle\left[1;\frac{Z}{\beta};1;\beta r_{c}\right] \\
\langle TV \rangle\left[1;Z;\beta; r_{c} \right]  = \beta^{4} \ \ \langle TV \rangle\left[1;\frac{Z}{\beta};1 ;\beta r_{c} \right], \ \ \ \ 
\langle VT \rangle\left[1;Z;\beta; r_{c} \right]  = \beta^{4} \ \ \langle VT \rangle\left[1;\frac{Z}{\beta};1;\beta r_{c}\right] \\
\end{aligned}
\end{equation}

Now, using Eq.~(\ref{eq:24}) into Eq.~(\ref{eq:8}), the multipole OS can have the generalized form,
\begin{equation}\label{eq:30}
f^{(k)}_{ni}\left[1;Z;\beta;r_{c}\right] = \left( \frac{f^{(k)}_{ni}\left[1;\frac{Z}{\beta};1;\beta r_{c}\right]}{\beta^{(2k-2)}} \right).
\end{equation}
This implies that, dipole OS is invariant under this scaling transformation. However, higher order ($k>1$) OS
depend on $\beta$. Again some straightforward mathematical manipulation gives the modified expression of $\alpha^{(k)}_{i}(\mathrm{bound})$ as,
\begin{equation}\label{eq:31}
\alpha^{(k)}_{i}(\mathrm{bound})\left[1;Z;\beta;r_{c}\right]  =  
\left(\frac{\alpha^{(k)}_{i}(\mathrm{bound})\left[1;\frac{Z}{\beta};1;\beta r_{c}\right]}{\beta^{2(k+1)}}\right)
\end{equation} 
\end{enumerate}

Thus we have successfully converted the initial Hamiltonian, Eq.~(\ref{eq:2}) into two independent scaled Hamiltonians, \emph{viz.}, Eqs.~(\ref{eq:22}) and (\ref{eq:27}). 
Now, the connecting relations are,
\begin{equation}\label{eq:32}
\mathcal{E}_{n,\ell}\left[1;Z;\beta;r_{c}\right] = Z^{2} \mathcal{E}_{n,\ell}\left[1;1;\frac{\beta}{Z};Zr_{c}\right] =
\beta^{2} \ \ \mathcal{E}_{n,\ell}\left[1;\frac{Z}{\beta};1;\beta r_{c}\right].
\end{equation}
Some reorganization leads to the following, 
\begin{equation}\label{eq:33}
\frac{\mathcal{E}_{n,\ell}\left[1;1;\frac{\beta}{Z};Zr_{c}\right]}{\mathcal{E}_{n,\ell}\left[1;\frac{Z}{\beta};1;\beta r_{c}\right]} = \left(\frac{\beta}{Z}\right)^{2}.   
\end{equation}
The expectation values then satisfy the following relations,  
\begin{equation}\label{eq:34}
\langle V^{n} \rangle \left[1;Z;\beta;r_{c}\right] = Z^{2n} \langle V^{n} \rangle \left[1;1;\frac{\beta}{Z};Zr_{c}\right] =
\beta^{2n} \langle V^{n} \rangle \left[1;\frac{Z}{\beta};1;\beta r_{c}\right].
\end{equation}
A slight rearrangement of the above equation leads to, 
\begin{equation}\label{eq:35}
\frac{\langle V^{n} \rangle \left[1;1;\frac{\beta}{Z};Zr_{c}\right]}{\langle V^{n} \rangle \left[1;\frac{Z}{\beta};1;\beta r_{c}\right]} = \left(\frac{\beta}{Z}\right)^{2n}.   
\end{equation}
In case of kinetic energy, one gets, 
\begin{equation}\label{eq:36}
\langle T^{n} \rangle \left[1;Z;\beta;r_{c}\right] = Z^{2n} \langle T^{n} \rangle \left[1;1;\frac{\beta}{Z};Zr_{c}\right] =
\beta^{2n} \langle T^{n} \rangle \left[1;\frac{Z}{\beta};1;\beta r_{c}\right].
\end{equation}
which, upon rearrangement, gives, 
\begin{equation}\label{eq:37}
\frac{\langle T^{n} \rangle \left[1;1;\frac{\beta}{Z};Zr_{c}\right]}{\langle T^{n} \rangle \left[1;\frac{Z}{\beta};1;\beta r_{c}\right]} = \left(\frac{\beta}{Z}\right)^{2n}.   
\end{equation}
The multipole OS accordingly becomes,
\begin{equation}\label{eq:38}
f^{(k)}_{ni} \left[1;Z;\beta;r_{c}\right] = \frac{f^{(k)}_{ni}\left[1;1;\frac{\beta}{Z};Zr_{c}\right]}{Z^{2(k-1)}} =
\frac{f^{(k)}_{ni} \left[1;\frac{Z}{\beta};1;\beta r_{c}\right]}{\beta^{2(k-1)}}.
\end{equation}
which can be recast to yield, 
\begin{equation}\label{eq:39}
\frac{f^{(k)}_{ni}\left[1;1;\frac{\beta}{Z};Zr_{c}\right]}{f^{(k)}_{ni} \left[1;\frac{Z}{\beta};1;\beta r_{c}\right]} = 
\left(\frac{Z}{\beta}\right)^{2(k-1)}.
\end{equation}
Finally, the polarizabilities are connected as,
\begin{equation}\label{eq:40}
\alpha^{(k)}_{i}(\mathrm{bound})\left[1;Z;\beta;r_{c}\right] = \frac{\alpha^{(k)}_{i}(\mathrm{bound})\left[1;1;\frac{\beta}{Z};Zr_{c}\right]}{Z^{2(k+1)}} =
\frac{\alpha^{(k)}_{i}(\mathrm{bound})\left[1;\frac{Z}{\beta};1;\beta r_{c}\right]}{\beta^{2(k+1)}}.
\end{equation}
This can be written in the following form,  
\begin{equation}\label{eq:41}
\frac{\alpha^{(k)}_{i}(\mathrm{bound})\left[1;1;\frac{\beta}{Z};Zr_{c}\right]}{\alpha^{(k)}_{i} (\mathrm{bound})\left[1;\frac{Z}{\beta};1;\beta r_{c}\right]} = \left(\frac{Z}{\beta}\right)^{2(k+1)}.
\end{equation}

The foregoing discussion thus shows that, a connection formula, as below, can be derived among three Hamiltonians, corresponding to the 
SE in Eqs.~(\ref{eq:2}), (\ref{eq:22}) and (\ref{eq:27}), \emph{viz.},
\begin{equation}\label{eq:42}
H\left(1;1;\frac{\beta}{Z};Zr_{c};r_{1}\right) \leftrightarrow H(1;Z;\beta;r_{c};r) \leftrightarrow  H\left(1;\frac{Z}{\beta};1;\beta r_{c};r_{2}\right)
\end{equation}
The above equation signifies that, performing the calculation at a particular ($Z,\beta$) pair, one can evaluate the properties of other pair of 
$(Z,\beta)$ (connected by scaling), without solving the SE. These are derived for any two-parameter potentials. These relations are applicable in 
all the three potentials used for the plasma characteristics in Sec.~II.C. In WCP, ECSCP and SCP, $\beta$ becomes $\lambda_{1}, \lambda_2, 
\sigma =\left(\frac{Z-Ne}{2R^{3}}\right)^{\frac{1}{4}}$ respectively. Some representative numerical results $(\mathcal{E}_{n,\ell}, f^{(1)}_{ns \rightarrow 2p},
\alpha^{(1)}_{ns})$ for these three Hamiltonians (connecting WCP, ECSCP, SCP) have been provided in Table~V of Appendix~B. 

\begin{figure}                         %%%Fig. 2, CHA
\begin{minipage}[c]{0.48\textwidth}\centering
\includegraphics[scale=0.72]{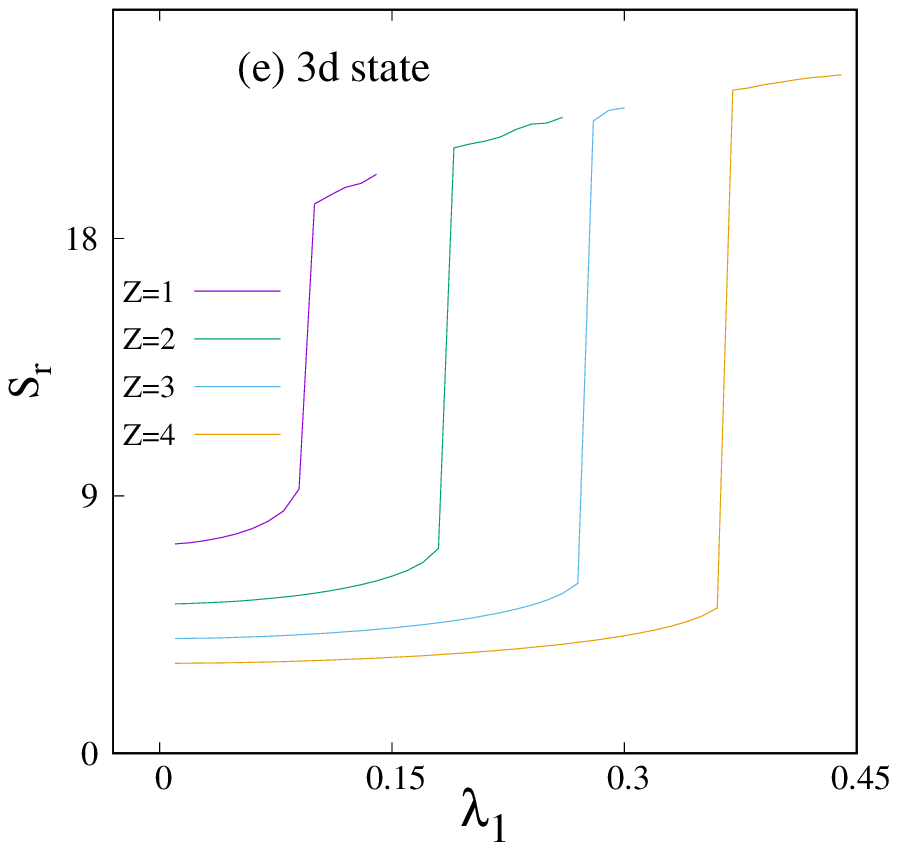}
\end{minipage}%
\begin{minipage}[c]{0.48\textwidth}\centering
\includegraphics[scale=0.72]{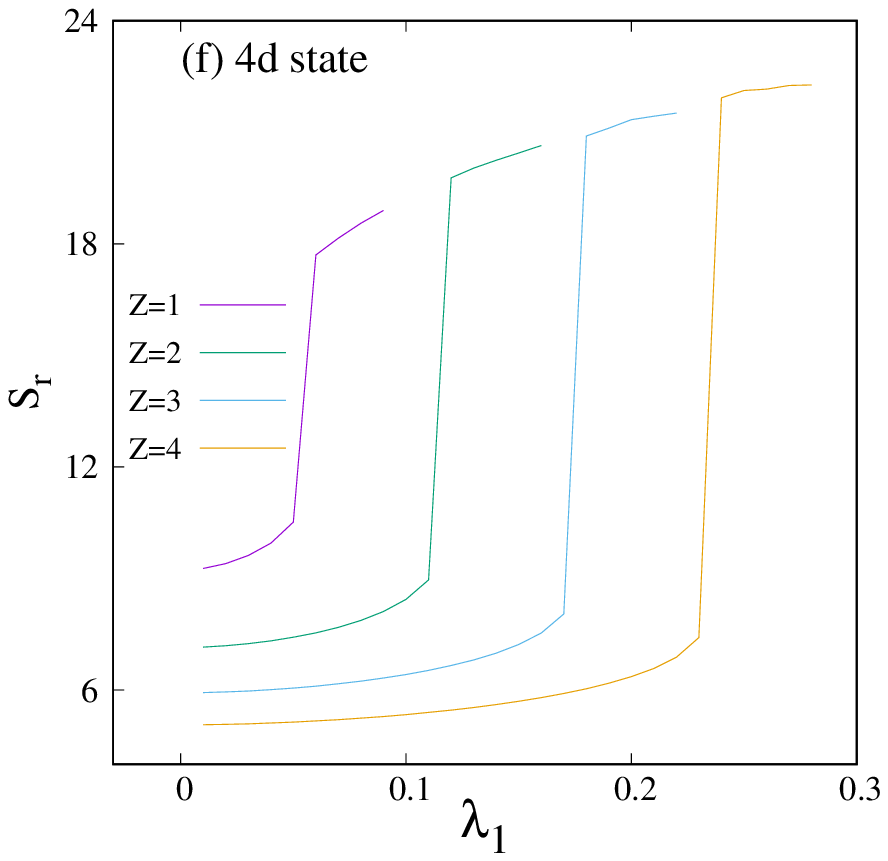}
\end{minipage}%
\vspace{1mm}
\begin{minipage}[c]{0.48\textwidth}\centering
\includegraphics[scale=0.72]{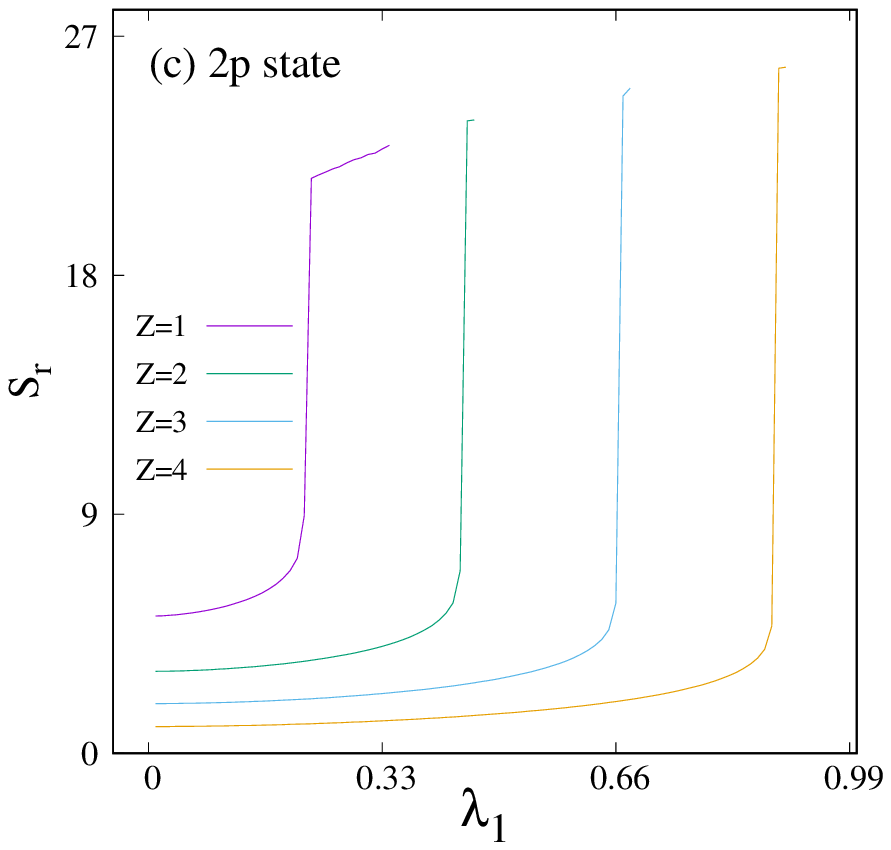}
\end{minipage}%
\begin{minipage}[c]{0.48\textwidth}\centering
\includegraphics[scale=0.72]{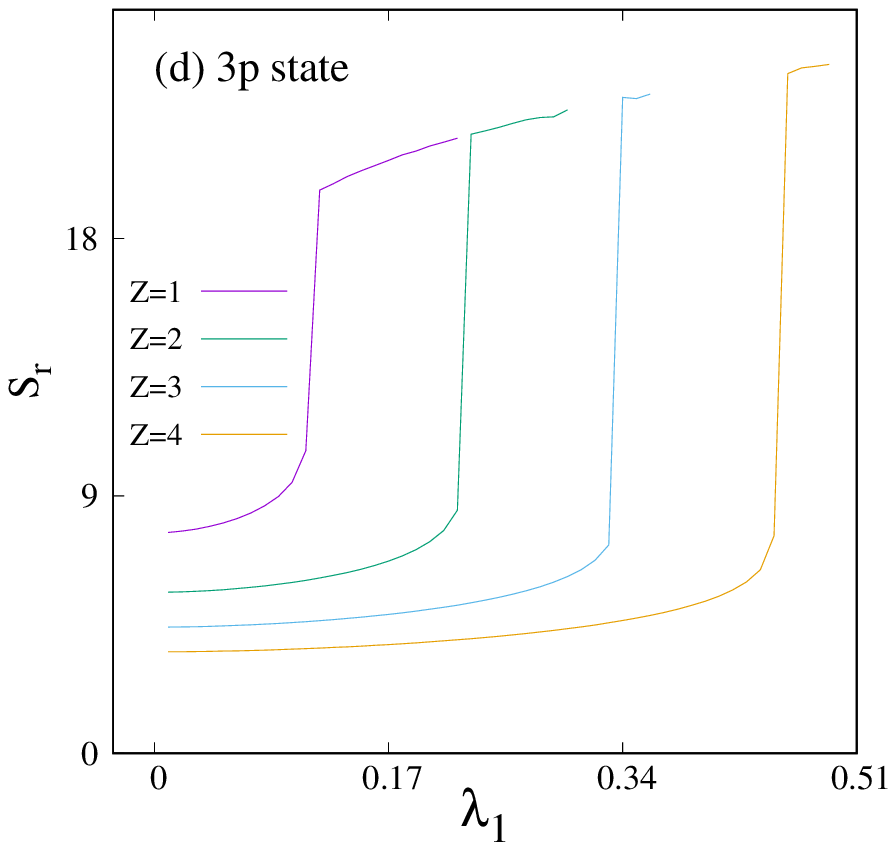}
\end{minipage}%
\vspace{1mm}
\begin{minipage}[c]{0.48\textwidth}\centering
\includegraphics[scale=0.72]{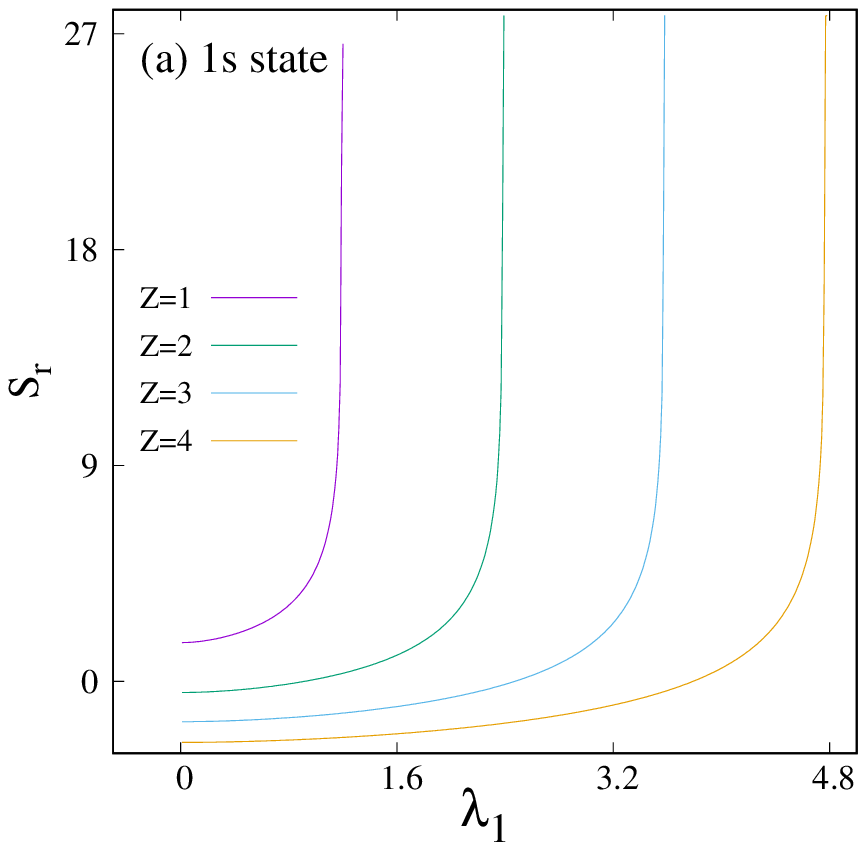}
\end{minipage}%
\begin{minipage}[c]{0.48\textwidth}\centering
\includegraphics[scale=0.72]{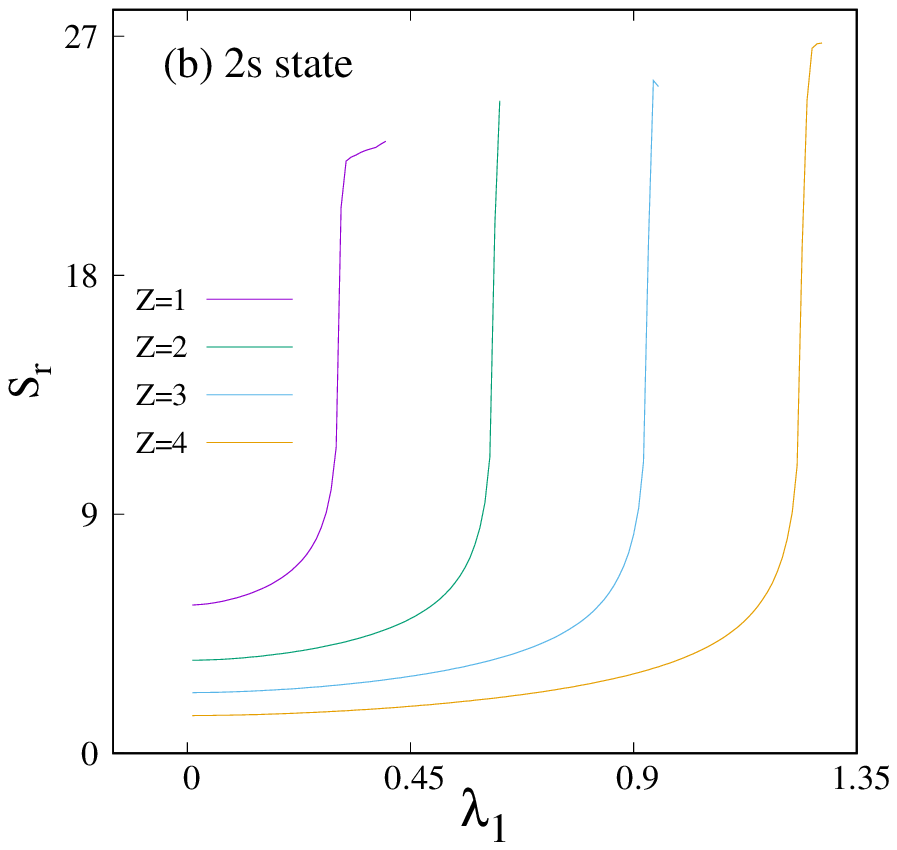}
\end{minipage}%
\caption{Plot of $S_{r}$ as function of $\lambda_{1}$ in WCP for (a) $1s$ (b) $2s$ (c) $2p$ (d) $3p$ (e) $3d$, and (f) $4d$ states at four selected 
values of $Z$, namely, $1,~2,~3,~4$. See text for details.}
\end{figure} 

\section{Result and Discussion}
In this section at first, we will discuss the critical screening constant in WCP and ECSCP. Then, the usefulness and efficacy of VT will be verified 
for WCP, ECSCP and SCP successively. Next, we report the multipole OS and polarizabilities for all these three potentials. Pilot calculations are done 
for $1s$ and $2s$ states choosing $Z=2$. Of course, employing the scaling relations of Eqs.~(\ref{eq:23}-{\ref{eq:26}}), one an easily extract the result 
for other $Z$ values. For ease of convenience, we have adopted the following notation. Use of $\lambda$ in the text implies \emph{both} 
$\lambda_{1},\lambda_{2}$, while explicit use of $\lambda_{1}$ or $\lambda_{2}$ refers to WCP and ECSCP only.  
   
\subsection{Critical screening constant in WCP and ECSCP}
In WCP and ECSCP (at $r_{c} \rightarrow \infty$), the number of bound states reduces with rise in screening parameter $\lambda$. Several attempts 
were made to estimate the characteristic value of $\lambda$ at which a particular state vanishes. Accurate numerical results are available up to $6h$ 
states of H-atom in WCP \cite{stubbins93,roy16a} and ECSCP \cite{diaz91,roy16a,singh83}. Further, in \cite{montgomery18}, the relation between this critical constant 
$\lambda^{(c)}_{n,\ell}(Z)$ and $Z$ was derived for ground state in WCP. These values are determined by applying the sign-change 
argument in energy. In stead of that, here, we have applied a simple density-based technique to ascertain these points in WCP and ECSCP. For that purpose, 
Shannon entropy ($S_{r}=-\int \rho(r) \ln \rho(r) r^{2} \mathrm{d}r$) \cite{mukherjee18} has been employed. Based on this study, a uniform relation 
between these two quantities ($\lambda^{(c)}_{n,\ell}(Z)$ and $Z$) is offered. This may be applied to an arbitrary state. Furthermore, a similar relation 
is also obtained by employing the scaling concept and some empirical idea (see below). 

\begin{figure}                         %%%Fig. 3, CHA
\begin{minipage}[c]{0.48\textwidth}\centering
\includegraphics[scale=0.72]{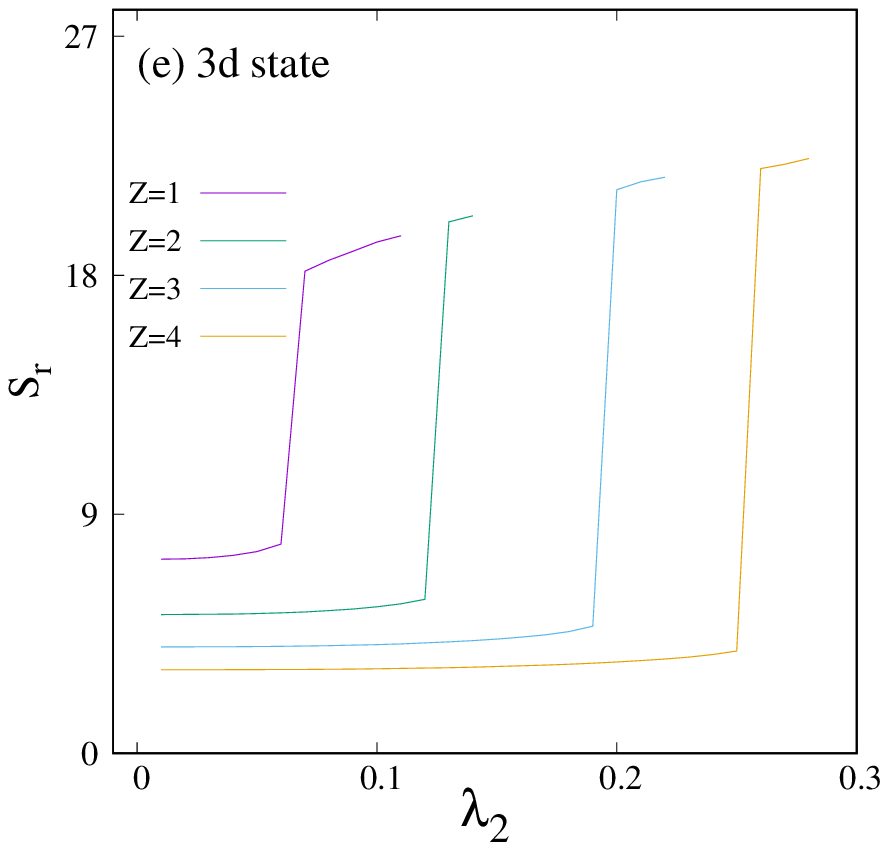}
\end{minipage}%
\begin{minipage}[c]{0.48\textwidth}\centering
\includegraphics[scale=0.72]{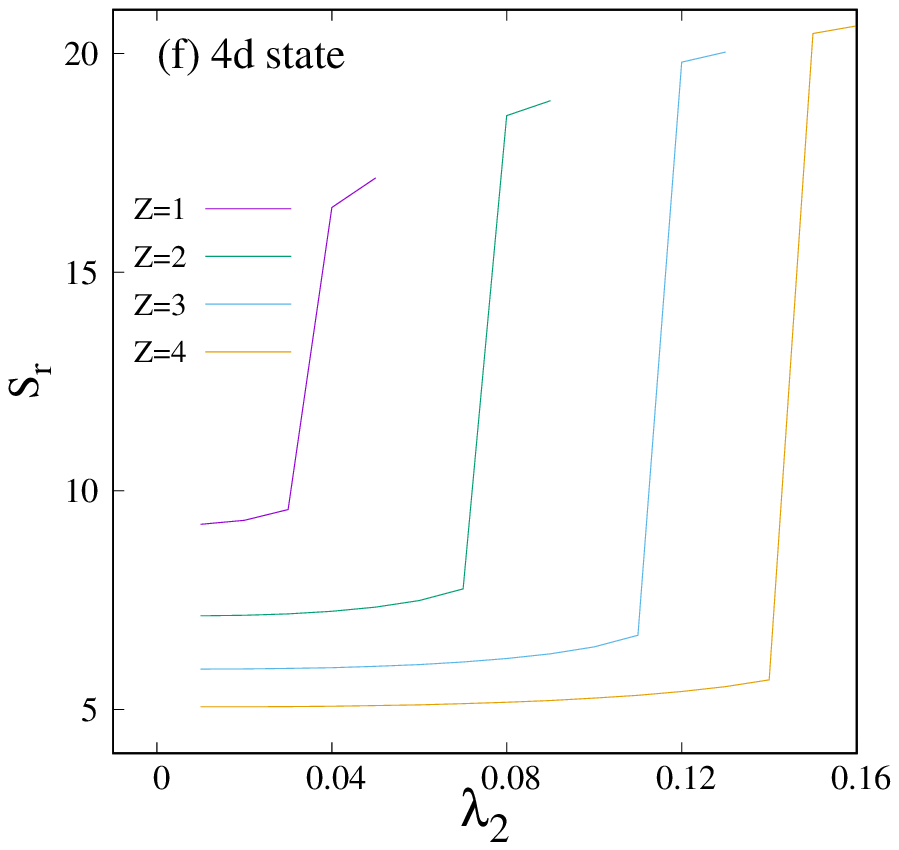}
\end{minipage}%
\vspace{1mm}
\begin{minipage}[c]{0.48\textwidth}\centering
\includegraphics[scale=0.72]{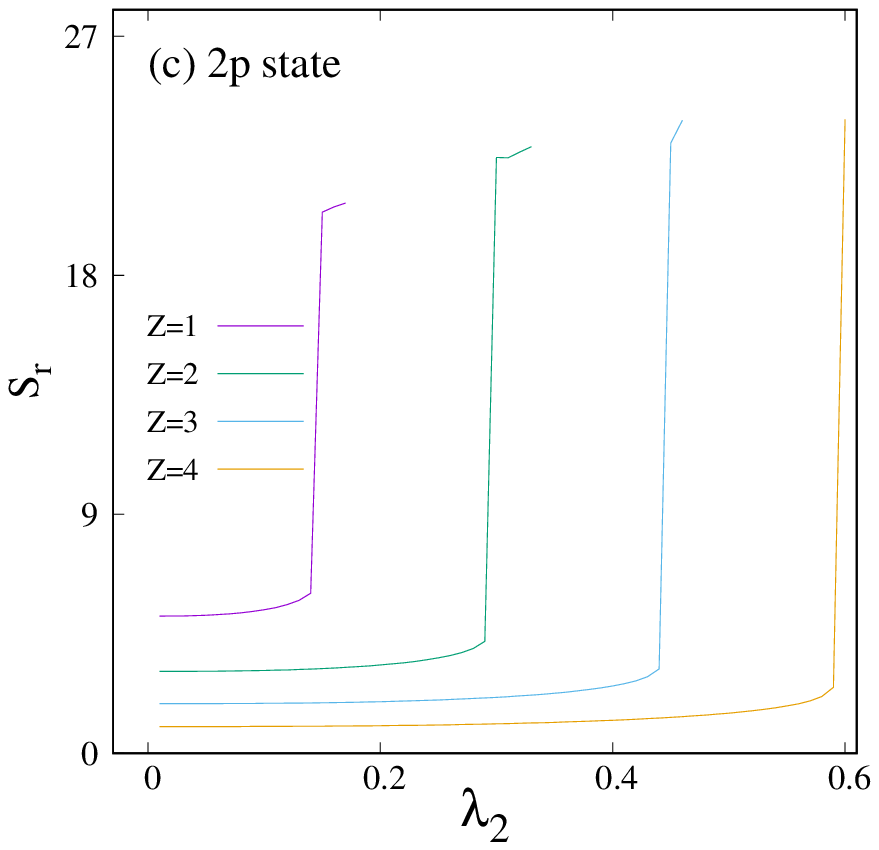}
\end{minipage}%
\begin{minipage}[c]{0.48\textwidth}\centering
\includegraphics[scale=0.72]{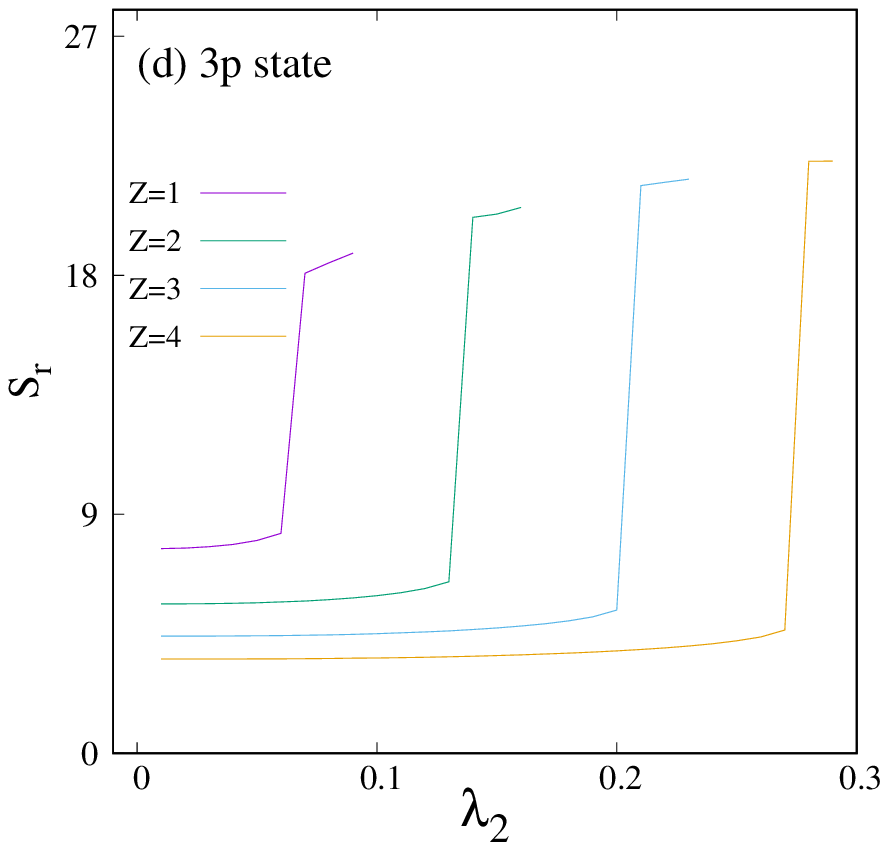}
\end{minipage}%
\vspace{1mm}
\begin{minipage}[c]{0.48\textwidth}\centering
\includegraphics[scale=0.72]{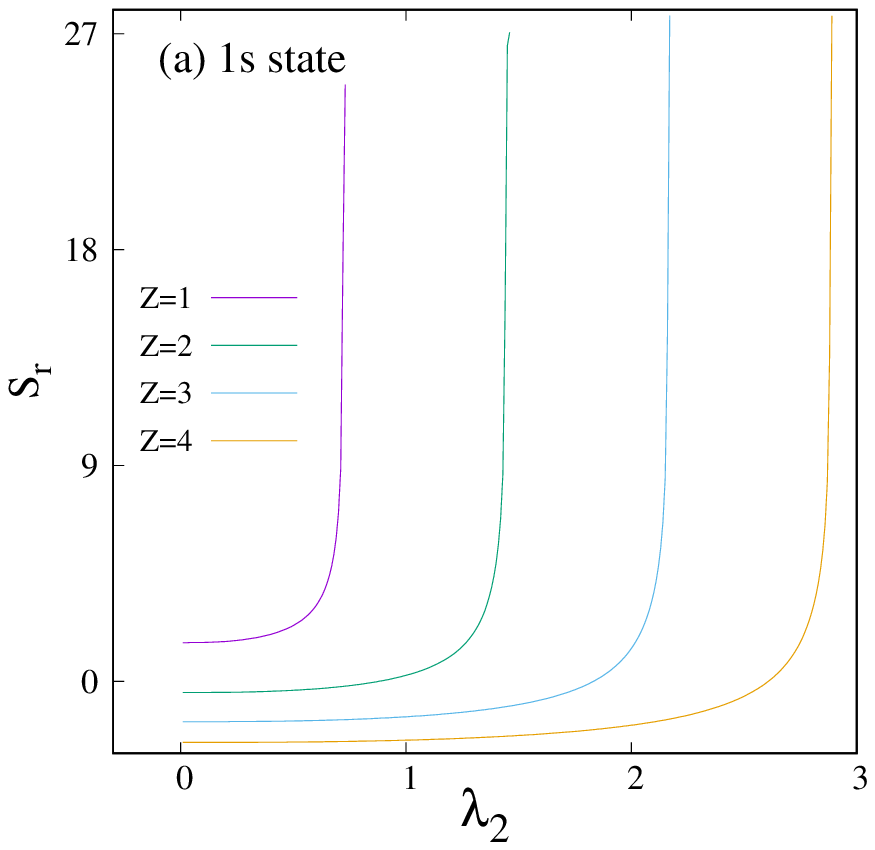}
\end{minipage}%
\begin{minipage}[c]{0.48\textwidth}\centering
\includegraphics[scale=0.72]{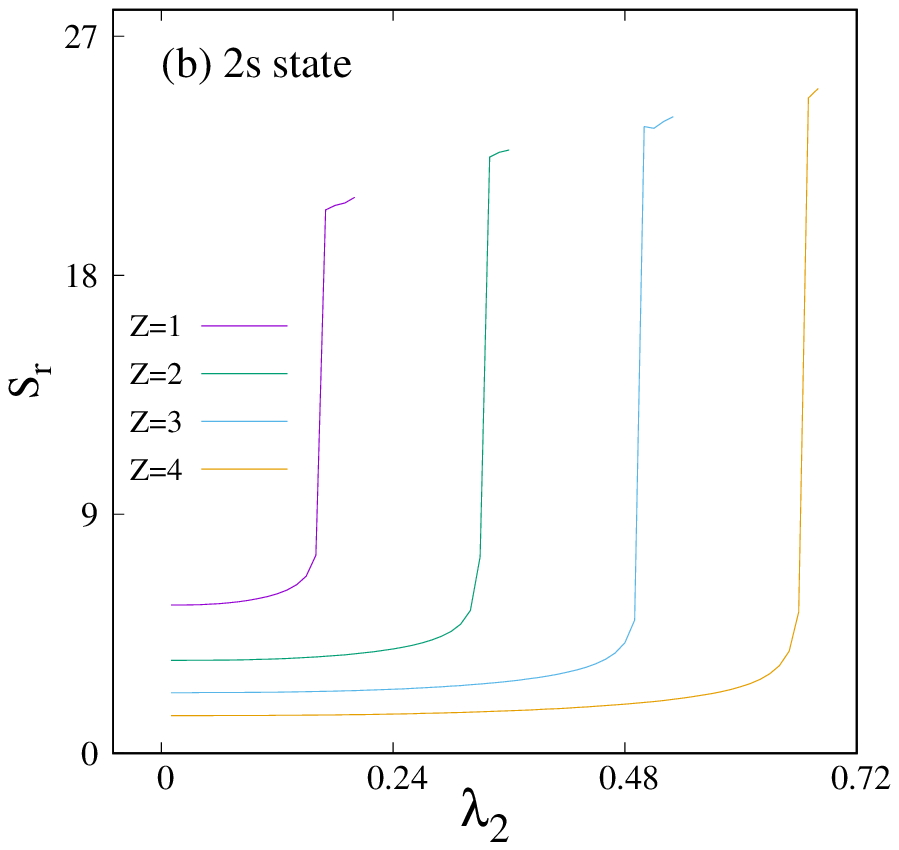}
\end{minipage}%
\caption{Plot of $S_{r}$ as function of $\lambda_{2}$ in ECSCP for (a) $1s$ (b) $2s$ (c) $2p$ (d) $3p$ (e) $3d$, and (f) $4d$ states at four selected 
values of $Z$, namely, $1,~2,~3,~4$. See text for details.}
\end{figure}  

The calculated $S_{r}$, as a function of $\lambda_{1}$ for first two states of each $\ell=0-2$ are displayed in Fig.~2. Panels (a)-(f) represent
$1s,2s,2p,3p,3d,4d$ states respectively. In each of these panels one can see equi-spaced curves corresponding to $Z=1-4$. At a fixed $Z$, in each of 
these states there occurs a sudden jump in $S_{r}$ at a characteristics $\lambda_{1}$. Therefore, $S_{r}$ can indicate the critical point, at which a 
particular state vanishes. Further, at a certain $Z$, $S_{r}$ increases with $\lambda_{1}$. It means that with decrease in $D$ confinement effect 
weakens. Conversely, with rise in $T$ this effect predominates. Analogous plots are supplied in Fig.~3(a)-3(f) for ECSCP, involving same 6 states 
of Fig.~2. The qualitative behavior of $S_{r}$ in WCP and ECSCP remains quite similar. In each state, a stiff increase in $S_{r}$ occurs at 
certain $\lambda_2$ value. Interestingly, with rise in $Z$, this $S_{r}$ vs $\lambda_{2}$ curve gets right shifted. Further, these curves are place 
equidistant from each other. From the above, it is clear that, $S_{r}$ can be used in determining critical screening constant in a given potential. 
Note that, in both potentials, for a given state, the ratio of screening constant and $Z$ is a constant, because, the four curves remain evenly 
separated. Depending upon these outcomes one can derive an empirical relation between $\lambda_{n,\ell}$ and Z.   

\begingroup           %%Table 1, critical screening
\squeezetable
\begin{table}
\caption{$\lambda^{(c)}_{n,\ell}$ for H-like ion for $1s,2s,2p,3p,3d,4d$ states in WCP, ECSCP. See text for details.}
\centering
\begin{ruledtabular}
\begin{tabular}{llllllll}
\multicolumn{4}{c}{WCP}  & \multicolumn{4}{c}{ECSCP}   \\
\cline{1-4} \cline{5-8}
$Z$   & State & \ \ $\lambda^{(c)}_{n,\ell}$  & \ \ \ \ \ $\mathcal{E}_{n,\ell}$ & $Z$  & State & \ \ $\lambda^{(c)}_{n,\ell}$ & \ \ \ \ \ $\mathcal{E}_{n,\ell}$  \\
\cline{1-4} \cline{5-8} 
1   & $1s$ &  1.1856$^{\ddag,\dagger}$  &  $-$0.00000656   & 1 & $1s$ &     0.7196$^{\S}$   &   $-$0.00000531  \\
2   &      &  2.3712  &  $-$0.00002650  & 2 & $1s$ &     1.4384    &  $-$0.00002124    \\
3   &      &  3.5573  &  $-$0.00005964  & 3 & $1s$ &     2.1576    &  $-$0.00004779    \\
4   &      &  4.7410  &  $-$0.00010265  & 4 & $1s$ &     2.8756    &  $-$0.00008496                \\
\cline{1-4} \cline{5-8}
1   & $2s$ &  0.3063$^{\ddag}$  &  $-$0.00000995   & 1 & $2s$ &  0.1664$^{\S}$  &  $-$0.00000552  \\
2   &      &  0.6124  &  $-$0.00003960  & 2 & $2s$ &    0.3328           &  $-$0.00002206  \\
3   &      &  0.9195  &  $-$0.00008970  & 3 & $2s$ &    0.4992         &  $-$0.00004965   \\
4   &      &  1.2254  &  $-$0.00015925  & 4 & $2s$ &    0.6656         &  $-$0.00008826  \\
\cline{1-4} \cline{5-8}
1   & $2p$ &  0.2206$^{\ddag}$  &  $-$0.00000723   & 1 & $2p$ &   0.1482$^{\S}$   & $-$0.00000234  \\
2   &      &  0.4404  &  $-$0.00002860  & 2 & $2p$ &  0.2964       &  $-$0.00000937 \\
3   &      &  0.6606  &  $-$0.00006341  & 3 & $2p$ &  0.4446       &  $-$0.00002109  \\
4   &      &  0.8821  &  $-$0.00011341  & 4 & $2p$ &  0.5928       &  $-$0.00003749 \\
\cline{1-4} \cline{5-8}
1   & $3p$ &  0.1126$^{\ddag}$  &  $-$0.00000701  & 1 & $3p$ &  0.0687$^{\S}$ & $-$0.00000488  \\
2   &      &  0.2254  &  $-$0.00002854  & 2 & $3p$ &   0.1374      &  $-$0.00001950   \\
3   &      &  0.3381  &  $-$0.00006371  & 3 & $3p$ &   0.2061      &  $-$0.00004388  \\
4   &      &  0.4504  &  $-$0.00011208  & 4 & $3p$ &   0.2748      &  $-$0.00007801   \\
\cline{1-4} \cline{5-8}
1   & $3d$ &  0.0914$^{\ddag}$  &  $-$0.00000878   & 1 & $3d$ &  0.0635$^{\S}$  &  $-$0.00001937   \\
2   &      &  0.1826  &  $-$0.00003614  & 2 & $3d$ &   0.1271      &  $-$0.00007787   \\
3   &      &  0.2739  &  $-$0.00008030  & 3 & $3d$ &   0.1907      &  $-$0.00017251   \\
4   &      &  0.3653  &  $-$0.00012718  & 4 & $3d$ &   0.2542      &  $-$0.00031150   \\
\cline{1-4} \cline{5-8}
1   & $4d$ &  0.0581$^{\ddag}$  &  $-$0.00000974  & 1 & $4d$ & 0.0374$^{\S}$ &  $-$0.00000260 \\
2   &      &  0.1161  &  $-$0.00003951  & 2 & $4d$ & 0.0748        &  $-$0.00001041 \\
3   &      &  0.1741  &  $-$0.00008364  & 3 & $4d$ & 0.1122        &  $-$0.00002342  \\
4   &      &  0.2321  &  $-$0.00016672  & 4 & $4d$ & 0.1496        &  $-$0.00004164 \\
\end{tabular}
\end{ruledtabular}
\begin{tabbing}
$^{\dagger}$Literature result of $\lambda^{(c)}_{1,0}$ \cite{zhu20}: 1.190612421. \\
$^{\ddag}$Literature results of $\lambda^{(c)}_{n,\ell}$ \cite{roy16a,diaz91}: (a) $\lambda^{(c)}_{1s}=1.190610$ (b) $\lambda^{(c)}_{2s}=0.310199$ 
(c) $\lambda^{(c)}_{2p}=0.220216$ 
(d) $\lambda^{(c)}_{3p}=0.112710$ \\
(e) $\lambda^{(c)}_{3d}=0.091345$ (f) $\lambda^{(c)}_{4d}=0.058105$. \\
$^{\S}$Literature results of $\lambda^{(c)}_{n,\ell}$ \cite{roy16a,singh83,diaz91}: (a) $\lambda^{(c)}_{1s}=0.720524$ (b) $\lambda^{(c)}_{2s}=0.166617$ (c) $\lambda^{(c)}_{2p}=0.148205$ 
(d) $\lambda^{(c)}_{3p}=0.068712$ \\ 
(e) $\lambda^{(c)}_{3d}=0.063581$ (f) $\lambda^{(c)}_{4d}=0.037405$.
\end{tabbing}
\end{table}  
\endgroup    
 
Both in WCP and ECSCP, the Hamiltonian in \emph{free condition} is scaled as,                  
\begin{equation}\label{eq:43}
H(Z;\lambda) \rightarrow H\left(1;\frac{\lambda}{Z}\right).  
\end{equation}
Similarly energy in a definite $(n,\ell)$ state is scaled as,
\begin{equation}\label{eq:44}
\mathcal{E}_{n,\ell}(Z;\lambda) = Z^{2} \mathcal{E}_{n,\ell}\left(1;\frac{\lambda}{Z}\right)
\end{equation}
Therefore, one can easily write the following relations for both WCP and ECSCP cases, 
\begin{equation}\label{eq:45}
\begin{aligned}
\frac{\lambda^{(c)}_{n,\ell}}{Z} \approx \lambda^{(c)}_{n,\ell}(Z=1), \\
\lambda^{(c)}_{n,\ell}(Z) \approx Z \ \lambda^{(c)}_{n,\ell}(Z=1). 
\end{aligned}
\end{equation}

This relation in Eq.~(\ref{eq:45}) is in excellent agreement with those achieved by computing $S_{r}$ in WCP and ECSCP. Representative numerical 
results are provided in Table~I for $Z=1-4$ involving the same six states of Figs.~1 and 2, in WCP and ECSCP. These critical parameters are compared 
with available reference results (for $Z=1$). which shows very good matching in both WCP \cite{roy16a,diaz91} and ECSCP \cite{roy16a,singh83,diaz91}. However, 
to the best of our knowledge, no such data are reported till date for $Z > 1$. The critical points from sign change argument also complement the outcomes achieved 
by employing the information entropy concept. This shows that $S_r$ may act as an efficient indicator for finding critical points and may be utilized 
in future. As expected, the tabular results strongly recommend the proposition of Eq.~(\ref{eq:45}) in both WCP and ECSCP. For the sake of completeness, 
$\lambda^{(c)}_{n,\ell}$ are computed for all the remaining states corresponding to $\ell=5$ $(3s,4s,4p,4f,5s,5p,5d,5f,5g)$. They are reported in
Table~VI of Appendix~C, along with the appropriate references.

\begingroup           %%Table 2, virial theorem 
\squeezetable
\begin{table}
\caption{$\mathcal{E}_{n,\ell}, \left(\Delta V_{n,\ell}\right)^{2}, \left(\Delta T_{n,\ell}\right)^{2}, \langle T \rangle_{n,\ell}
\langle V \rangle_{n,\ell}-\langle TV \rangle_{n,\ell}, \langle T \rangle_{n,\ell}\langle V \rangle_{n,\ell}-\langle VT \rangle_{n,\ell}$ of $1s,2s$ states in 
WCP, ECSCP and SCP, choosing $Z=2$, at six different sets of $(\lambda_{1}, r_{c})$, $(\lambda_{2}, r_{c})$ 
and $r_{c}$ respectively.}
\centering
\begin{ruledtabular}
\begin{tabular}{llllllll}
\multicolumn{8}{c}{WCP}     \\
\hline
 &  &  $\lambda_{1}=0.1$   &  $\lambda_{1}=0.1$  & $\lambda_{1}=0.5$  &
$\lambda_{1}=1$ & $\lambda_{1}=1.5$ & $\lambda_{1}=0.45$ \\
State & Quantity  &  $r_{c}=0.1$ & $r_{c}=0.5$ & $r_{c}=0.5$ & $r_{c}=1$ & $r_{c}=5$ & $r_{c}=\infty$ \\
\hline
   & $\mathcal{E}_{1,0}$  & 444.47894213 & 9.69364280 & 10.43995746 & 1.13262338 & $-$0.22737500 & $-$1.23411551 \\
   & $\left(\Delta V_{1,0}\right)^{2}$ & 1285.99378348 & 71.83641411 & 71.53062216  & 26.32910366 & 8.20218577 & 15.00733998 \\
1s & $\left(\Delta T_{1,0}\right)^{2}$ & 1285.99378348 & 71.83641411 & 71.53062216 &  26.32910366 & 8.20218577 & 15.00733998 \\
   & $\langle T \rangle_{1,0}\langle V \rangle_{1,0}-\langle TV \rangle_{1,0}$ & 1285.99378355 & 71.83641411 & 71.53062216 & 26.32910366 & 8.20218577 & 15.00733998 \\
   & $\langle T \rangle_{1,0}\langle V \rangle_{1,0}-\langle VT \rangle_{1,0}$ & 1285.99378355 & 71.83641411 & 71.53062216 & 26.32910366 & 8.20218577 & 15.00733998 \\
\hline
   & $\mathcal{E}_{2,0}$  & 1911.60619014 & 66.47853464 & 67.22135372 & 14.89326554 & 0.38477218 & $-$0.02806813 \\
   & $\left(\Delta V_{2,0}\right)^{2}$ & 3787.39749470 & 180.91460373 & 180.32580299 & 53.43300740 & 3.14499616 & 1.08218497 \\
2s & $\left(\Delta T_{2,0}\right)^{2}$ & 3787.39749470 & 180.91460373 & 180.32580299 & 53.43300740 & 3.14499616 & 1.08218497 \\
   & $\langle T \rangle_{2,0}\langle V \rangle_{2,0}-\langle TV \rangle_{2,0}$ & 3787.39749467 & 180.91460373 & 180.32580300 & 53.43300740 & 3.14499616 & 1.08218497 \\
   & $\langle T \rangle_{2,0}\langle V \rangle_{2,0}-\langle VT \rangle_{2,0}$ & 3787.39749467 & 180.91460373 & 180.32580300 & 53.43300740 & 3.14499616 & 1.08218497 \\
\hline
\multicolumn{8}{c}{ECSCP}   \\
\hline
 &  &  $\lambda_{2}=0.1$   &  $\lambda_{2}=0.1$  & $\lambda_{2}=0.5$  &
$\lambda_{2}=1$ & $\lambda_{2}=1.5$ & $\lambda_{2}=0.25$ \\
State & Quantity  &  $r_{c}=0.1$ & $r_{c}=0.5$ & $r_{c}=0.5$ & $r_{c}=1$ & $r_{c}=5$ & $r_{c}=\infty$ \\
\hline
   & $\mathcal{E}_{1,0}$  & 444.47943354 &  9.69592190 & 10.49107011 & 1.39032540 & 0.07291645 & $-$1.50671442 \\
   & $\left(\Delta V_{1,0}\right)^{2}$ & 1286.00324892 & 71.84954944 & 71.80723304 & 27.20420371 & 4.95684246 & 15.91293469 \\
1s & $\left(\Delta T_{1,0}\right)^{2}$ & 1286.00324892 & 71.84954944 & 71.80723304 & 27.20420371 & 4.95684246 & 15.91293469 \\
   & $\langle T \rangle_{1,0}\langle V \rangle_{1,0}-\langle TV \rangle_{1,0}$ & 1286.00324885 & 71.84954944 & 71.80723304 & 27.20420371 & 4.95684246 & 15.91293469 \\
   & $\langle T \rangle_{1,0}\langle V \rangle_{1,0}-\langle VT \rangle_{1,0}$ & 1286.00324885 & 71.84954944 & 71.80723304 & 27.20420371 & 4.95684246 & 15.91293469 \\
\hline
   & $\mathcal{E}_{2,0}$  & 1911.60668730 & 66.48097112 & 67.27483408  & 15.15621209 & 0.49330671 & $-$0.07314818 \\
   & $\left(\Delta V_{2,0}\right)^{2}$ & 3787.42042508 & 180.93992940 & 180.85424331 & 54.56130508 & 4.89088914  & 2.26239187 \\
2s & $\left(\Delta T_{2,0}\right)^{2}$ & 3787.42042508 & 180.93992940 & 180.85424331 & 54.56130508 & 4.89088914 & 2.26239187 \\
   & $\langle T \rangle_{2,0}\langle V \rangle_{2,0}-\langle TV \rangle_{2,0}$ & 3787.42042492 & 180.93992940 & 180.85424331 & 54.56130508 & 4.89088914 & 2.26239187 \\
   & $\langle T \rangle_{2,0}\langle V \rangle_{2,0}-\langle VT \rangle_{2,0}$ & 3787.42042492 & 180.93992940 & 180.85424331 & 54.56130508 & 4.89088914 & 2.26239187 \\
\hline
\multicolumn{8}{c}{SCP}   \\
\hline
State & Quantity &  $r_{c}=0.1$  &  $r_{c}=0.5$  & $r_{c}=1$  &
$r_{c}=2$ & $r_{c}=5$ & $r_{c}=10$ \\
\hline
   & $\mathcal{E}_{1,0}$  & 471.50566905 & 14.98747298 & 2.27917566 & $-$0.50537037 & $-$1.40602867 & $-$1.70075051 \\
   & $\left(\Delta V_{1,0}\right)^{2}$ & 1207.82025521 & 67.49838821 & 26.33780873 & 16.31778893 & 15.91807847 & 15.98870703 \\
1s & $\left(\Delta T_{1,0}\right)^{2}$ & 1207.82025521 & 67.49838821 & 26.33780873 & 16.31778893 & 15.91807847 & 15.98870703 \\
   & $\langle T \rangle_{1,0}\langle V \rangle_{1,0}-\langle TV \rangle_{1,0}$ & 1207.82025512 & 67.49838821 & 26.33780873 & 16.31778893 & 15.91807847 & 15.98870703 \\
   & $\langle T \rangle_{1,0}\langle V \rangle_{1,0}-\langle VT \rangle_{1,0}$ & 1207.82025512 & 67.49838821 & 26.33780873 & 16.31778893 & 15.91807847 & 15.98870703 \\
\hline
   & $\mathcal{E}_{2,0}$  & 1938.19369550 & 71.63098684 & 15.97749590 & 3.00469070 & 0.09030651 & $-$0.21064190 \\
   & $\left(\Delta V_{2,0}\right)^{2}$ & 3589.04237047 & 172.06135205 & 53.23498447 & 18.29571481 & 3.69613370 & 2.88000313 \\
2s & $\left(\Delta T_{2,0}\right)^{2}$ & 3589.04237047 & 172.06135205 & 53.23498447 & 18.29571481 & 3.69613370 & 2.88000313 \\
   & $\langle T \rangle_{2,0}\langle V \rangle_{2,0}-\langle TV \rangle_{2,0}$ & 3589.04237064 & 172.06135205 & 53.23498447 & 18.29571481 & 3.69613370 & 2.88000313 \\
   & $\langle T \rangle_{2,0}\langle V \rangle_{2,0}-\langle VT \rangle_{2,0}$ & 3589.04237064 & 172.06135205 & 53.23498447 & 18.29571481 & 3.69613370 & 2.88000313 \\
\end{tabular}
\end{ruledtabular}
\end{table}  
\endgroup  
 
\subsection{Virial-like Theorem}
As mentioned in Sec.~II.A, the conventional VT is not satisfied in confined condition. Recently \cite{mukherjee19}, a virial-like expression is derived  
and successfully applied in H-atom trapped in various confined environment \cite{mukherjee19}. It is found that, at the end, the perturbing potential 
does not appear in the final expression. In this subsection, we are probing this theorem in the context of WCP, ECSCP and SCP successively.

In WCP, the necessary expectation values will take the form,
\begin{equation}\label{eq:46}
\begin{aligned}
\langle T V \rangle_{n,\ell} & = \left\langle T \left(-\frac{Z}{r}e^{(-\lambda_{1}r)} \right) \right\rangle_{n,\ell}, \ \ \   
\langle V T \rangle_{n,\ell}  = \left\langle \left(-\frac{Z}{r}e^{(-\lambda_{1}r)} \right) T \right\rangle_{n,\ell}, \\
\langle V^{2} \rangle_{n,\ell} & = \left\langle \frac{Z^{2}}{r^{2}}e^{(-2\lambda_{1}r)} \right\rangle_{n,\ell}, \ \ \   
\langle V \rangle_{n,\ell}  = \left\langle -\frac{Z}{r}e^{(-\lambda_{1}r)}  \right\rangle_{n,\ell}.
\end{aligned}            
\end{equation} 
Now, applying the expression of Eq.~(\ref{eq:46}) in Eq.~(\ref{eq:4}) we obtain,
\begin{equation}\label{eq:47}
\begin{aligned}
\langle T^{2} \rangle_{n,\ell} - \langle T \rangle^{2}_{n,\ell}  =  (\Delta T_{n,\ell})^{2} = \langle V^{2} \rangle_{n,\ell} - \langle V \rangle^{2}_{n,\ell} = (\Delta V_{n,\ell})^{2} \\
=\left\langle  \frac{Z^{2}}{r^{2}}e^{(-2\lambda_{1}r)} \right\rangle_{n,\ell} - \left\langle \frac{Z}{r}e^{(-\lambda_{1}r)} \right\rangle^{2}_{n,\ell} \\
= \langle T \rangle_{n,\ell} \left\langle -\frac{Z}{r}e^{(-\lambda_{1}r)} \right\rangle_{n,\ell}-
\left\langle T \left(-\frac{Z}{r}e^{(-\lambda_{1}r)} \right) \right\rangle_{n,\ell}.
\end{aligned}
\end{equation}
The relevant expectation values in ECSCP are expressed as,
\begin{equation}\label{eq:48}
\begin{aligned}
\langle T V \rangle_{n,\ell} & = \left\langle T \left(-\frac{Z}{r}e^{(-\lambda_{2}r)} \cos  \lambda_{2}r  \right) \right\rangle_{n,\ell}, \ \ \   
\langle V T \rangle_{n,\ell}  = \left\langle \left(-\frac{Z}{r}e^{(-\lambda_{1}r)} \cos  \lambda_{2}r \right) T \right\rangle_{n,\ell}, \\
\langle V^{2} \rangle_{n,\ell} & = \left\langle \frac{Z^{2}}{r^{2}}e^{(-2\lambda_{2}r)} \cos^{2} \lambda_{2}r \right\rangle_{n,\ell}, \ \ \   
\langle V \rangle_{n,\ell}  = \left\langle -\frac{Z}{r}e^{(-\lambda_{2}r)} \cos  \lambda_{2}r \right\rangle_{n,\ell}.
\end{aligned}            
\end{equation} 
Now, substituting the results of Eq.~(\ref{eq:48}) in Eq.~(\ref{eq:4}) we achieve,
\begin{equation}\label{eq:49}
\begin{aligned}
\langle T^{2} \rangle_{n,\ell} - \langle T \rangle^{2}_{n,\ell}  =  (\Delta T_{n,\ell})^{2} = \langle V^{2} \rangle_{n,\ell} - \langle V \rangle^{2}_{n,\ell} = (\Delta V_{n,\ell})^{2} \\
=\left\langle  \frac{Z^{2}}{r^{2}}e^{(-2\lambda_{2}r)} \cos^{2} \lambda_{2}r \right\rangle_{n,\ell} - \left\langle \frac{Z}{r}e^{(-\lambda_{2}r)} \cos  \lambda_{2}r \right\rangle^{2}_{n,\ell} \\
= \langle T \rangle_{n,\ell} \left\langle -\frac{Z}{r}e^{(-\lambda_{2}r)} \cos \lambda_{2}r \right\rangle_{n,\ell}-
\left\langle T \left(-\frac{Z}{r}e^{(-\lambda_{2}r)} \right) \cos \lambda_{2}r \right\rangle_{n,\ell}.
\end{aligned}
\end{equation}
In SCP, the respective expectation values are manifested as,
\begin{equation}\label{eq:50}
\begin{aligned}
\langle T V \rangle_{n,\ell} & = \left\langle T \left[-\frac{Z}{r}+\frac{(Z-N_{e})}{R}\left(3-\left(\frac{r}{R}\right)^{2}\right)\right] \right\rangle_{n,\ell}, \\ 
\langle V T \rangle_{n,\ell} &  = \left\langle \left[-\frac{Z}{r}+\frac{(Z-N_{e})}{R}\left(3-\left(\frac{r}{R}\right)^{2}\right)\right] T \right\rangle_{n,\ell}, \\
\langle V^{2} \rangle_{n,\ell} & = \left\langle \left[-\frac{Z}{r}+\frac{(Z-N_{e})}{R}\left(3-\left(\frac{r}{R}\right)^{2}\right)\right]^{2} \right\rangle_{n,\ell}, \\
\langle V \rangle_{n,\ell} & = \left\langle \left[-\frac{Z}{r}+\frac{(Z-N_{e})}{R}\left(3-\left(\frac{r}{R}\right)^{2}\right)\right]  \right\rangle_{n,\ell}.
\end{aligned}            
\end{equation} 
Finally, engaging the outcome of Eq.~(\ref{eq:50}) in Eq.~(\ref{eq:4}) becomes,   
\begin{equation}\label{eq:51}
\begin{aligned}
\langle T^{2} \rangle_{n,\ell} - \langle T \rangle^{2}_{n,\ell}  = (\Delta T_{n,\ell})^{2}  = \langle V^{2} \rangle_{n,\ell} - \langle V \rangle^{2}_{n,\ell}  = (\Delta V_{n,\ell})^{2} \\
=\left\langle \left[-\frac{Z}{r}+\frac{(Z-N_{e})}{R}\left(3-\left(\frac{r}{R}\right)^{2}\right)\right]^{2} \right\rangle_{n,\ell} - 
\left\langle \left[-\frac{Z}{r}+\frac{(Z-N_{e})}{R}\left(3-\left(\frac{r}{R}\right)^{2}\right)\right]  \right\rangle_{n,\ell}^{2} \\
= \langle T \rangle_{n,\ell} \left\langle \left[-\frac{Z}{r}+\frac{(Z-N_{e})}{R}\left(3-\left(\frac{r}{R}\right)^{2}\right)\right]  \right\rangle_{n,\ell}-
\left\langle T \left[-\frac{Z}{r}+\frac{(Z-N_{e})}{R}\left(3-\left(\frac{r}{R}\right)^{2}\right)\right] \right\rangle_{n,\ell}.
\end{aligned}
\end{equation}

\begingroup           %%Table 3, dipole os
\squeezetable
\begin{table}
\caption{$f^{(1)}$ values for WCP, ECSCP (in free and confined conditions) and SCP involving $ns \rightarrow 2p$, $ns \rightarrow 3p$ ($n=1,2$) transitions. See text for details.}
\centering
\begin{ruledtabular}
\begin{tabular}{l|llllllll}
Transition &  \multicolumn{6}{c|}{Confined WCP}  &   \multicolumn{2}{c}{Free WCP} \\
 \cline{2-7} \cline{8-9}
	& $\lambda_{1}$ & $r_{c}=0.1$ & $r_{c}=0.5$ & $r_{c}=1$ & $r_{c}=2$ & \multicolumn{1}{c|}{$r_{c}=5$} & $\lambda_{1}$ & $r_{c}=\infty$ \\
\cline{1-1} \cline{2-7} \cline{8-9}
$1s \rightarrow 2p$   &  $0.1$  &   0.97072714   & 0.98455633 & 0.99105667  &  0.92744965   & \multicolumn{1}{c|}{0.48674542}   &  $0.1$  & 0.40181907  \\
		      &  $0.5$  &   0.97072657   & 0.98450970 & 0.99101296  &  0.93172951   & \multicolumn{1}{c|}{0.42593118}   &  $0.2$  & 0.36301391  \\
		      &  $1  $  &   0.97072481   & 0.98437662 & 0.99088526  &  0.94270974   & \multicolumn{1}{c|}{0.42487213}   &  $0.3$  & 0.29859664  \\
		      &  $2.2$  &   0.97071618   & 0.98380464 & 0.99014361  &  0.97196847   & \multicolumn{1}{c|}{0.84116523}   &  $0.4$  & 0.19333749  \\       
$1s \rightarrow 3p$   &  $0.1$  &   0.02145207   & 0.00772756 &   0.00000194        & 0.04896547    & \multicolumn{1}{c|}{0.30906124} &  $0.01$   & 0.07892729 \\
		      &  $0.5$  &   0.02145255   & 0.00776480 &   0.00000008       & 0.04498399    &  \multicolumn{1}{c|}{0.32619988}  &  $0.05$   & 0.07536052 \\
		      &  $1  $  &   0.02145402   & 0.00787227 &   0.00002302       & 0.03498548    &  \multicolumn{1}{c|}{0.07783255}  &  $0.1$    & 0.06581437 \\
		      &  $2.2$  &   0.02146127   & 0.00834459 &   0.00047966       & 0.00951569    &  \multicolumn{1}{c|}{0.26193927}  &  $0.2$    & 0.02982086 \\
$2s \rightarrow 2p$   &  $0.1$  & $-$0.59617944 & $-$0.60825425  & $-$0.61188356  & $-$0.54000701  &  \multicolumn{1}{c|}{$-$0.06993817}  & $0.1$   & 0.01961263 \\
		      &  $0.4$  &  $-$0.59617891 & $-$0.60820263  & $-$0.61167657  & $-$0.54121728 &  \multicolumn{1}{c|}{0.00417437}  &   $0.2$   & 0.07522974 \\
		      &  $0.5$  &  $-$0.59617859 & $-$0.60817302  & $-$0.61156439  & $-$0.54200502 &  \multicolumn{1}{c|}{0.03181610}  &   $0.3$   & 0.17737202 \\
		      &  $1  $  &  $-$0.59617598 & $-$0.60794575  & $-$0.61078004  & $-$0.54873672 &  \multicolumn{1}{c|}{0.07905123}  &   $0.4$   & 0.37896055      \\
$2s \rightarrow 3p$   &  $0.1$  &   1.53239528   & 1.56032134     & 1.57779183     &  1.51296821   &  \multicolumn{1}{c|}{0.96212776}  &   $0.01$  & 0.43399889      \\
		      &  $0.4$  &   1.53239452   & 1.56024714     & 1.57830084     &  1.51388585   &  \multicolumn{1}{c|}{0.90876197}  &   $0.05$  & 0.41594460      \\
		      &  $0.5$  &   1.53239406   & 1.56020449     & 1.57797157     &  1.51453142   &  \multicolumn{1}{c|}{0.88783614}  &   $0.1$   & 0.36639711      \\
		      &  $1  $  &   1.53239034   & 1.55987605     & 1.57652076     &  1.52043617   &  \multicolumn{1}{c|}{0.85542392}  &   $0.2$   & 0.17105455      \\
\hline
Transition &  \multicolumn{6}{c|}{Confined ECSCP}   &   \multicolumn{2}{c}{Free ECSCP} \\
 \cline{2-7} \cline{8-9}
	& $\lambda_{2}$ & $r_{c}=0.1$ & $r_{c}=0.5$ & $r_{c}=1$ & $r_{c}=2$ & \multicolumn{1}{c|}{$r_{c}=5$}  & $\lambda_{2}$ & $r_{c}=\infty$ \\
\cline{1-1} \cline{2-7} \cline{8-9}
$1s \rightarrow 2p$   &  $0.1$  & 0.97072717  & 0.98455833      & 0.99105855  & 0.92726266   & \multicolumn{1}{c|}{0.49020012}  &  $0.05$  & 0.41541265                     \\
		      &  $0.5$  & 0.97072715  & 0.98455130      & 0.99103661  & 0.92859601   & \multicolumn{1}{c|}{0.39261746}  &  $0.1$   & 0.41059123                     \\
		      &  $1  $  & 0.97072703  & 0.98450670      & 0.99091497  & 0.93723099   & \multicolumn{1}{c|}{0.38873757}  &  $0.2$   & 0.37680897      \\
		      &  $1.4$  & 0.97072679  & 0.98442680      & 0.99070952  & 0.95086935   & \multicolumn{1}{c|}{0.72484192}  &  $0.25$  & 0.33815629      \\
$1s \rightarrow 3p$   &  $0.1$  & 0.02145205  & 0.00772596      & 0.00000215  & 0.04914057   & \multicolumn{1}{c|}{0.25938284}  &  $0.01$  & 0.07908337      \\
		      &  $0.5$  & 0.02145206  & 0.00773097      & 0.00000111  & 0.04772537   & \multicolumn{1}{c|}{0.32705934}  &  $0.05$  & 0.07727923      \\
		      &  $1$    & 0.02145216  & 0.00776310      & 0.00000195  & 0.03909999   & \multicolumn{1}{c|}{0.33012451}  &  $0.1$   & 0.06672974      \\
		      &  $1.4$  & 0.02145234  & 0.00782141      & 0.00003015  & 0.02603893   & \multicolumn{1}{c|}{0.12347883}  &  $0.12$  & 0.05778373      \\
$2s \rightarrow 2p$   &  $0.05$ & $-$0.59617948 & $-$0.60825787 & $-$0.61189909  & $-$0.53993757  & \multicolumn{1}{c|}{$-$0.07746440} &  $0.05$  & 0.00099592      \\
		      &  $0.1$  & $-$0.59617948 & $-$0.60825776 & $-$0.61189792  & $-$0.53993008  & \multicolumn{1}{c|}{$-$0.07524706} &  $0.1$   & 0.00719122                     \\
		      &  $0.5$  & $-$0.59617945 & $-$0.60824322 & $-$0.61176097  & $-$0.53953695  & \multicolumn{1}{c|}{0.07707677}    &  $0.2$   & 0.05235875      \\
		      &  $1  $  & $-$0.59617924 & $-$0.60815247 & $-$0.61102508  & $-$0.54115043  & \multicolumn{1}{c|}{0.16873325}    &  $0.25$  & 0.10793827      \\
$2s \rightarrow 3p$   &  $0.05$ & 1.53239533   & 1.56032654     & 1.57832534     & 1.51292362     & \multicolumn{1}{c|}{0.96729789}  &  $0.01$  & 0.43478301      \\
		      &  $0.1$  & 1.53239533   & 1.56032638     & 1.57832365     & 1.51291209     & \multicolumn{1}{c|}{0.96572414}  &  $0.05$  & 0.42653220      \\
		      &  $0.5$  & 1.53239529   & 1.56030638     & 1.57812514     & 1.51213836     & \multicolumn{1}{c|}{0.84864260}  &  $0.1$   & 0.37668282      \\
		      &  $1  $  & 1.53239501   & 1.56018160     & 1.57705831     & 1.51229530     & \multicolumn{1}{c|}{0.76798664}  &  $0.12$  & 0.33241753      \\
\hline
\multicolumn{9}{c}{SCP}     \\
\hline
Transition & $r_{c}=0.1$ & $r_{c}=0.2 $ & $r_{c}=0.5$ & $r_{c}=1$ & $r_{c}=2$ & $r_{c}=2.5$ & $r_{c}=5$  & $r_{c}=10$ \\
\hline
	$1s \rightarrow 2p$   &  0.97051035   &  0.97420550 & 0.98379490  & 0.99067302  & 0.92958863  & 0.84910611  & 0.46356524  & 0.40514594  \\
	$1s \rightarrow 3p$   &  0.02161960   &  0.01795866 & 0.00826516  & 0.00002030  & 0.04648691  & 0.10750364  & 0.27699337  & 0.10047172  \\
	$2s \rightarrow 2p$   & $-$0.59580794 &  $-$0.59894650 & $-$0.60667433 & $-$0.60974041 & $-$0.53810301 & $-$0.45106011 & $-$0.03771967 & 0.01415902       \\
	$2s \rightarrow 3p$   & 1.53188858    &  1.53916195 & 1.55815448  & 1.57522845    & 1.51008687    & 1.41740733  & 0.93835324  & 0.51932973       \\

\end{tabular}
\end{ruledtabular}
\end{table}  
\endgroup  

The upper segment of Table~II represents results for WCP in $1s$ and $2s$ states at six different sets of $\{\lambda_{1}, r_{c}\}$ values, 
namely $(0.1,0.1),(0.1,0.5),(0.5,0.5),(1,1),(1.5,5),(0.45,\infty)$. In all these cases, Eq.~(\ref{eq:4}) of Sec.~II.A is corroborated. 
More importantly, in a given state, at a fixed $r_{c}$, energy increases with $\lambda_{1}$. Similarly, at a certain $\lambda_{1}$, 
it declines with rise in $r_{c}$. In the middle portion, the corresponding outcomes are tabulated for ECSCP at six chosen 
$(\lambda_{2}, r_{c})$ values, \emph{viz.}, $(0.1,0.1),(0.1,0.5),(0.5,0.5),(1,1),(1.5,5),(0.25,\infty)$. Again, these data support 
the conclusion drawn from Eq.~(\ref{eq:4}). Like the WCP, here also energy enhances with $\lambda_{2}$ at fixed $r_c$, and diminishes
with $r_{c}$ at a specific $\lambda_2$. In the bottom part, numerical data about the validity of VT in the context of SCP are presented. 
Like the earlier two cases, this also satisfies Eq.~(\ref{eq:4}). It may be mentioned that, a few attempts were made before to establish 
such a theorem in confined condition (that includes plasma environment), by means of Hellmann-Feynman theorem and conventional VT 
\cite{katriel12,montgomery18}. There, the mathematical form of the expression changes from system to system; the present form, on the 
other hand, provides a uniform mathematical expression \emph{irrespective of the system of interest}.

\begin{figure}                         %%%Fig. 4, wcp
\begin{minipage}[c]{0.48\textwidth}\centering
\includegraphics[scale=0.78]{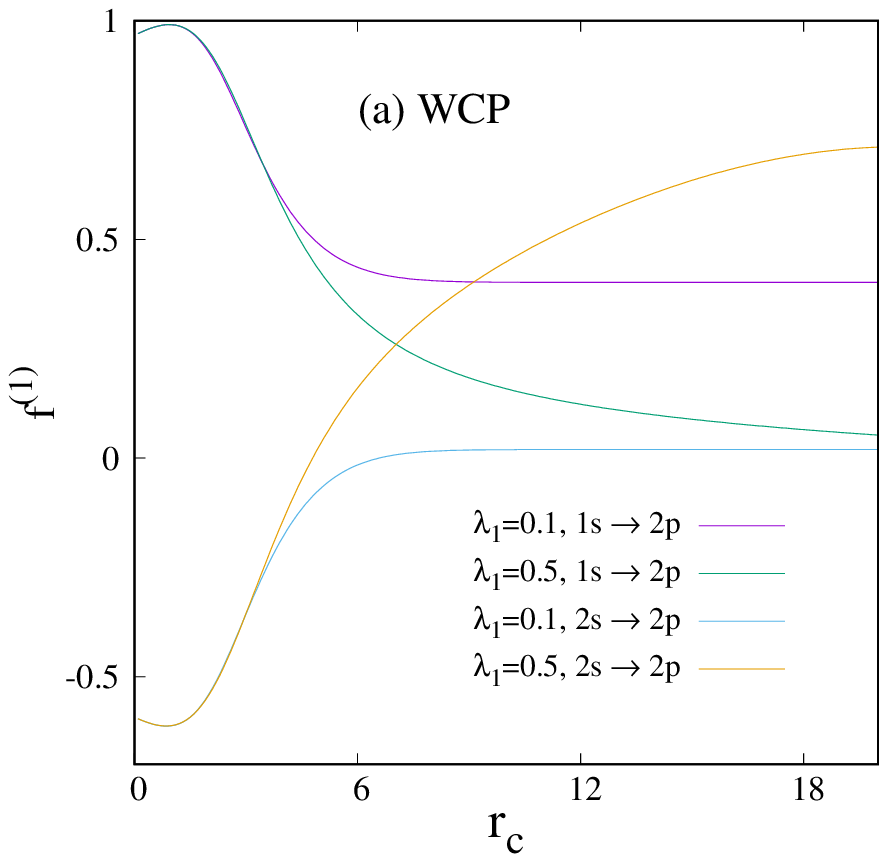}
\end{minipage}%
\begin{minipage}[c]{0.48\textwidth}\centering
\includegraphics[scale=0.78]{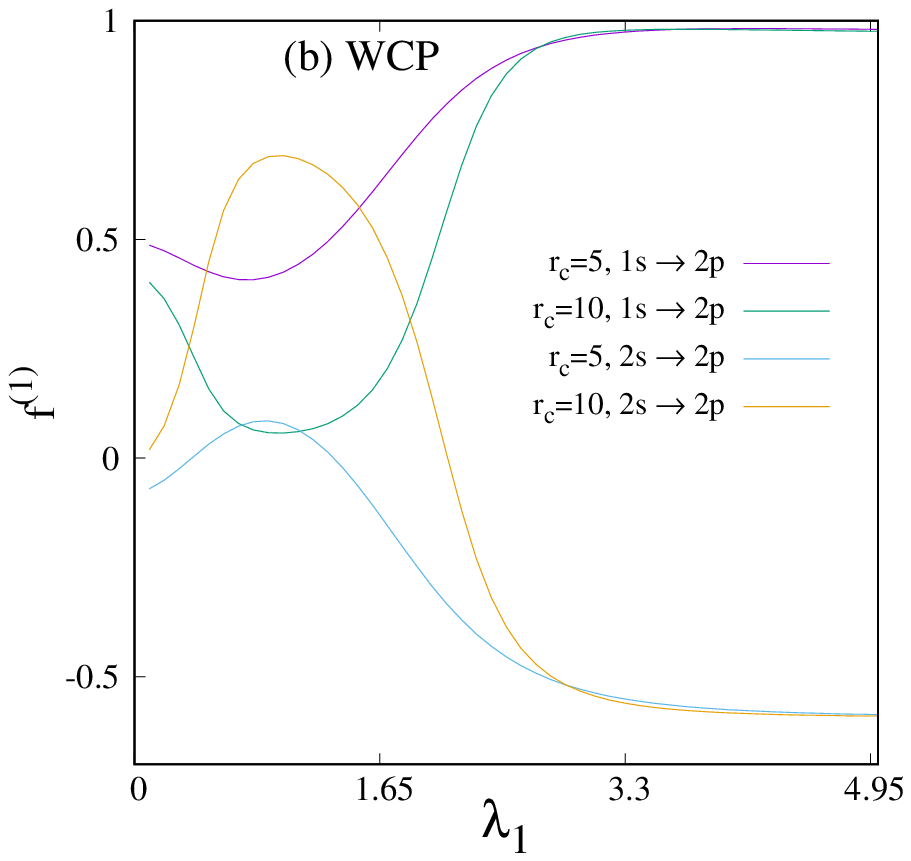}
\end{minipage}%
\caption{$f^{(1)}_{ns \rightarrow 2p} (n=1,2)$ for WCP. Panel (a) gives $r_{c}$ variation at two selected $\lambda_{1}$ (0.1, 0.5), while panel 
(b) shows $\lambda_{1}$ variation at two different $r_{c}$ (5, 10). See text for details.}
\end{figure}  

\subsection{Multipole oscillator strengths and polarizabilities} 
In the following discussion, $Z=2$ is chosen; that means, in SCP, $\beta$ only depends on $r_{c}$. Hence, 
in SCP, the results are provided with respect to variation of $r_{c}$ only. It may be recalled from Sec.~II.C that in SCP, 
$Z$ is required to be greater than 1. That is why, we have selected $Z=2$ in stead of 1, for all three environments. 
Note that, results for $Z=1$ in \emph{free} WCP and 
ECSCP were also calculated. They are found to be in consonance with available literature (see, e.g., \cite{zhu20, jiao21}, and 
references therein). In this work, the primary focus, however, lies on \emph{confined} plasma systems. The multipole OS sum rule given 
in Eq.~(\ref{eq:a}) were estimated in both $1s$ and $2s$ states, involving all four $k$ ($k=1,2,3,4$). In both \emph{free} and 
\emph{confined} conditions, this equation was obeyed. Further, this sum rule remains invariant under scaling transformations.         

\begin{figure}                         %%%Fig. 5, ecscp
\begin{minipage}[c]{0.48\textwidth}\centering
\includegraphics[scale=0.72]{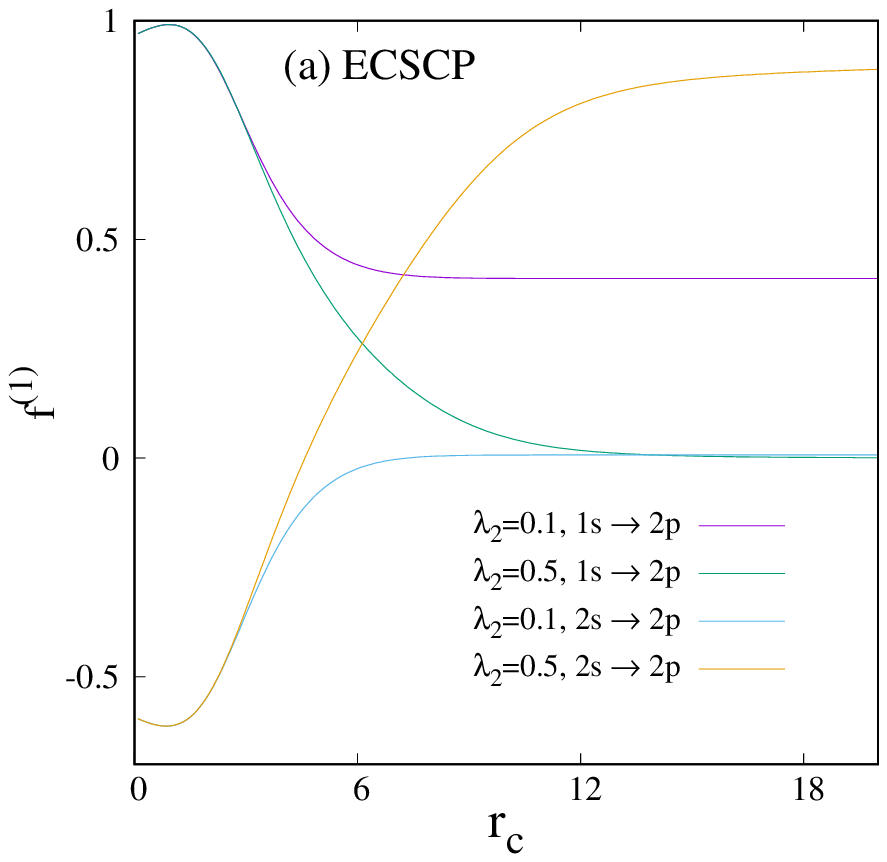}
\end{minipage}%
\begin{minipage}[c]{0.48\textwidth}\centering
\includegraphics[scale=0.72]{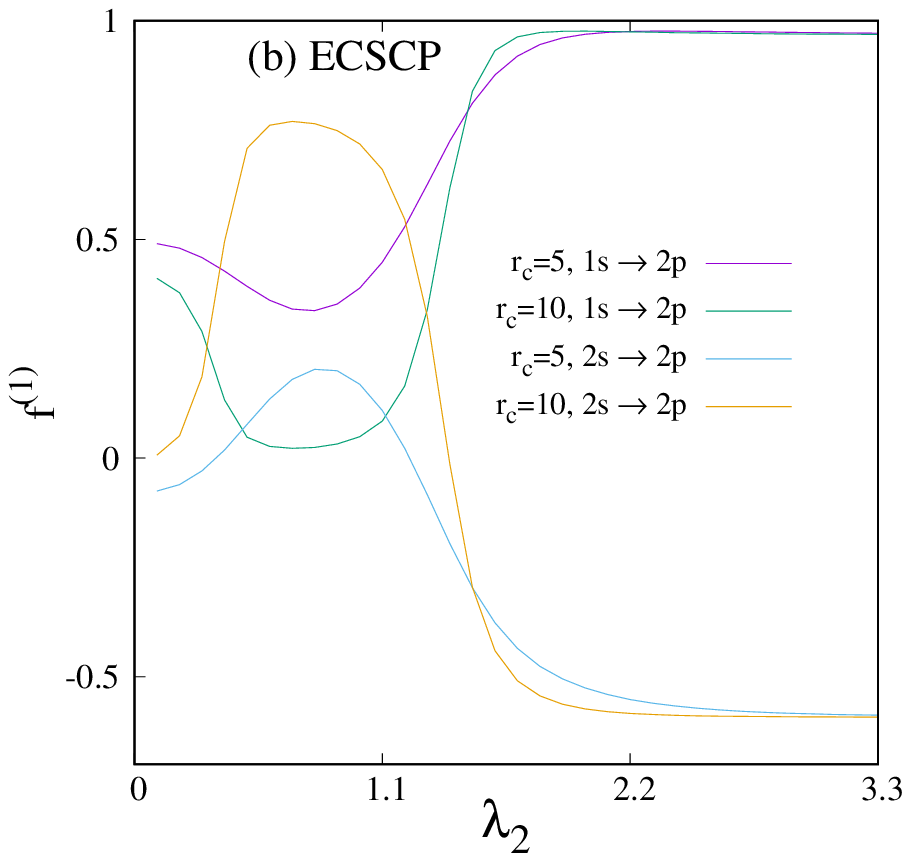}
\end{minipage}%
\caption{$f^{(1)}_{ns \rightarrow 2p} (n=1,2)$ for ECSCP. Panel (a) gives $r_{c}$ variation at two selected $\lambda_{2}$ (0.1, 0.5), 
while panel (b) shows $\lambda_{2}$ variation at two different $r_{c}$ (5, 10). See text for details.}
\end{figure}  

\begingroup           %%Table 4, dipole polarizability
\squeezetable
\begin{table}
\caption{$\alpha^{(1)}$ for WCP, ECSCP (in free and confined conditions) and SCP in $1s$ and $2s$ states.}
\centering
\begin{ruledtabular}
\begin{tabular}{l|lllllllll}
    &  \multicolumn{7}{c|}{Confined WCP}  &   \multicolumn{2}{c}{Free WCP} \\
 \cline{2-8} \cline{9-10}
	State  & $\lambda_{1}$ & $r_{c}=0.1$ & $r_{c}=0.5$ & $r_{c}=1$ & $r_{c}=2$ & $r_{c}=3$ & \multicolumn{1}{c|}{$r_{c}=5$} & $\lambda_{1}$ & $r_{c}=\infty$ \\
\cline{1-1} \cline{2-8} \cline{9-10}
$1s$   &  $0.1$  & 0.00000348 & 0.00179958  & 0.02141842 & 0.14911950 & 0.25573621 &  \multicolumn{1}{c|}{0.28428984}   &  $0.05$ & 0.2820913 \\
       &  $0.5$  & 0.00000348 & 0.00180142  & 0.02159981 & 0.15905913 & 0.29929691 &  \multicolumn{1}{c|}{0.35670443}   &  $0.1$  & 0.2845122 \\
       &  $1$    & 0.00000348 & 0.00180662  & 0.02207243 & 0.18422797 & 0.43081698 &  \multicolumn{1}{c|}{0.66907982}   &  $0.2$  & 0.2937360   \\
       &  $2$    & 0.00000348 & 0.00182404  & 0.02347293 & 0.25795381 & 0.96822978 &  \multicolumn{1}{c|}{4.64009044}   &  $0.25$ & 0.3004862   \\       
       &  $2.5$  & 0.00000348 & 0.00183517  & 0.02427643 & 0.29812432 & 1.30252206 &  \multicolumn{1}{c|}{8.89513209}   &  $0.3$  & 0.3086925  \\
       &  $3$    & 0.00000348 & 0.00184738  & 0.02509476 & 0.33616289 & 1.60942186 &  \multicolumn{1}{c|}{12.66519620}  &  $0.4$  & 0.3297730  \\
$2s$   &  $0.1$  & 0.000000784 & 0.00027589  & $-$0.00105258   & $-$0.30696765  & $-$4.32610957 & \multicolumn{1}{c|}{$-$150.21895140}  &  $0.05$  & 4035.9536  \\
       &  $0.5$  & 0.000000784 & 0.00027635  & $-$0.00105258   & $-$0.30187923  & $-$4.62797003 & \multicolumn{1}{c|}{503.53816745}  &  $0.1$      & 1161.9265 \\
       &  $1$    & 0.000000784 & 0.00027768  & $-$0.00081299   & $-$0.27873557  & $-$4.40335922 & \multicolumn{1}{c|}{287.82230501}  &  $0.2$      & 419.04688 \\
       &  $2$    & 0.000000784 & 0.00028251  & $-$0.00020805   & $-$0.19039680  & $-$2.24216995 & \multicolumn{1}{c|}{$-$53.18407330}  &  $0.25$   & 339.24125 \\
       &  $2.5$  & 0.000000784 & 0.00028579  & 0.00017337      & $-$0.14358283  & $-$1.42113266 & \multicolumn{1}{c|}{$-$18.92337897}  &  $0.3$    & 312.66238 \\
       &  $3$    & 0.000000784 & 0.00028955 &  0.00058151      & $-$0.10232148  & $-$0.87783191 & \multicolumn{1}{c|}{$-$8.40883662}  &  $0.4$     & 393.94908 \\
\hline
 &  \multicolumn{7}{c|}{Confined ECSCP}   &   \multicolumn{2}{c}{Free ECSCP} \\
 \cline{2-8} \cline{9-10}
	State  & $\lambda_{2}$ & $r_{c}=0.1$ & $r_{c}=0.5$ & $r_{c}=1$ & $r_{c}=2$ & $r_{c}=3$  & \multicolumn{1}{c|}{$r_{c}=5$}  & $\lambda_{2}$ & $r_{c}=\infty$ \\
\cline{1-1} \cline{2-8} \cline{9-10}
$1s$  &  $0.1$  & 0.00000348 & 0.00179950 & 0.02141041 & 0.14867869 & 0.25393187   &  \multicolumn{1}{c|}{0.28156914}   &  $0.01$  & 0.2812505    \\
      &  $0.5$  & 0.00000348 & 0.00179980 & 0.02146640 & 0.15382762 & 0.28118411   &  \multicolumn{1}{c|}{0.33101794}   &  $0.05$  & 0.2813193    \\
      &  $1$    & 0.00000348 & 0.00180171 & 0.02178057 & 0.18029000 & 0.45239114   &  \multicolumn{1}{c|}{0.91168796}   &  $0.1$   & 0.2817742    \\
      &  $1.25$ & 0.00000348 & 0.00180359 & 0.02206630 & 0.20378040 & 0.65045235   &  \multicolumn{1}{c|}{2.57374006}   &  $0.2$   & 0.2850410    \\       
      &  $1.4$  & 0.00000348 & 0.00180507 & 0.02227983 & 0.22117395 & 0.81774044   &  \multicolumn{1}{c|}{4.89195099}   &  $0.25$  & 0.2883457   \\
$2s$  &  $0.1$  & 0.00000078 & 0.00027587 & $-$0.00105523 & $-$0.30712416 & $-$4.30664314  &  \multicolumn{1}{c|}{$-$136.34737101}  &  $0.01$   & 89914.38   \\
      &  $0.5$  & 0.00000078 & 0.00027591 & $-$0.00104609 & $-$0.30960540 & $-$4.85239904  &  \multicolumn{1}{c|}{230.50804606}  &  $0.05$   & 20470.3544   \\
      &  $1$    & 0.00000078 & 0.00027620 & $-$0.00098316 & $-$0.30418590 & $-$5.36027549  &  \multicolumn{1}{c|}{171.45198632}  &  $0.1$    & 2954.0860   \\
      &  $1.25$ & 0.00000078 & 0.00027650 & $-$0.00091626 & $-$0.28685821 & $-$4.42905058  &  \multicolumn{1}{c|}{$-$831.87504495}  &  $0.2$    & 546.52109   \\
      &  $1.4$  & 0.00000078 & 0.00027674 & $-$0.00086180 & $-$0.27036124 & $-$3.63524261  &  \multicolumn{1}{c|}{$-$91.86677448}  &  $0.25$   & 386.92234   \\
\hline
\multicolumn{9}{c}{SCP}     \\
\hline
 State  & $r_{c}=0.1$ & $r_{c}=0.2 $ & $r_{c}=0.5$ & $r_{c}=1$ & $r_{c}=2$ & $r_{c}=2.5$ & $r_{c}=3$ & $r_{c}=5$  & $r_{c}=10$ \\
\hline
$1s$   & 0.00000349 & 0.00005369 & 0.00183246 & 0.022259 & 0.160406 & 0.232152 & 0.273467 & 0.288320 & 0.282136 \\
$2s$   & 0.00000078 & 0.00001131 & 0.00028033 & 0.011415 & 0.133172 & 0.275261 & 0.486354 & 2.354053 & 8.230057
\end{tabular}
\end{ruledtabular}
\end{table}  
\endgroup  

The OS, in practice, measures the probability of transition between an initial to a final state. The dipole OS for first two 
$\ell =0$ states of WCP, ECSCP, SCP are presented in top, middle and bottom portions of Table~III respectively. These changes do not seem to be 
straight forward. At $\lambda_1 \rightarrow 0$ (WCP) and $\lambda_2 \rightarrow 0$ (ECSCP), these results coalesce to FHA. On the other side, 
OS in SCP approach FHA 
in the limit of $r_{c} \rightarrow \infty$. The selection rule is $\Delta \ell = \pm 1$; therefore, only p-wave states are permitted 
as final states. In all three occasions, these are provided for $ns \rightarrow mp$ ($n=1,2; m=2,3$) states, in both \emph{free} and 
\emph{confined} conditions. In the \emph{first two} plasma conditions, $f^{(1)}_{1s \rightarrow 2p}$ lowers at strong confinement regime 
($r_{c} \le 1$), with rise in screening constant, keeping $r_{c}$ fixed. But in low-moderate $r_{c}$ $(1,2)$ it increases with 
$\lambda$. However, at $r_{c}=5$, it reduces to attain a minimum and then grows gradually. Interestingly, 
in free condition, it again declines with advancement of $\lambda$. On the other hand, in either of the plasmas, at a fixed $\lambda$, 
it increases with $r_{c}$, then reaches a maximum and eventually falls off. The positions of the maxima do not change with $\lambda$. 
In SCP also, $f^{(1)}_{1s \rightarrow 2p}$ imprints a similar behavior; initially gains to reach a maximum and then declines. It can 
thus be stated that, in all these three plasmas, there is an optimum $T$ (refer to Sec.~II.C) at which the probability of transition attains a 
maximum. Moreover, with growth in $r_{c}$ and $T$, the plasma tail effect predominates. At $r_{c}=0.1$ and $0.5$, only minor changes occur in 
$f^{(1)}_{1s \rightarrow 3p}$ with progress in $\lambda$, in both WCP and ECSCP. However, at $r_{c}=1$, though the values are significantly small, 
but nevertheless there appears a minimum in $f^{(1)}_{1s \rightarrow 3p}$ versus $\lambda$ plots, in both plasmas. At $r_{c}=2$, it 
decays with growth in $\lambda$. Further, at $r_{c}=5$, there appears a maximum in $f^{(1)}_{1s \rightarrow 3p}$ against $\lambda_{2}$ plot in 
ECSCP. However, in the similar plot for WCP, one finds a maximum followed by a minimum. On the contrary, at a fixed $\lambda$, in WCP and 
ECSCP with rise in $r_{c}$, $f^{(1)}_{1s \rightarrow 3p}$ decreases to reach a minimum and then increases. But in SCP, at first, there occurs 
a minimum followed by a maximum. Thus, with rise in $T$, the probability of transition from $1s$ to $3p$ decreases initially in all 
these three potentials, and increases thereafter. It is noticed that, $f^{(1)}_{1s \rightarrow mp} (m=2,3)$ in \emph{free} WCP and ECSCP reduces
with progress in $\lambda$.  

\begin{figure}                    %%%Fig. 6, ecscp
\begin{minipage}[c]{0.48\textwidth}\centering
\includegraphics[scale=0.78]{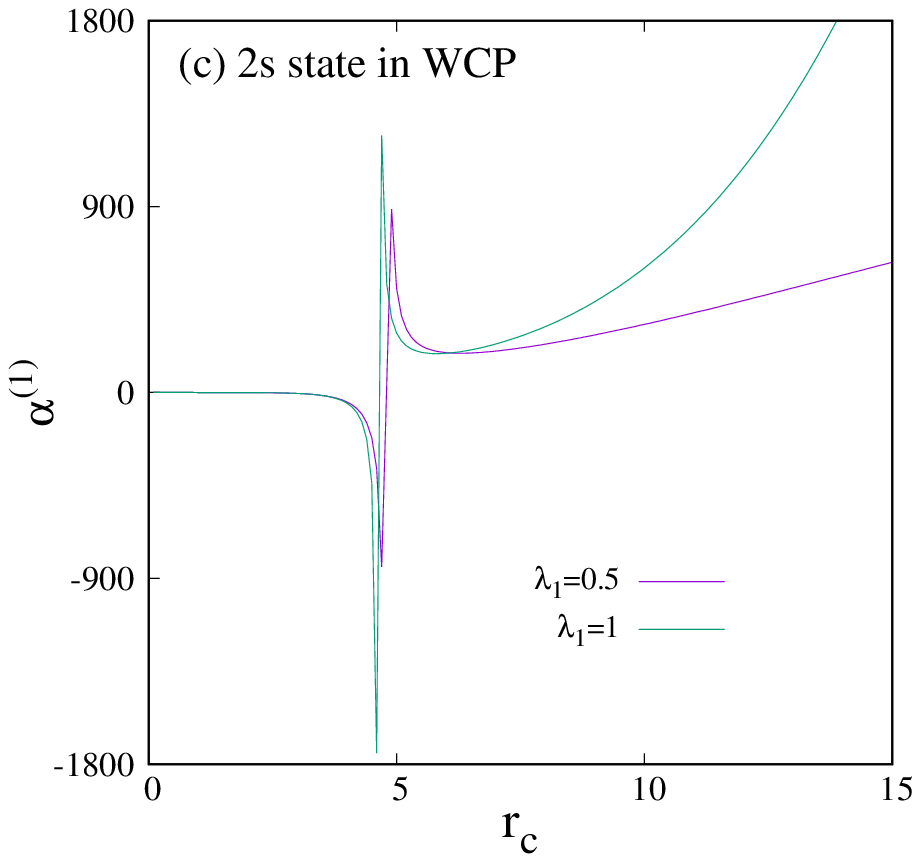}
\end{minipage}%
\begin{minipage}[c]{0.48\textwidth}\centering
\includegraphics[scale=0.78]{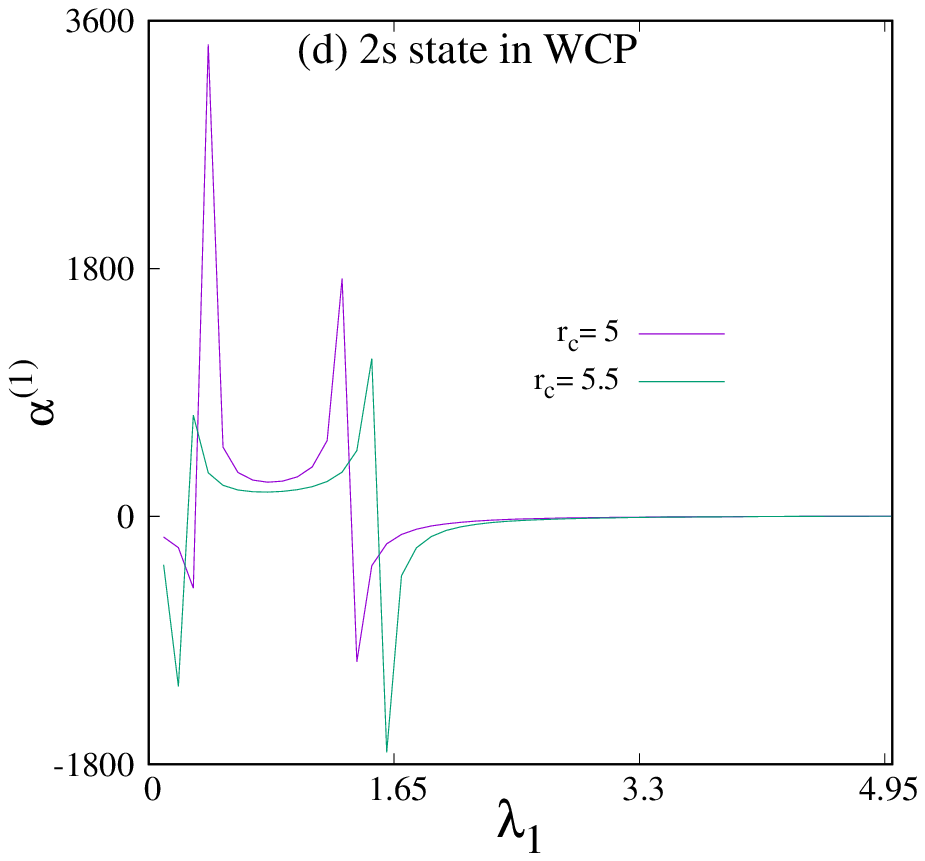}
\end{minipage}%
\vspace{1mm} 
\begin{minipage}[c]{0.48\textwidth}\centering
\includegraphics[scale=0.78]{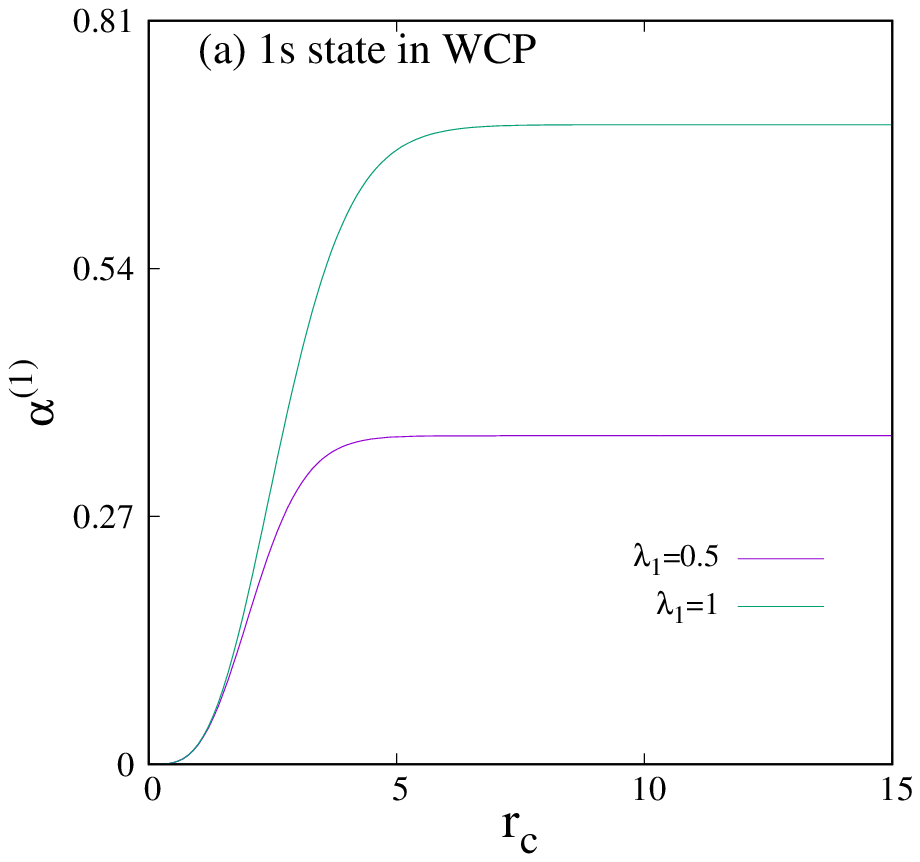}
\end{minipage}%
\begin{minipage}[c]{0.48\textwidth}\centering
\includegraphics[scale=0.78]{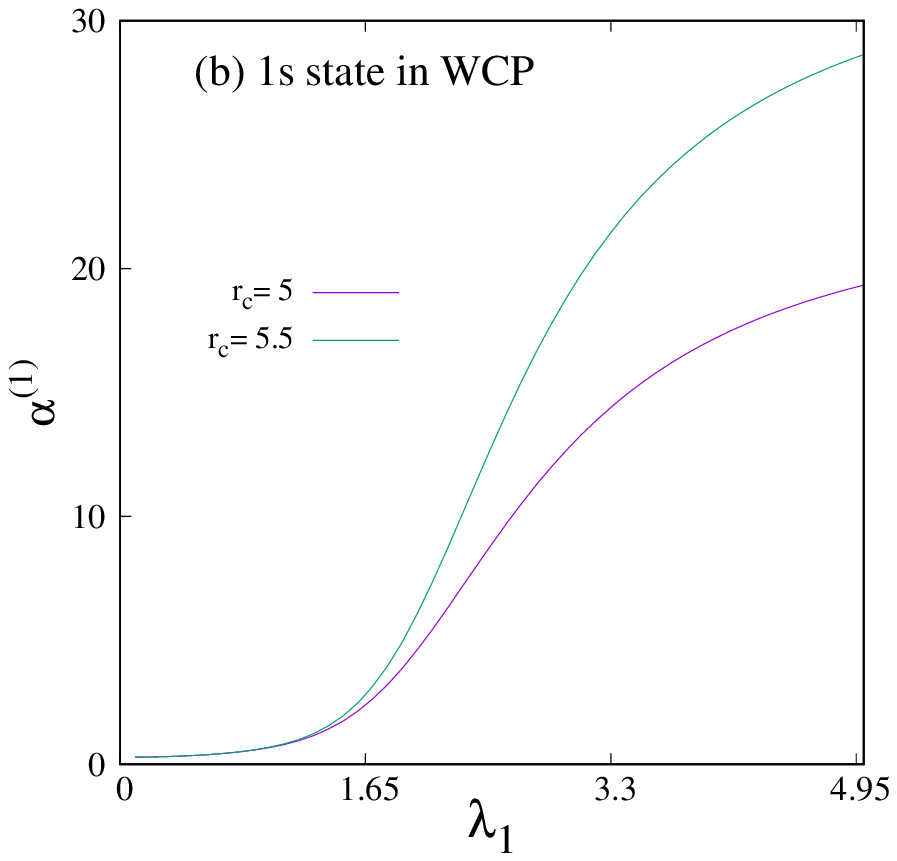}
\end{minipage}%
\caption{$\alpha^{(1)}$ in $1s, 2s$ states in WCP. In panels (a)-(b), $r_{c}$ variation at two selected $\lambda_{1}$ (0.5, 1), and in 
panels (c)-(d), $\lambda_{1}$ variation at two different $r_{c}$ (5, 5.5). See text for details.}
\end{figure}

\begin{figure}                         %%%Fig. 7, CHA
\begin{minipage}[c]{0.48\textwidth}\centering
\includegraphics[scale=0.78]{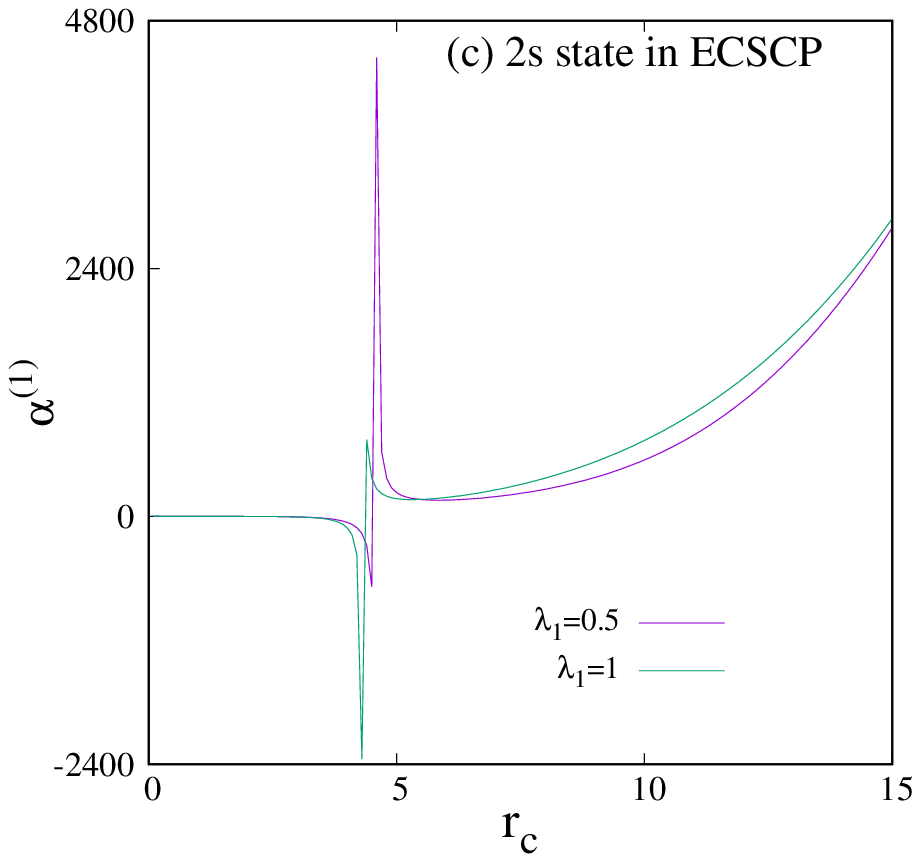}
\end{minipage}%
\begin{minipage}[c]{0.48\textwidth}\centering
\includegraphics[scale=0.78]{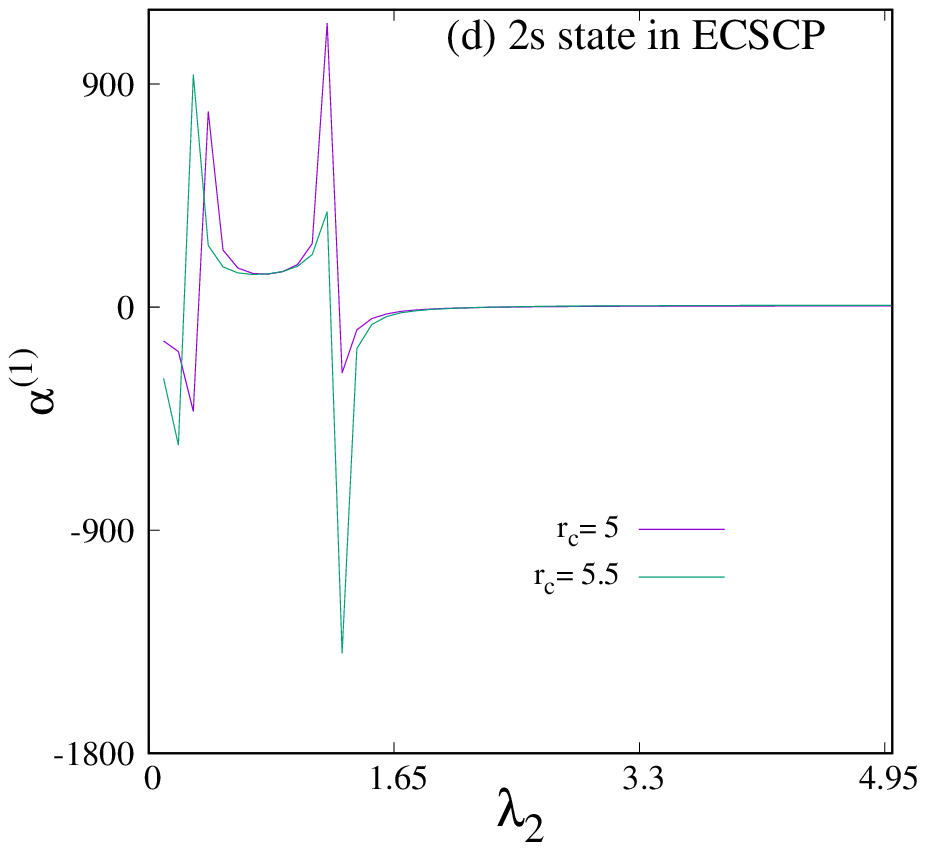}
\end{minipage}%
\vspace{1mm}
\begin{minipage}[c]{0.48\textwidth}\centering
\includegraphics[scale=0.78]{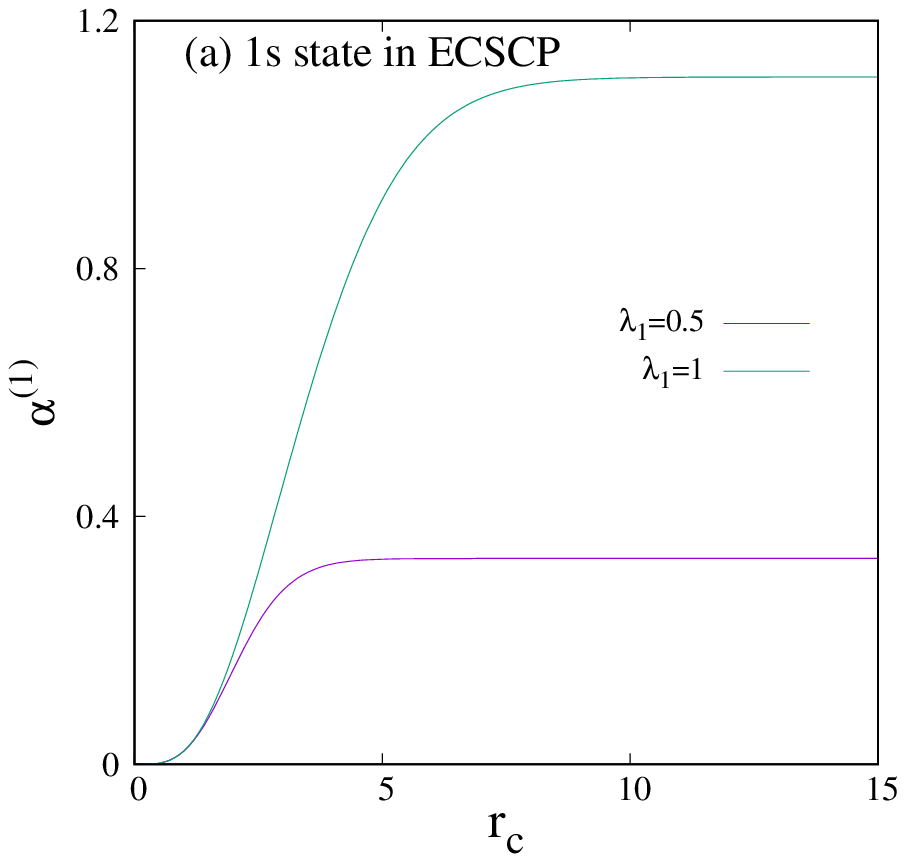}
\end{minipage}%
\begin{minipage}[c]{0.48\textwidth}\centering
\includegraphics[scale=0.78]{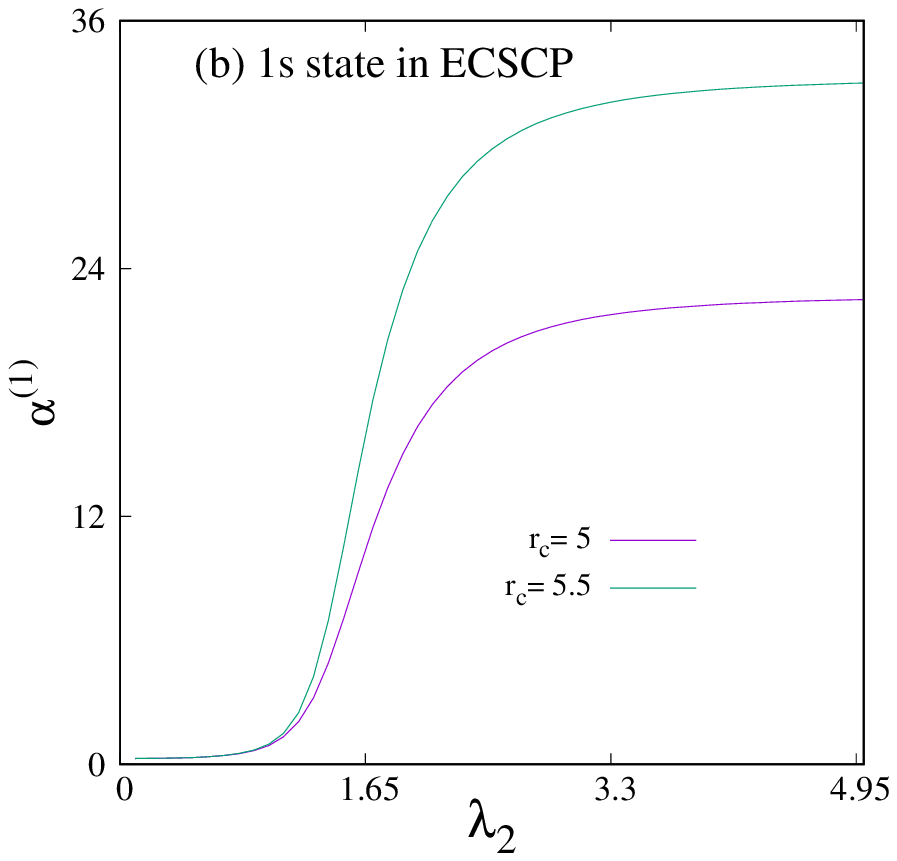}
\end{minipage}%
\caption{$\alpha^{(1)}$ for ECSCP in $1s$ and $2s$ states. Panels (a)-(b): $r_{c}$ variation at two selected $\lambda_{1}$ 
(0.5, 1); panels (c)-(d): $\lambda_{1}$ variation at two different $r_{c}$ (5, 5.5). See text for details.}
\end{figure} 

Now, the focus is on $2s$ states. Like the previous case, here also non-trivial variations are recorded in their changes with $r_{c}$ and $\lambda$.
In this case, the occurrence of a negative sign in $f^{(1)}_{2s \rightarrow 2p}$ indicates emission. 
In WCP and ECSCP, at a fixed $r_{c}$ in strong confinement region ($r_c \le 2$), it remains almost unchanged with changes in $\lambda$. At 
this low $r_{c}$ region, emission occurs between these two states for all the $\lambda$ considered. However, in $r_{c}=5$, emission 
happens at lower values of $\lambda$. Thus, at this particular $r_{c}$, there appears a crossover between $\mathcal{E}_{2s}$ and 
$\mathcal{E}_{2p}$, with progress in $\lambda$. The $2s$ to $3p$ transition provides absorption spectrum. Similar to 
$f^{(1)}_{2s \rightarrow 2p}$, at strong confinement zone (at a fixed $r_{c}$), nominal changes occur in $f^{(1)}_{2s \rightarrow 3p}$ 
in both WCP and ECSCP. The same, however, at $r_{c}=5$, decreases with rise in $\lambda$. At a fixed $\lambda$, with rise in $r_{c}$, it
advances to reach a maximum and then decays. Similarly, in SCP also, one gets a maximum with increase in $r_{c}$, at fixed $\lambda$. One 
observes that, in both WCP and ECSCP, $f^{(1)}_{2s \rightarrow mp} (m=2,3)$ decays with growth in $\lambda$. 
The above results of Table~III are graphically shown in Figs.~IV and V for \emph{confined WCP} and \emph{confined ECSCP} respectively. 
Thus $f^{(1)}_{ns \rightarrow 2p} (n=1,2)$ is plotted as function of (a) $r_{c}$ at fixed $\lambda$ and (b) $\lambda$ at given $r_{c}$ 
in these plasma conditions. Two representative $\lambda$ (5, 10) and $r_c$ (0.1, 0.5) are chosen to illustrate these. There are certain 
similarities in the qualitative nature of these plots in two left panels, (a) of Figs.~IV and V, as well as two right panels, (b).   
From panels (a) of these figures, one notices that, for both $\lambda$ values, starting from a non-zero positive number, 
$f^{(1)}_{1s \rightarrow 2p}$ grows to a moderate extent, to reach a maximum at a lower $r_c$, and then sharply falls until converging 
to the respective free system. However, $f^{(1)}_{2s \rightarrow 2p}$ starts from a small negative number, then lowers to a slight 
extent to attain a minimum, and finally accelerates rapidly to reach the corresponding free limit, in both WCP and ECSCP (also shown in 
left panels). Next, panels (b) shows, $f^{(1)}_{1s \rightarrow 2p}$ gradually falls to a minimum from an initial positive number with 
progress in $\lambda$ and thereafter grows until arriving at the free limit. As $r_c$ progresses, the plots display a well-like 
behavior with a flatter minimum, without any significant change in the positions of these minima. On the other hand, 
$f^{(1)}_{2s \rightarrow 2p}$ (again from panel (b)), initially shows a tendency to reach a maximum (which flattens with rise in $r_c$) 
followed by a sharp fall to attain the FHA limit. All these patterns are not necessarily evident from the table, as it offers only few entries 
to minimize the space. Thus one sees that $1s$ and $2s$ states maintain a complementary nature in Figs.~IV and V. 

\begin{figure}                         %%%Fig. 8, CHA
\begin{minipage}[c]{0.5\textwidth}\centering
\includegraphics[scale=0.78]{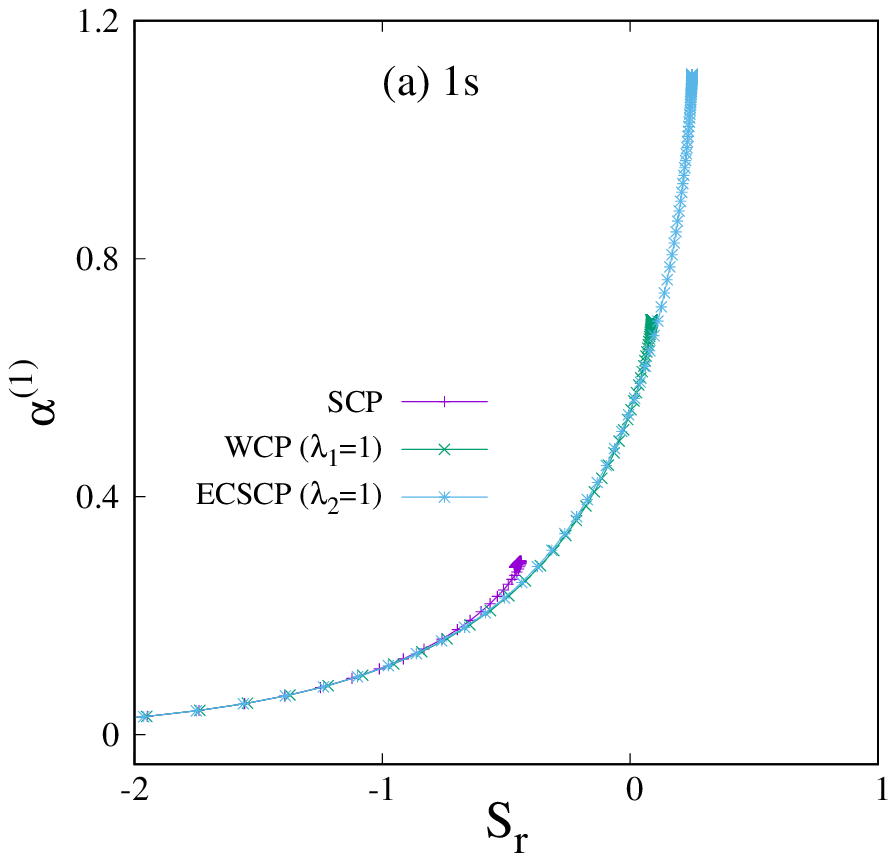}
\end{minipage}%
\begin{minipage}[c]{0.5\textwidth}\centering
\includegraphics[scale=0.78]{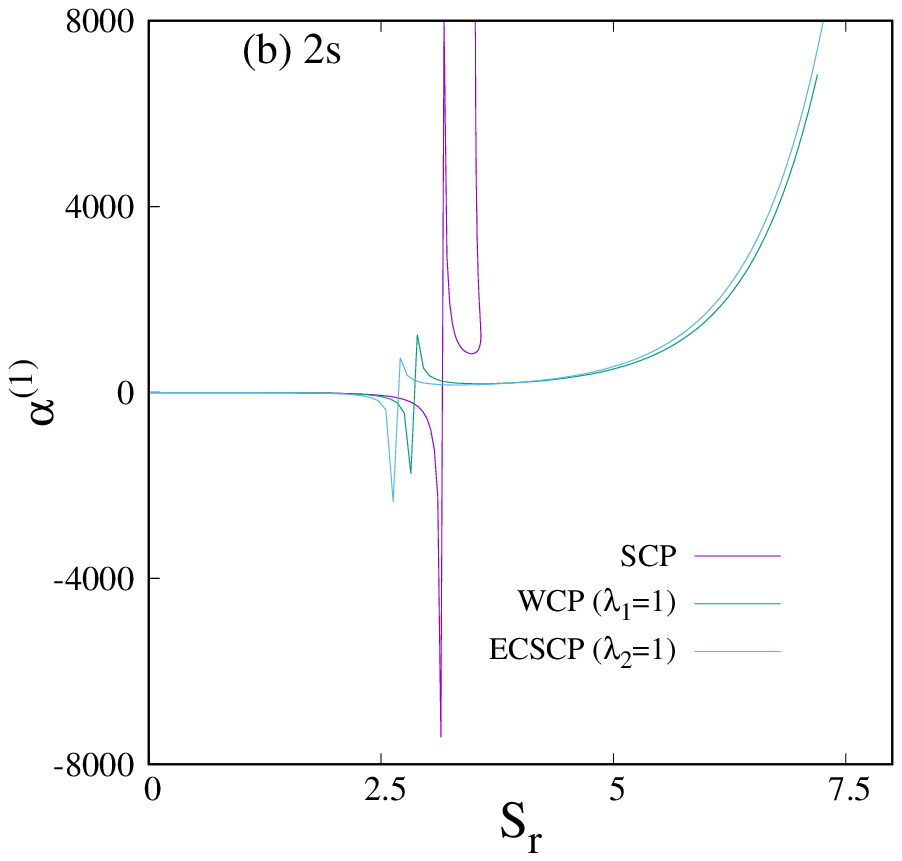}
\end{minipage}%
\caption{Change in $\alpha^{(1)}$ with $S_{r}$ in SCP, WCP ($\lambda_{1}=1$), ECSCP ($\lambda_{2}$=1) involving (a) $1s$ and 
(b) $2s$ states. See text for details.}
\end{figure}

Now, Table~IV  presents dipole polarizabilities, $\alpha^{(1)}$, in $1s$ and $2s$ for all the three plasmas. It retains the arrangement 
pattern of Table~III; so the top, middle and bottom portion contain results of WCP, ECSCP and SCP respectively. However, the chosen 
$\lambda$'s differ from Table~III. At lower $r_c$'s ($\le 0.5$) covered, $\alpha^{(1)}_{ns}$ is quite small and remains 
practically unaltered with changes in $\lambda$. Similarly, in SCP also, it is rather small. In $r_{c} > 0.5$ region, however, 
$\alpha^{(1)}_{1s}$ continually increases with $\lambda$, for a fixed $r_{c}$. Further, at a specific $\lambda$, it progresses with 
$r_{c}$. In essence, it is concluded that, in $1s$ state, with relaxation in confinement (increase in $T$) $\alpha^{(1)}$ enhances. 
However, in $2s$ state, $\alpha^{(1)}$ does not maintain the regular feature of ground state. Thus, $\alpha^{(1)}_{2s}$ at $r_c=0.5$ 
is higher compared to its counterpart in $r_{c}=0.1$, for all $\lambda$'s. In WCP and ECSCP, it progresses with $\lambda$ at a 
constant $r_c$. At $r_{c}=1$, in WCP $\alpha^{(1)}_{2s}$ attains (-)ve value at lower $\lambda_{1}$; with rise in $\lambda_{1}$ 
it generally grows and eventually becomes (+)ve towards the end. In contrast, in ECSCP it remains (-)ve for all the $\lambda_2$ 
considered, and slowly increases as we descend down the column. Further, at $r_{c}=2,~3$, in both WCP and ECSCP, it reflects 
(-)ve value but, overall, advances with rise in $\lambda$. Interestingly, however, at $r_{c}=5$, in either WCP or ECSCP, it starts 
from an initial (-)ve value at lower $\lambda$, then escalates to a (+)ve, followed by a drop to attain certain (-)ve value again. 
These results have prompted us to investigate the behavior of $\alpha^{(1)}_{ns}$  as function of $\lambda$, keeping $r_{c}$ fixed 
at $5$ and $5.5$ in their corresponding plots (see Figs.~6 and 7, later). However, in SCP, $\alpha^{(1)}_{ns}$ smoothly increases 
from a small number to reach a maximum and finally merge to FHA results (0.282136 and 7.5002 for $1s$ and $2s$). In free WCP and 
ECSCP, $\alpha^{(1)}_{1s}$ accelerates while $\alpha^{(1)}_{2s}$ reduces with growth in $\lambda$. These are demonstrated in the 
last two columns.

\begin{figure}                         %%%Fig. 9, CHA
\begin{minipage}[c]{0.33\textwidth}\centering
\includegraphics[scale=0.58]{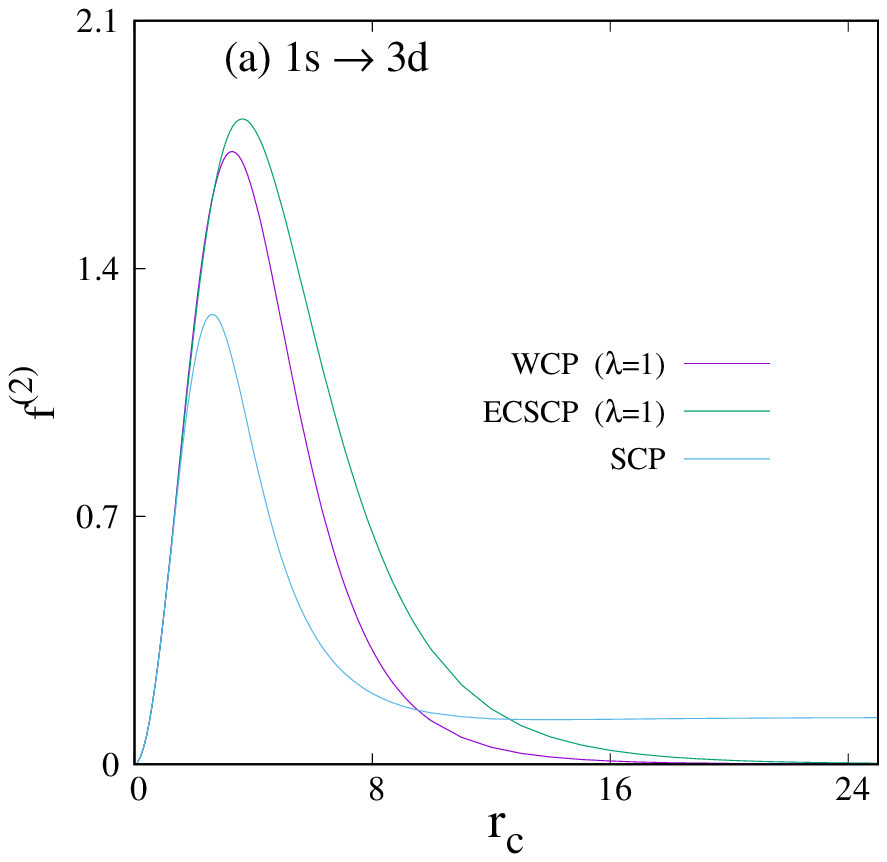}
\end{minipage}%
\begin{minipage}[c]{0.33\textwidth}\centering
\includegraphics[scale=0.58]{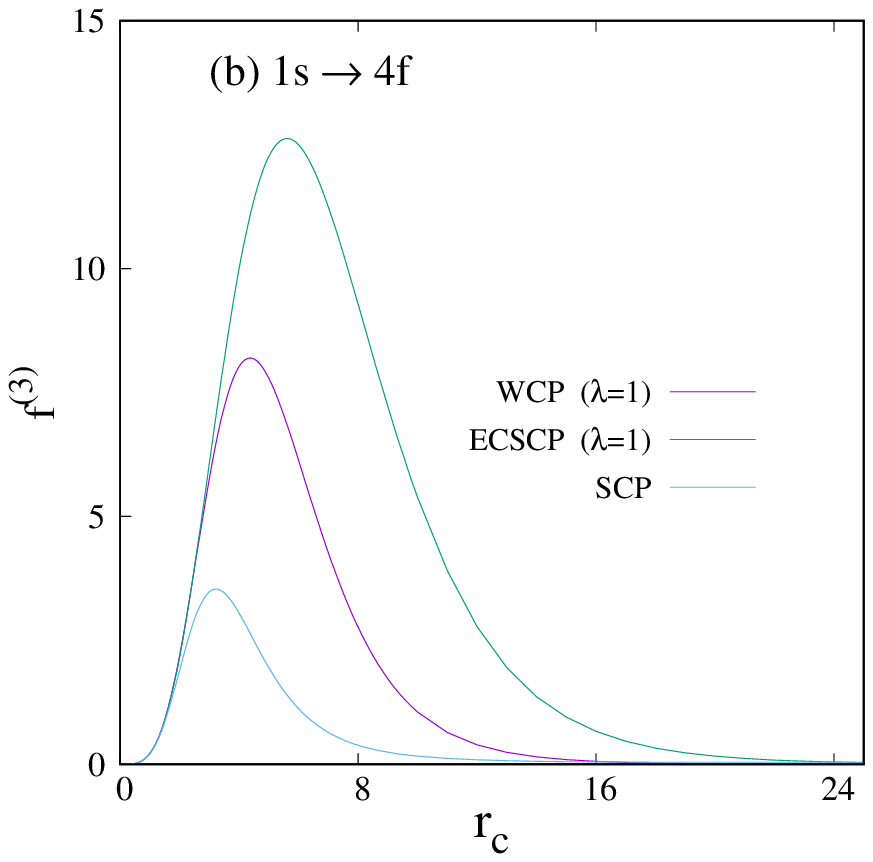}
\end{minipage}%
\begin{minipage}[c]{0.33\textwidth}\centering
\includegraphics[scale=0.58]{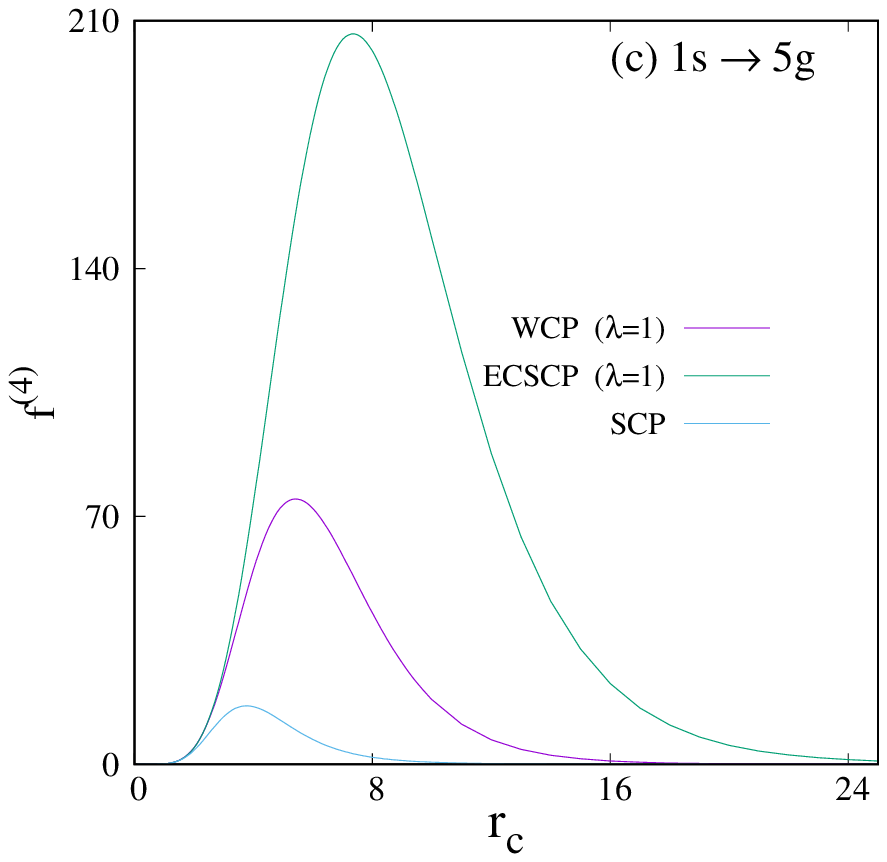}
\end{minipage}%
\caption{Changes in (a) $f^{(2)}_{(1s \rightarrow 3d)}$ (b) $f^{(3)}_{(1s \rightarrow 4f)}$ and (c) $f^{(4)}_{(1s \rightarrow 5g)}$ 
with $r_{c}$, in WCP ($\lambda_{1}=1$), ECSCP ($\lambda_{2}$=1) and SCP. See text for details.}
\end{figure}

The above $\alpha^{(1)}_{ns}$ results of Table~IV are depicted graphically in Figs.~6 and 7. Thus two lower left panels (a) 
suggest that, at a fixed $\lambda$ (0.5, 1), $\alpha^{(1)}_{1s}$ steadily progresses with $r_{c}$ until converging to 
free limit. The two lower right panels (b) show that, at fixed $r_c$ (5, 5.5), $\alpha^{(1)}_{1s}$ advances, 
initially slowly, but later sharply with $\lambda$, and then reach the FHA limit. It is observed that, in 
either WCP or ECSCP, the numerical value of $\alpha^{(1)}_{1s}$ at $r_{c}=5.5$ remains higher compared to that at $r_{c}=5$. 
Similarly, the top rows of these figures provide respective plots for $2s$ in WCP and ECSCP. From two lower 
right panels (c), it is inferred that, for a given $\lambda$ (0.5, 1), $\alpha^{(1)}_{2s}$ records some abrupt fall to a high 
(-)ve at certain $r_c$, followed by a dramatic shoot-up to a high (+)ve in a spike-like fashion, then again a drop and 
eventually steady growth, thus giving rise to one maximum and minimum. On the contrary, at $r_{c}=5$ or 5.5, in two top right 
panels (d), it proceeds through two spike-like features with change of sign in between high (-)ve to high (+)ve, passing through 
two maxima and minima. This complex pattern may occur due to a sign change in various energy states.              

From the above discussion it appears that, the impact of confinement on $\alpha^{(1)}_{ns}$ $(n=1,2)$ is thought provoking. 
In order to probe it further, it would be interesting to invoke Shannon entropy. It is well known that, $S_r$ is an efficient 
measure of confinement \cite{mukherjee18, mukherjee18a}; with increase of confinement strength $S_r$ decreases, while enhancing
with its relaxation. Therefore $\alpha^{(1)}$ has been plotted as function of $S_{r}$ in both $1s, 2s$ states of all these three 
plasmas. In WCP and ECSCP $\lambda$ is kept fixed at $1$. Panel (a) in Fig.~8 signifies that, in all these three occasions,
$\alpha^{(1)}_{1s}$ progress with $S_{r}$. But the same for $2s$ in panel (b) shows a behavior that is not so straightforward. 
Therefore an in-depth analysis would be highly desirable.          

\begin{figure}                         %%%Fig. 10, CHA
\begin{minipage}[c]{0.33\textwidth}\centering
\includegraphics[scale=0.58]{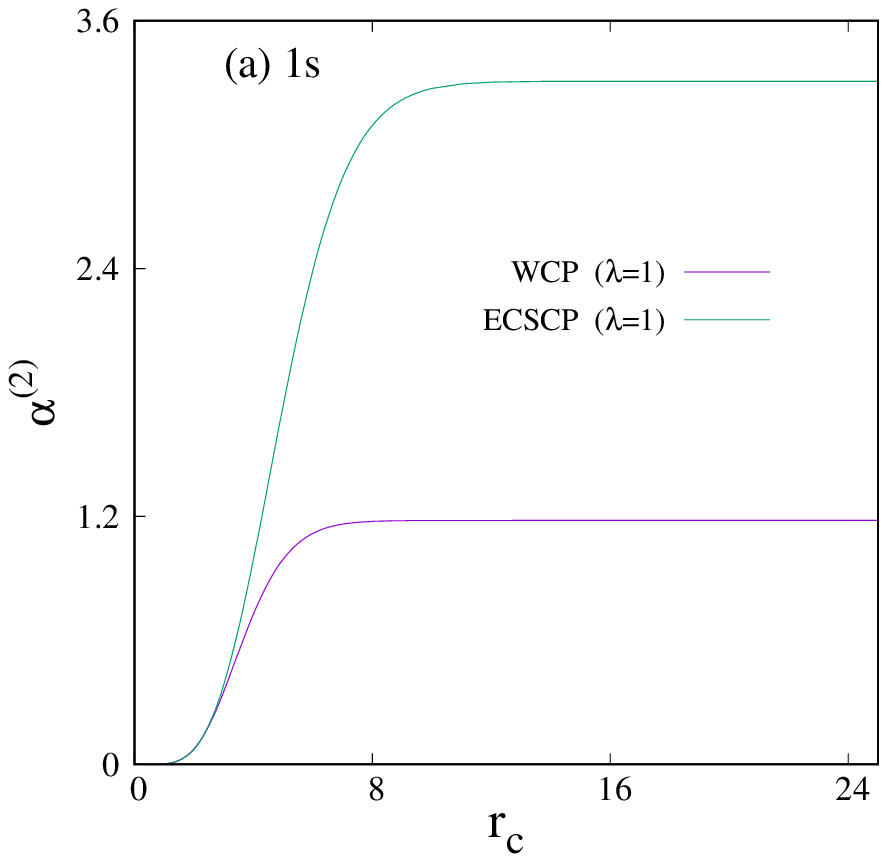}
\end{minipage}%
\begin{minipage}[c]{0.33\textwidth}\centering
\includegraphics[scale=0.58]{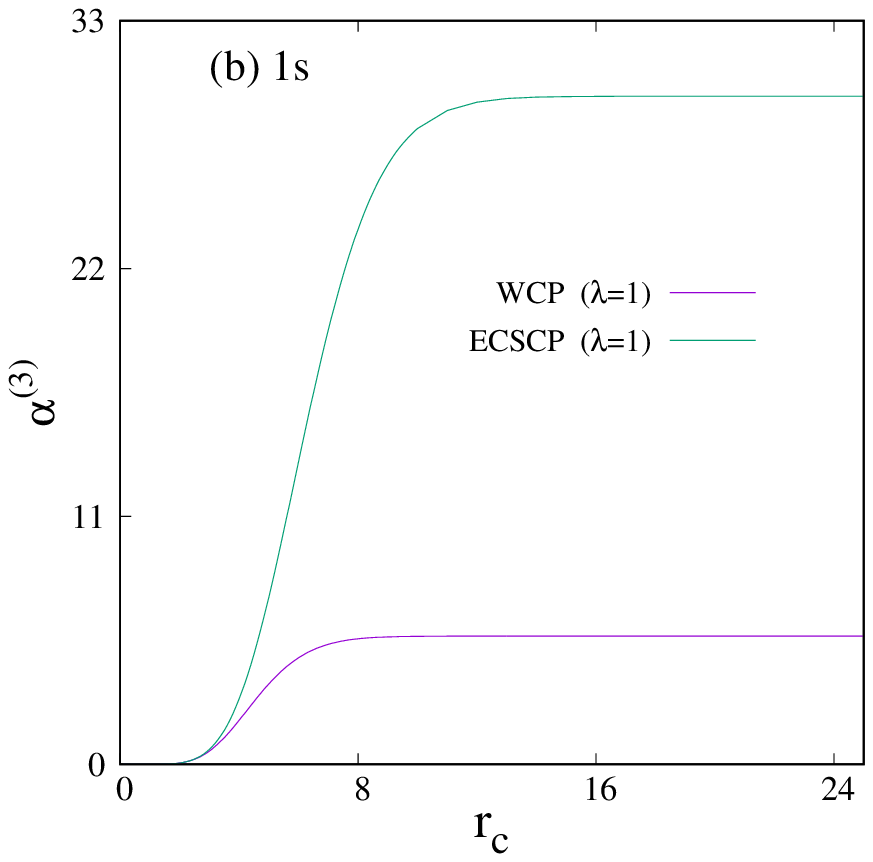}
\end{minipage}%
\begin{minipage}[c]{0.33\textwidth}\centering
\includegraphics[scale=0.58]{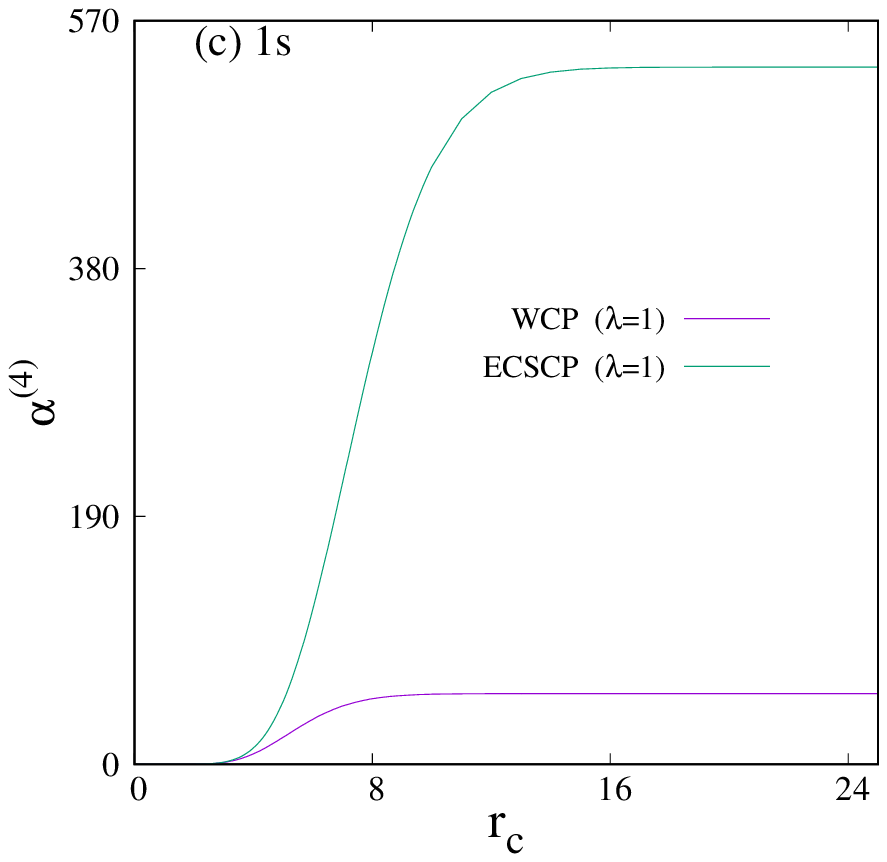}
\end{minipage}%
\caption{Changes in (a) $\alpha^{(2)}$ (b) $\alpha^{(3)}$ and (c) $\alpha^{(4)}$, with $r_{c}$, involving $1s$ state of WCP 
($\lambda_{1}=1$) and ECSCP ($\lambda_{2}$=1). See text for details.}
\end{figure}

At last, some sample results are now presented for quadrupole, octupole and hexadecapole OS, as well as the polarizabilities 
involving WCP, ECSCP and SCP. The selection rules for these three different transitions are $\Delta \ell \pm 2,3$ and $4$ 
respectively. To illustrate the qualitative features, we offer a cross section of these, while detailed results will be published 
elsewhere. Figure~9 imprints the variation of $f^{(2)}, f^{(3)}, f^{(4)}$ respectively, as function of $r_{c}$, for the three 
potentials, in panels (a)-(c), for $1s \rightarrow 3d$, $1s \rightarrow 4f$ and $1s \rightarrow 5g$ transitions. In WCP and ECSCP 
$\lambda$ was chosen to be $1$. For all three potentials, OS rises with $r_{c}$, then attains a maximum and finally reach
the free values. This features holds true for all the higher order OS. That means, there exists a characteristic $r_{c}$ at which the 
probability of concerned transition is maximum. Similarly, the left, middle and right panels of Fig.~10 displays changes in  
$\alpha^{(2)}, \alpha^{(3)}, \alpha^{(4)}$, with $r_{c}$ in $1s$ state for WCP and ECSCP. In both plasmas, $\alpha^{(k)}$ continually 
increases until reaching a constant value corresponding to the free system. The analogous SCP plots are qualitatively similar, and thus 
omitted. 

\section{conclusion}
Multipole (up to order 4) OS and polarizabilities are probed for H-like ions in WCP, ECSCP and SCP. In first two cases, investigation 
is done in both \emph{free} and \emph{confined} conditions. The connection between $T$ and $r_{c}$ is proposed and analyzed. It is 
found that, the plasma tail effect can be controlled by introducing this confinement. Two generalized 
scaling ideas are derived connecting $Z$ and $\lambda$ separately. The relation between these two independent ideas is also achieved. 
Starting from a given Hamiltonian and using these designed relations, one can easily extract results for a series of Hamiltonian.
A new $S_r$-driven technique is designed to determine $\lambda^{(c)}_{n,\ell}$ for both WCP and ECSCP in free environment accurately, 
where it shoots up stiffly. Further, using $S_r$-based results, and this scaling idea, a generalized relation between $\lambda^{(c)}_{n,\ell}$ 
and $Z$ is proposed, which is applicable to an arbitrary state. The applicability of a recently proposed virial-like theorem has been verified 
to the plasma systems studied here. Results are also presented in \emph{free} WCP and ECSCP. A detailed investigation of these 
spectroscopic properties for $\ell \ne 0$ states would be highly desirable. The influence of plasma 
screening effect on two-photon transition amplitude, photoionization cross section also need to be explored in confined condition. 
Other information-theoretic quantities like Fisher information, Onicescu energy, complexity, mutual and relative information, etc.,
are required to be examined. Exploration of Hellmann-Feynman theorem in the context of \emph{confined} plasma is necessary. Similar 
calculation in Helium plasmas may provide vital insight about the effect of confinement on many-electron plasmas.    

\section{Acknowledgement}
Financial support from BRNS, India (sanction order: 58/14/03/2019-BRNS/10255) is gratefully acknowledged. NM thanks CSIR, New Delhi, India, for a Senior Research Associateship (Pool No. 9033A).  

\appendix
\section{Analytical forms of $f^{(k)}$ and $\alpha^{(k)}$ in FHA}
Analytical expression of dipole polarizabilities in FHA was reported in \cite{tanner83} for $1s$ state. In this appendix, we provide
the $2^{k}$-pole OS ($k=1,4$) and respective polarizabilities for FHA, in both $1s,2s$ states. 

The closed form expressions of $f^{(1)}_{(1s \rightarrow np)}(Z)$ and $f^{(1)}_{(2s \rightarrow np)}(Z)$ are obtained as,
\begin{equation}\label{eq:52}
\begin{aligned}
f^{(1)}_{(1s \rightarrow np)}(Z) & = \frac{2^{8}}{3Z^{7}}  \ n^{5} \ \frac{(n-1)^{(2n-4)}}{(n+1)^{(2n+4)}}, \\
f^{(1)}_{(2s \rightarrow np)}(Z) & = \frac{2^{15}}{3Z^{7}}  \ n^{5} \  (n^{2}-1) \ \frac{(n-2)^{(2n-5)}}{(n+2)^{(2n+5)}}.
\end{aligned}
\end{equation}
Now, applying Eq.~(\ref{eq:52}) in Eq.~(\ref{eq:6}), one easily obtains $\alpha^{(1)}_{i}(\mathrm{bound})(Z)$ for $1s$ and $2s$ states 
of FHA. They take the following forms, 
\begin{equation}\label{eq:53}
\begin{aligned}
\alpha^{(1)}_{1s}(\mathrm{bound})(Z) & = \sum^{n}_{i=2} \frac{2^{10}}{3Z^{9}}  \ i^{9} \ \frac{(i-1)^{(2i-6)}}{(i+1)^{(2i+6)}}, \\
\alpha^{(1)}_{2s}(\mathrm{bound})(Z) & = \sum^{n}_{i=2} \frac{2^{21}}{3Z^{9}}  \ i^{9} \  (i^{2}-1) \ \frac{(i-2)^{(2i-7)}}{(i+2)^{(2i+7)}}.
\end{aligned}
\end{equation} 
$f^{(2)}_{(1s \rightarrow nd)}(Z)$ and $f^{(2)}_{(2s \rightarrow nd)}(Z)$ are expressed as,
\begin{equation}\label{eq:54}
\begin{aligned}
f^{(2)}_{(1s \rightarrow nd)}(Z) & = \frac{2^{12}}{5Z^{9}}  \ n^{7} \ (n^{2}-4) \ \frac{(n-1)^{(2n-6)}}{(n+1)^{(2n+6)}}, \\
f^{(2)}_{(2s \rightarrow nd)}(Z) & = \frac{2^{27}}{5Z^{9}}  \ n^{7} \  (n^{2}-1) \ \frac{(n-2)^{(2n-9)}}{(n+2)^{(2n+9)}}.
\end{aligned}
\end{equation}
Invoking Eq.~(\ref{eq:54}) in Eq.~(\ref{eq:6}), one gets $\alpha^{(2)}_{i}(\mathrm{bound})(Z)$ in $1s$ and $2s$ states of FHA
as follows, 
\begin{equation}\label{eq:55}
\begin{aligned}
\alpha^{(2)}_{1s}(\mathrm{bound})(Z) & = \sum^{n}_{i=3}\frac{2^{12}}{5Z^{11}}  \ i^{11} \ (i^{2}-4) \ \frac{(i-1)^{(2i-8)}}{(i+1)^{(2i+8)}}, \\
\alpha^{(2)}_{2s}(\mathrm{bound})(Z) & = \sum^{n}_{i=3}\frac{2^{33}}{5Z^{11}}  \ i^{11} \  (i^{2}-1) \ \frac{(i-2)^{(2i-10)}}{(i+2)^{(2i+10)}}.
\end{aligned}
\end{equation}
The analytical expressions for $f^{(3)}_{(1s \rightarrow nf)}(Z)$ and $f^{(3)}_{(2s \rightarrow nf)}(Z)$ are presented as,
\begin{equation}\label{eq:56}
\begin{aligned}
f^{(3)}_{(1s \rightarrow nf)}(Z) & = \frac{9}{7} \frac{2^{12}}{Z^{11}}  \ n^{9} \ (n^{2}-9)(n^{2}-4) \ \frac{(n-1)^{(2n-8)}}{(n+1)^{(2n+8)}}, \\
f^{(3)}_{(2s \rightarrow nf)}(Z) & = \frac{9}{7} \frac{2^{27}}{5Z^{11}}  \ n^{9} \  (n^{2}-9)(n^{2}+4)(n^{2}-1) \ \frac{(n-2)^{(2n-10)}}{(n+2)^{(2n+10)}}.
\end{aligned}
\end{equation}
Doing some mathematical manipulation after substituting Eq.~(\ref{eq:56}) in Eq.~(\ref{eq:6}), yields $\alpha^{(3)}_{i}(\mathrm{bound})(Z)$ for 
$1s$ and $2s$ states of FHA as below, 
\begin{equation}\label{eq:57}
\begin{aligned}
\alpha^{(3)}_{1s}(\mathrm{bound})(Z) & = \sum^{n}_{i=4}\frac{9}{7} \frac{2^{14}}{Z^{13}}  \ i^{13} \ (i^{2}-9)(i^{2}-4) \ \frac{(i-1)^{(2i-10)}}{(i+1)^{(2i+10)}}, \\
\alpha^{(3)}_{2s}(\mathrm{bound})(Z) & = \sum^{n}_{i=4}\frac{9}{7} \frac{2^{33}}{Z^{13}}  \ i^{13} \ (i^{2}-9)(i^{4}+4)(i^{2}-1) \ \frac{(i-2)^{(2i-12)}}{(i+2)^{(2i+12)}}.
\end{aligned}
\end{equation}
Finally, $f^{(4)}_{(1s \rightarrow ng)}(Z)$ and $f^{(4)}_{(2s \rightarrow ng)}(Z)$ are manifested as,
\begin{equation}\label{eq:58}
\begin{aligned}
f^{(4)}_{(1s \rightarrow ng)}(Z) & = \frac{2^{18}}{9Z^{13}}  \ n^{11} \ (n^{2}-16)(n^{2}-9)(n^{2}-4) \ \frac{(n-1)^{(2n-10)}}{(n+1)^{(2n+10)}}, \\
f^{(4)}_{(2s \rightarrow ng)}(Z) & = \frac{2^{39}}{9Z^{13}}  \ n^{11} \ (n^{2}-16)(n^{2}-9)(n^{2}+2)^{2}(n^{2}-1) \ \frac{(n-2)^{(2n-12)}}{(n+2)^{(2n+12)}}.
\end{aligned}
\end{equation}
By replacing Eq.~(\ref{eq:58}) in Eq.~(\ref{eq:6}) one may extract $\alpha^{(4)}_{i}(\mathrm{bound})(Z)$ for $1s, 2s$ with the form,
\begin{equation}\label{eq:59}
\begin{aligned}
\alpha^{(4)}_{1s}(\mathrm{bound})(Z) & = \sum^{n}_{i=5}\frac{2^{20}}{9Z^{15}}  \ i^{15} \ (i^{2}-16)(i^{2}-9)(i^{2}-4) \ \frac{(i-1)^{(2i-12)}}{(i+1)^{(2i+12)}}, \\
\alpha^{(4)}_{2s}(\mathrm{bound})(Z) & = \sum^{n}_{i=5}\frac{2^{45}}{9Z^{15}}  \ i^{15} \ (i^{2}-16)(i^{2}-9)(i^{2}+2)^{2}(i^{2}-1) \ \frac{(i-2)^{(2i-14)}}{(i+2)^{(2i+14)}}.
\end{aligned}
\end{equation}

\section{Some selected results using scaling concept}
Here, we demonstrate the derived relations presented in Sec.II.D. Table~V, imprints some sample results obtained by the 
proposed scaling concept. Here, we have used these formulas to connect $Z$, $\lambda$ and $r_{c}$. However, it can be applied and 
extended to any Hamiltonian. 

\begingroup           %%Table 5, dipole polarizability
\squeezetable
\begin{table}
\caption{$\mathcal{E}_{n,0}, f^{(1)}_{ns \rightarrow 2p}, \alpha^{(1)}_{ns}$ $(n=1,2)$ values for three Hamiltonians, given in 
Eqs.~(\ref{eq:42}).}
\centering
\begin{ruledtabular}
\begin{tabular}{l|ll|ll|ll|ll|ll|ll}
 \multicolumn{13}{c}{WCP}    \\
\hline
 & \multicolumn{4}{c|}{$H\left(1,1,\frac{\lambda_{1}}{Z},Zr_{c},r_{1}\right)$}  & 
\multicolumn{4}{c|}{$H\left(1,\frac{Z}{\lambda_{1}},1,\lambda_{1}r_{c},r_{2}\right)$}  &
\multicolumn{4}{c}{$H\left(1,Z,\lambda_{1},r_{c},r\right)$} \\   
\hline
$\frac{\lambda_{1}}{Z}$=2 &  \multicolumn{2}{c|}{$H\left(1,1,2,1,r_{1}\right)$} & \multicolumn{2}{c|}{$H\left(1,1,2,2,r_{1}\right)$} & 
\multicolumn{2}{c|}{$H\left(1,0.5,1,2,r_{2}\right)$} & \multicolumn{2}{c|}{$H\left(1,0.5,1,4,r_{2}\right)$} &
\multicolumn{2}{c|}{$H\left(1,1,2,1,r\right)$} & \multicolumn{2}{c}{$H\left(1,2,4,1,r\right)$}  \\
%$r_{c}=1$  &   &  &  &   &    &     &   & &  &  & &  \\
\cline{2-13} 
$r_{c}=1$  & $\lambda_{1}=2$ & $Z=1$ & $\lambda_{1}=4$ & $Z=2$ & $\lambda_{1}=2$ & $Z=1$ & $\lambda_{1}=4$ & $Z=2$ & $\lambda_{1}=2$ & $Z=1$ & $\lambda_{1}=4$ & $Z=2$ \\
\hline
& $I$ & $II$ & $I$ & $II$ & $I$ & $II$ & $I$ & $II$ & $I$ & $II$ & $I$ & $II$ \\
\hline
$\mathcal{E}_{1,0}$ & 3.6923 & 3.6923  & 0.8644 & 0.8644  & 0.9230 & 0.9230  & 0.2161 & 0.2161 & 3.6923 & 3.6923 & 3.4576 & 3.4576  \\ 
$f^{(1)}_{1s \rightarrow 2p}$ & 0.9825 & 0.9825  & 0.9877 & 0.9877  & 0.9825 & 0.9825  & 0.9877 & 0.9877 & 0.9825 & 0.9825 & 0.9877 & 0.9877  \\
$\alpha^{(1)}_{1s}$ & 0.02998  & 0.02998 & 0.42689 & 0.42689 & 0.47968 & 0.47968 & 6.83026 & 6.83026 & 0.02998 & 0.02998 & 0.02668 & 0.02668  \\
\hline
$\mathcal{E}_{2,0}$           & 17.8794    & 17.8794   & 4.2884    & 4.2884  & 4.4698    & 4.4698  & 1.07212 & 1.07212 & 17.8794    & 17.8794   & 17.1538 & 17.1538  \\ 
$f^{(1)}_{2s \rightarrow 2p}$ & $-$0.6051  & $-$0.6051 & $-$0.6039 & $-$0.6039 & $-$0.6051 & $-$0.6051 & $-$0.6039 & $-$0.6039 & $-$0.6051  & $-$0.6051 & $-$0.6039 & $-$0.6039 \\
$\alpha^{(1)}_{2s}$  & 0.00477  & 0.00477 & 0.02271 & 0.02271 & 0.07632 & 0.07632 & 0.36349 & 0.36349 & 0.00477 & 0.00477 & 0.00142 & 0.00142   \\
\hline
 \multicolumn{13}{c}{ECSCP}    \\
\hline
 & \multicolumn{4}{c|}{$H\left(1,1,\frac{\lambda_{1}}{Z},Zr_{c},r_{1}\right)$}  & 
\multicolumn{4}{c|}{$H\left(1,\frac{Z}{\lambda_{1}},1,\lambda_{1}r_{c},r_{2}\right)$}  &
\multicolumn{4}{c}{$H\left(1,Z,\lambda_{1},r_{c},r\right)$} \\   
\hline
$\frac{\lambda_{1}}{Z}$=2 &  \multicolumn{2}{c|}{$H\left(1,1,2,1,r_{1}\right)$} & \multicolumn{2}{c|}{$H\left(1,1,2,2,r_{1}\right)$} & 
\multicolumn{2}{c|}{$H\left(1,0.5,1,2,r_{2}\right)$} & \multicolumn{2}{c|}{$H\left(1,0.5,1,4,r_{2}\right)$} &
\multicolumn{2}{c|}{$H\left(1,1,2,1,r\right)$} & \multicolumn{2}{c}{$H\left(1,2,4,1,r\right)$}  \\
\cline{2-13} 
$r_{c}$=1  & $\lambda_{1}=2$ & $Z=1$ & $\lambda_{1}=4$ & $Z=2$ & $\lambda_{1}=2$ & $Z=1$ & $\lambda_{1}=4$ & $Z=2$ & $\lambda_{1}=2$ & $Z=1$ & $\lambda_{1}=4$ & $Z=2$ \\
\hline
& $I$ & $II$ & $I$ & $II$ & $I$ & $II$ & $I$ & $II$ & $I$ & $II$ & $I$ & $II$ \\
\hline
$\mathcal{E}_{1,0}$           & 4.00195 & 4.00195 & 1.07647 & 1.07647 & 1.00048 & 1.00048 & 0.29612 & 0.29612 & 4.00195 & 4.00195 & 4.30589 & 4.30589   \\ 
$f^{(1)}_{1s \rightarrow 2p}$ & 0.98265 & 0.98265 & 0.98488 & 0.98488 & 0.98265 & 0.98265 & 0.98488 & 0.98488 & 0.98265 & 0.98265 & 0.98488 & 0.98488   \\
$\alpha^{(1)}_{1s}$           & 0.02998 & 0.02998 & 0.46172 & 0.46172 & 0.47981 & 0.47981 & 7.38750 & 7.38750 & 0.02998 & 0.02998 & 0.02885 & 0.02885   \\
\hline
$\mathcal{E}_{2,0}$           & 18.1544     & 18.1544    & 4.47386  & 4.47386      & 4.53860    & 4.53860    & 1.11846    & 1.11846    & 18.15440 & 18.15440     & 17.89546 & 17.89546   \\ 
$f^{(1)}_{2s \rightarrow 2p}$ & $-$0.6047  & $-$0.6047 & $-$0.5971 & $-$0.5971 & $-$0.6047 & $-$0.60477 & $-$0.5971 & $-$0.5971 & $-$0.6047 & $-$0.6047 & $-$0.6047 & $-$0.6047 \\
$\alpha^{(1)}_{2s}$           & 0.00466    & 0.00466   & 0.02914 & 0.02914 & 0.07454 & 0.07454 & 0.46631 & 0.46631 & 0.00466 & 0.00466 & 0.00182 & 0.00182   \\
\hline
 \multicolumn{13}{c}{SCP}    \\
\hline
 & \multicolumn{4}{c|}{$H\left(1,1,\left(\frac{\sigma}{Z}\right)^{4},Zr_{c},r_{1}\right)$}  & 
\multicolumn{4}{c|}{$H\left(1,\frac{Z}{\sigma},1,\sigma r_{c},r_{2}\right)$}  &
\multicolumn{4}{c}{$H\left(1,Z,\sigma^{4},r_{c},r\right)$} \\   
\hline
  &  \multicolumn{2}{c|}{$H\left(1,1,\frac{1}{16},2,r_{1}\right)$} & \multicolumn{2}{c|}{$H\left(1,1,1,3B,r_{1}\right)$} & 
\multicolumn{2}{c|}{$H\left(1,2,1,1,r_{2}\right)$} & \multicolumn{2}{c|}{$H\left(1,1,1,3B,r_{2}\right)$} &
\multicolumn{2}{c|}{$H\left(1,2,1,1,r\right)$} & \multicolumn{2}{c}{$H\left(1,3,3,B,r\right)$}  \\
\cline{2-13} 
  & $\sigma=1$ & $Z=2$ & $\sigma=3$ & $Z=3$ & $\sigma=1$ & $Z=2$ & $\sigma=3$ & $Z=3$ & $\sigma=1$ & $Z=2$ & $\sigma=3$ & $Z=3$ \\
\hline
& $I$ & $II$ & $I$ & $II$ & $I$ & $II$ & $I$ & $II$ & $I$ & $II$ & $I$ & $II$ \\
\hline
$\mathcal{E}_{1,0}$           & 0.56979  & 0.56979 & 5.64694 & 5.64694 & 2.27917 & 2.27917 & 5.64694 & 5.64694 & 2.27917 & 2.27917 & 50.64694 & 50.64694 \\ 
$f^{(1)}_{1s \rightarrow 2p}$ & 0.99067  & 0.99067 & 0.98176 & 0.98176 & 0.99067 & 0.99067 & 0.98176 & 0.98176 & 0.99067 & 0.99067 & 0.98176  &  0.98176  \\
$\alpha^{(1)}_{1s}$           & 0.35614  & 0.35614 & 0.01769 & 0.01769 & 0.02226 & 0.02226 & 0.01769 & 0.01769 & 0.02226 & 0.02226 & 0.00021 &  0.00021  \\
\hline
$\mathcal{E}_{2,0}$           & 3.99437   & 3.99437   & 24.2876 & 22.2876 & 15.97749  & 15.97749  & 24.2876 & 24.2876 & 15.97749  & 15.97749  & 218.5886 & 218.5886  \\ 
$f^{(1)}_{2s \rightarrow 2p}$ & $-$0.6097 & $-$0.6097 & $-$0.6048 & $-$0.6048 & $-$0.6097 & $-$0.6097 & $-$0.6048 & $-$0.6048 & $-$0.6097 & $-$0.6097 & $-$0.6048 & $-$0.6048   \\
$\alpha^{(1)}_{2s}$           & $-$0.0149 & $-$0.0149 & 0.00296 & 0.00296 & $-$0.0009 & $-$0.0009 & 0.00296 & 0.00296 & $-$0.0009 & $-$0.0009 & 0.000036 & 0.000036   \\
\end{tabular}
\end{ruledtabular}
\end{table}    

The top, middle and bottom portions present results for WCP, ECSCP and SCP respectively. In all three cases, columns 
$\{2,3\}, \{6,7\}, \{10,11\}$ and $\{4,5\}, \{8,9\}, \{12,13\})$ form two separate groups. Here, due to lack of space, we restrict 
our calculation using three Hamiltonians. However, one can extend the number of such Hamiltonians in a given group by using this 
formulation. Interestingly, one can extract the results for all members of a particular group just by performing calculations for 
any one Hamiltonian belonging to that group. The symbols have following meanings.
Firstly, $\alpha^{(1)}_{ns}$ represents the bound-state polarizability. $I$ signifies analytical results obtained by employing 
(a) Eqs.~(\ref{eq:23}),(\ref{eq:28}),(\ref{eq:33}) for $\mathcal{E}_{n,\ell}$, (b) Eqs.~(\ref{eq:25}),(\ref{eq:30}),(\ref{eq:40}) 
for $f^{(1)}_{js \rightarrow np}$, (c) Eqs.~(\ref{eq:26}),(\ref{eq:31}),(\ref{eq:41}) for $\alpha^{(1)}_{ns}$. $II$ indicates 
numerical results calculated by using the Hamiltonian directly. And finally, $B=\left(\frac{2}{81}\right)^{\frac{1}{3}}$.

%%% start from here
\section{$\lambda^{(c)}_{n,\ell}$ values for higher states in WCP and ECSCP}
The critical screening,  $\lambda^{(c)}_{n,\ell}$, of WCP and ECSCP for $Z=1-4$, in the $3s,4s,4p,4f,5s,5p,5d,5f,5g$ states are 
produced in Table~VI. 
  
\begingroup           %%Table 6, critical screening
\squeezetable
\begin{table}[htp]
\caption{$\lambda^{(c)}_{n,\ell}$ for H-like ions $(Z=1-4)$ for $3s,4s,4p,4f,5s,5p,5d,5f,5g$ states in WCP, ECSCP.}
\centering
\begin{ruledtabular}
\begin{tabular}{llllllll}
\multicolumn{4}{c}{WCP}  & \multicolumn{4}{c}{ECSCP}   \\
\cline{1-4} \cline{5-8}
$Z$   & State & \ \ $\lambda^{(c)}_{n,\ell}$  & \ \ \ \ \ $\mathcal{E}_{n,\ell}$ & $Z$  & State & \ \ $\lambda^{(c)}_{n,\ell}$ & \ \ \ \ \ $\mathcal{E}_{n,\ell}$  \\
\cline{1-4} \cline{5-8}
1   & $3s$ & 0.13656$^{\dag}$  &  $-$0.00000013  & 1 & $3s$ & 0.289685$^{\ddag}$  &  $-$0.00000005  \\
2   & $3s$ & 0.27614 &  $-$0.00000013  & 2 & $3s$ & 0.217247 &  $-$0.00000014  \\
3   & $3s$ & 0.41563 &  $-$0.00000012  & 3 & $3s$ & 0.144808 &  $-$0.00000026  \\
4   & $3s$ & 0.55510 &  $-$0.00000015  & 4 & $3s$ & 0.072366 &  $-$0.00000009  \\
\cline{1-4} \cline{5-8}
1   & $4s$ & 0.07636$^{\dag}$    &  $-$0.00000020 & 1 & $4s$ & 0.040407$^{\ddag}$ &  $-$0.00000015  \\
2   & $4s$ & 0.1554320 &  $-$0.00000001 & 2 & $4s$ & 0.080838 &  $-$0.00000009  \\
3   & $4s$ & 0.23433   &  $-$0.00000002 & 3 & $4s$ & 0.121266 &  $-$0.00000016   \\
4   & $4s$ & 0.31319   &  $-$0.00000005 & 4 & $4s$ & 0.161693 &  $-$0.00000077  \\
\cline{1-4} \cline{5-8}
1   & $4p$ & 0.06769$^{\dag}$   &  $-$0.00000058  & 1 & $4p$ & 0.03926$^{\ddag}$  &  $-$0.00000076  \\
2   & $4p$ & 0.13572  &  $-$0.00000025  & 2 & $4p$ & 0.078526 &  $-$0.00000116 \\
3   & $4p$ & 0.20363  &  $-$0.00000116  & 3 & $4p$ & 0.117789 &  $-$0.00000025  \\
4   & $4p$ & 0.271529 &  $-$0.00000076  & 4 & $4p$ & 0.157053 &  $-$0.00000058 \\
\cline{1-4} \cline{5-8}
1   & $4f$ & 0.04984$^{\dag}$   &  $-$0.00000024 & 1 & $4f$ & 0.035241$^{\ddag}$  & $-$0.00000016  \\
2   & $4f$ & 0.099662 &  $-$0.00000014 & 2 & $4f$ & 0.0704820 & $-$0.00000064   \\
3   & $4f$ & 0.149493 &  $-$0.00000031 & 3 & $4f$ & 0.1057237 & $-$0.00000005  \\
4   & $4f$ & 0.199324 &  $-$0.00000056 & 4 & $4f$ & 0.1409649 & $-$0.00000019   \\
\cline{1-4} \cline{5-8}
1   & $5s$ & 0.04822$^{\dag}$  &  $-$0.00000024 & 1 & $5s$ & 0.02578$^{\ddag}$  &  $-$0.00000016   \\
2   & $5s$ & 0.09921 &  $-$0.00000022 & 2 & $5s$ & 0.051569 &  $-$0.00000065   \\
3   & $5s$ & 0.14991 &  $-$0.00000006 & 3 & $5s$ & 0.077357 &  $-$0.00000065   \\
4   & $5s$ & 0.20054 &  $-$0.00000017 & 4 & $5s$ & 0.103145 &  $-$0.00000024   \\
\cline{1-4} \cline{5-8}
1   & $5p$ & 0.04471$^{\dag}$   &  $-$0.00000001 & 1 & $5p$ & 0.025313$^{\ddag}$ &  $-$0.00000039 \\
2   & $5p$ & 0.090253 &  $-$0.00000007 & 2 & $5p$ & 0.05063  &  $-$0.00000068 \\
3   & $5p$ & 0.125506 &  $-$0.00000001 & 3 & $5p$ & 0.075946 &  $-$0.00000083  \\
4   & $5p$ & 0.18071  &  $-$0.00000063 & 4 & $5p$ & 0.101262 &  $-$0.00000064 \\
\cline{1-4} \cline{5-8}
1   & $5d$ & 0.03996$^{\dag}$   &  $-$0.00000002 & 1 & $5d$ &  0.024499$^{\ddag}$ &  $-$0.00000037 \\
2   & $5d$ & 0.08004  &  $-$0.00000081 & 2 & $5d$ &  0.049    &  $-$0.00000001 \\
3   & $5d$ & 0.120072 &  $-$0.00000007 & 3 & $5d$ &  0.0735   &  $-$0.00000006  \\
4   & $5d$ & 0.160097 &  $-$0.00000002 & 4 & $5d$ &  0.098    &  $-$0.00000010 \\
\cline{1-4} \cline{5-8}
1   & $5f$ & 0.03538$^{\dag}$   & $-$0.00000055 & 1 & $5f$ & 0.023482$^{\ddag}$  &  $-$0.00000008 \\
2   & $5f$ & 0.070778 & $-$0.00000023 & 2 & $5f$ & 0.046964  &  $-$0.00000035 \\
3   & $5f$ & 0.106168 & $-$0.00000008 & 3 & $5f$ & 0.0704464 &  $-$0.00000012  \\
4   & $5f$ & 0.141557 & $-$0.00000038 & 4 & $5f$ & 0.0939286 &  $-$0.00000006 \\
\cline{1-4} \cline{5-8}
1   & $5g$ & 0.031343$^{\dag}$   & $-$0.00000006 & 1 & $5g$ & 0.022371$^{\ddag}$   &  $-$0.00000029 \\
2   & $5g$ & 0.062687  & $-$0.00000007 & 2 & $5g$ & 0.0447428  &  $-$0.00000007 \\
3   & $5g$ & 0.09403   & $-$0.00000056 & 3 & $5g$ & 0.0671140  &  $-$0.00000056  \\
4   & $5g$ & 0.125374  & $-$0.00000024 & 4 & $5g$ & 0.0894856  &  $-$0.00000026 \\
\end{tabular}
\end{ruledtabular}
\begin{tabbing}
$^{\dag}$Literature results of $\lambda^{(c)}_{n,\ell}$ \cite{roy16a,diaz91}: (a) $\lambda^{(c)}_{3s}=0.1394$ (b) $\lambda^{(c)}_{4s}=0.07882$ 
(c) $\lambda^{(c)}_{4p}=0.067885$ (d) $\lambda^{(c)}_{4f}=0.049831$ \\
(e) $\lambda^{(c)}_{5s}=0.05058$ (f) $\lambda^{(c)}_{5p}=0.045186$ (g) $\lambda^{(c)}_{5d}=0.040024$ (h) $\lambda^{(c)}_{5f}=0.035389$ (i) $\lambda^{(c)}_{5g}=0.031343$. \\
$^{\ddag}$Literature results of $\lambda^{(c)}_{n,\ell}$ \cite{roy16a,singh83,diaz91}: (a) $\lambda^{(c)}_{3s}=0.072436$ (b) $\lambda^{(c)}_{4s}=0.040427$ (c) $\lambda^{(c)}_{4p}=0.039263$ 
(d) $\lambda^{(c)}_{4f}=0.035241$ \\ 
(e) $\lambda^{(c)}_{5s}=0.025787$ (f) $\lambda^{(c)}_{5p}=0.025315$ (g) $\lambda^{(c)}_{5d}=0.024500$ (h) $\lambda^{(c)}_{5f}=0.023482$ (i) $\lambda^{(c)}_{5g}=0.022371$.
\end{tabbing}
\end{table}  
\endgroup
 
\bibliography{ref}
\bibliographystyle{unsrt} 
\end{document}